\documentclass[11pt, journal, draftclsnofoot, onecolumn]{IEEEtran}

\usepackage{amssymb,enumitem,bm}
\usepackage{braket}
\usepackage[table,xcdraw]{xcolor}
 \usepackage[normalem]{ulem}
\useunder{\uline}{\ul}{}
\usepackage{graphicx,epsf,epsfig,verbatim,amssymb,amsmath,array,cite,amsthm,subfigure,multicol,multirow}
\usepackage{hyperref}
\usepackage{algorithm}
\usepackage{algorithmic}
\usepackage{caption,bbm}
\usepackage{bbm}
\usepackage[T1]{fontenc}
\usepackage{tikz}
\usepackage[scr=dutchcal]{mathalfa}
\let\mathscr\mathbscr
 \usepackage[normalem]{ulem}
 
\newtheorem{theorem}{Theorem}
\newtheorem{proposition}{Proposition}
\newtheorem{lemma}{Lemma}
\newtheorem{corollary}{Corollary}
\newtheorem{fact}{Fact}
\newtheorem{definition}{Definition}

\newtheorem{remark}{Remark}

\def\QNISS{\mathcal{Q}(P_{XY}, \mathcal{U}, \mathcal{V})}
\def\Qpartition{\mathcal{P}(P_{XY},Q_U,Q_V)}
\def\FBoolQUV{\mathcal{F}_{X Y}}
\def\ENISS{\mathcal{E}(P_{XY},Q_U,Q_V)}

\def\fSu{\mathtt{f}_{\mathcal{S}}}
\def\fS{{f}_{\mathcal{S}}}

\newcommand\pmm{\{-1,1\}}

\newcommand{\prob}[1]{\PP\Big( #1 \Big) }

\ifdefined \set 
  \renewcommand{\set}[1]{\left\{#1\right\}}
\else
  \newcommand{\set}[1]{\left\{#1\right\}}
\fi

\def\<{\langle}
\def\>{\rangle}

\def\deq{\triangleq }

\ifdefined \norm \else
 \newcommand{\norm}[1]{\lVert#1\rVert}
\fi

\def\gS{{g}_{\mathcal{S}}}

\def\ps{\phi_\mathcal{S}}
\def\pS{\ps}

\def\L2{\mathcal{L}^2(\mathcal{X}, P_{X^d})}

\def\E_mu{\mathcal{E}_{\mu}(\epsilon')}

\global\long\def\RR{\mathbb{R}}

\global\long\def\NN{\mathbb{N}}

\global\long\def\EE{\mathbb{E}}
\global\long\def\PP{\mathbb{P}}

\global\long\def\11{\mathbbm{1}}

\newcommand{\bfx}{\mathbf{x}}

\newcommand{\CS}{\mathcal{S}}

\newcommand{\CU}{\mathcal{U}}
\newcommand{\CV}{\mathcal{V}}
\newcommand{\CX}{\mathcal{X}}
\newcommand{\CY}{\mathcal{Y}}

\newcommand\numberthis{\addtocounter{equation}{1}\tag{\theequation}}

\begin{document}
\title{On Non-Interactive Simulation of Distributed Sources with Finite Alphabets}
\author{Hojat Allah Salehi  and Farhad Shirani
\\
Knight Foundation School of Computing and Information Sciences, Florida International University, Miami, FL, $\{$hsalehi,fshirani$\}$@fiu.edu}
\date{}

\maketitle
\thispagestyle{empty} \pagestyle{plain}
\begin{abstract}%
This work presents a Fourier analysis framework for the non-interactive source simulation (NISS) problem. Two distributed agents observe a pair of sequences $X^d$ and $Y^d$ drawn according to a joint distribution $P_{X^dY^d}$. The agents aim to generate outputs $U=f_d(X^d)$ and $V=g_d(Y^d)$  with a joint distribution sufficiently close in total variation  to a target  distribution $Q_{UV}$. Existing works have shown that the NISS problem with finite-alphabet outputs is decidable. For the binary-output NISS, an upper-bound to the input complexity was derived which is $O(\exp\operatorname{poly}(\frac{1}{\epsilon}))$. In this work, the input complexity and algorithm design are addressed in several classes of NISS scenarios. For binary-output NISS scenarios with doubly-symmetric binary inputs, it is shown that the input complexity is $\Theta(\log{\frac{1}{\epsilon}})$, thus providing a super-exponential improvement in input complexity. An explicit characterization of the simulating pair of functions is provided. For general finite-input scenarios, a constructive algorithm is introduced that explicitly finds the  simulating functions $(f_d(X^d),g_d(Y^d))$. The approach relies on  a novel Fourier analysis framework. Various numerical simulations of NISS scenarios with IID inputs are provided. Furthermore, to illustrate the general applicability of the Fourier framework, several examples with non-IID inputs, including entanglement-assisted NISS and NISS with Markovian inputs are provided.
\end{abstract}

\section{Introduction}

A critical resource, often required in distributed protocols enabling multiparty coordination,  is the availability of \textit{common or correlated random bit-strings} at each agent's terminal \cite{rabin1983randomized,bogdanov2011extracting,witsenhausen1975sequences,hirschfeld1935connection,gebelein1941statistische,renyi1959measures,maurer1993secret,anantharam2007common,kamath2016non,nielsen1999conditions,li2020boolean}. A well-known example is Rabin's randomized Byzantine generals  (RBG) protocol \cite{rabin1983randomized}, which requires a common binary symmetric variable shared among the distributed agents during the ``lottery phase'', which is crucial for secure and effective coordination in the presence of malicious actors.  More broadly, distributed correlated randomness is necessary for randomized consensus protocols \cite{cachin2000random,rabin1983randomized}, proof-of-stake based blockchain \cite{bentov2016snow,gilad2017algorand}, scaling smart contracts \cite{das2018yoda}, anonymous communication \cite{goel2003herbivore}, private browsing \cite{dingledine2004tor},  publicly auditable auctions and
lottery \cite{bonneau2015bitcoin}, and cryptographic parameter
generation \cite{baigneres2015trap}, among other applications. 

Sharing common binary strings among distributed agents, which is required in many multiparty protocols, results in communication overhead and latency issues, and represents significant obstacles when scaling distributed computing paradigms to massive networks \cite{wang2020randchain,gilad2017algorand}. Consequently, 
a natural question is whether, instead of common binary strings,  one could use \textit{correlated} binary strings, acquired via distributed observations of correlated signals, and apply some form of \textit{correlation inducing} local processing to produce common randomness. Witsenhausen's converse result provides a negative answer by
showing that distributed agents cannot coordinate \textit{perfectly} in performing a non-trivial binary task in the absence of common randomness \cite{witsenhausen1975sequences}. That is, there is a fundamental limit on the correlation in the agents' actions, which is determined by the correlation among their distributed observations. 
This gives rise to the problem of Non-Interactive Source Simulation (NISS), which studies the set of joint distributions that can be simulated non-interactively by distributed agents observing correlated pairs of sequences \cite{gacs1973common,witsenhausen1975sequences,ghazi2016decidability,ghazi2017dimension,shirani2017correlation,Heidari2019,yu2022common,bhushan2023secure,sudan2019communication,shirani2023non}.

\begin{figure}[!t]
\centering 
\includegraphics[width=0.6\textwidth]{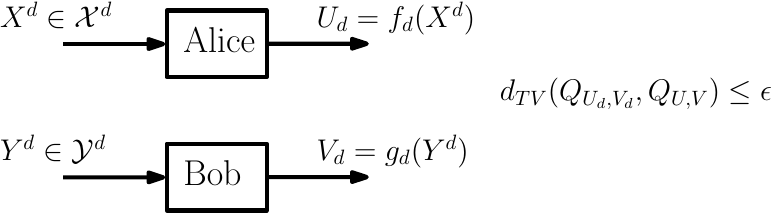}
\caption{The non-interactive source simulation scenario.
%The source sequences $X^d$ and $Y^d$ are IID with distributions $P_X$ and $P_Y$, respectively. . 
}
\label{fig:1}
\end{figure}

The NISS scenario considered in this work is shown in Figure \ref{fig:1}. Two distributed agents, Alice and Bob,  each observe a sequence of independent and identically distributed (IID) random variables,  $X^d\in \mathcal{X}^d$ and $Y^d\in \mathcal{Y}^d$, respectively, where $d\in \mathbb{N}$, $\mathcal{X},\mathcal{Y}$ are finite alphabets, and the underlying joint distribution is denoted by $P_{XY}$. The agents wish to non-interactively generate (simulate) random outputs $(U_d, V_d)\in \mathcal{U}\times\mathcal{V}$ with a joint distribution, $P_{U_dV_d}$, sufficiently close in total variation to a target distribution $Q_{UV}$.  The target distribution $Q_{UV}$ is said to be simulatable for an input distribution $P_{XY}$  if
\begin{equation}\label{eq:NISS}
\lim_{d\rightarrow \infty } \quad \min_{\substack{f_d:\mathcal{X}^d\to \mathcal{U}\\g_d:\mathcal{Y}^d\to \mathcal{V}}}d_{TV}(P_{U_dV_d},
Q_{UV})=0,
\vspace{-.1in}
\end{equation}
where $U_d\triangleq f_d(X^d), V_d\triangleq g_d(Y^d), d\in \mathbb{N}$, and $d_{TV}(\cdot,\cdot)$ is the total variation distance. The sequence of functions $(f_d,g_d)_{d\in \mathbb{N}}$ are called a \textit{simulating sequence of functions} for $Q_{UV}$ if the total variation distance between their output distribution $P_{U_dV_d}$ and the target distribution $Q_{UV}$ vanishes as $d$ grows asymptotically. An algorithm which takes the input and target distributions, $(P_{XY},Q_{UV})$, and finds the corresponding simulating sequence of functions, $(f_d,g_d)_{d\in \mathbb{N}}$, is called a \textit{simulating protocol}.

%i.e.,   $\abs{P_{UV}-Q_{UV}}_{TV}$.  
%is {close} to a target distribution $Q_{UV}$ with respect to an underlying distance measure, e.g., total variation.via 
%(possibly stochastic) functions $U=f(X^d)$ and $V=g(Y^d)$, for some $d\in \mathbb{N}$, such that the joint distribution $P_{UV}$ is {close} to a target distribution $Q_{UV}$ with respect to an underlying distance measure, e.g., total variation. 

%A natural question is whether given input distribution $P_{X,Y}$ and target distribution $Q_{UV}$, the target distribution can be produced in an NISS scenario with vanishing variational distance, i.e., whether a sequence of functions . In light of that, we are interested in characterizing the set of all feasible  target distributions; those $Q_{UV}$ for which \ref{eq:NISS} holds. %forsaid to be feasible if it can be produced from $(X^d,Y^d), d\in \mathbb{N}$ with vanishing total variation distance as $d$ grows asymptotically large. 

Prior works have investigated the following questions in the NISS problem:
\begin{itemize}[leftmargin=*]
    \item  \textbf{Decidability:} Is it possible for a Turing Machine to determine in finite time whether a target distribution $Q_{UV}$ can be simulated via a distributed input $(X^d,Y^d)\sim P_{X^d,Y^d}$? 
\item \textbf{Input Complexity:} given a desired total variation distance $\epsilon>0$, how many inputs samples $d\in \mathbb{N}$ are necessary to simulate the source? That is, to determine
\vspace{-.1in}
\[d_{\epsilon}\triangleq \min
\{d\in \mathbb{N}| \exists (f_d,g_d): U_d=f_d(X^d), V_d=g_d(Y^d), d_{TV}(P_{U_{d},V_{d}},
Q_{UV})<\epsilon.\}\]
\vspace{-.1in}
\item \textbf{Implementability:} Given $d\in \mathbb{N}$, an input distribution $P_{XY}$, and target distribution $Q_{UV}$, how to construct the simulating functions $(f^*_{d}(X^{d}),g^*_{d}(Y^{d}))$ to achieve the desired total variation distance? That is, to determine 
\[(f^*_{d},g^*_{d})\triangleq \arg\min_{(f_{d},g_{d})}d_{TV}(P_{U_{d},V_{d}},
Q_{UV}),\]
where $U_{d}\triangleq f_{d}(X^{d}),V_{d}\triangleq g_{d}(Y^{d})$.
\end{itemize}

These questions have been studied under various assumptions on input distributions and input and output alphabets.  Witsenhausen   \cite{witsenhausen1975sequences} studied  the scenario for doubly-symmetric binary-output NISS (i.e., $Q_U(1)=Q_V(1)=\frac{1}{2}$), as well as scenarios where $(U,V)$ are jointly Gaussian (i.e., $Q_{UV}$ is a Gaussian measure on $\RR^2$), and derived necessary and sufficient conditions on the correlation coefficients of $(U,V)$ and $(X,Y)$  under which $Q_{UV}$  is simulatable for $P_{XY}$. In both cases, it was shown that any joint distribution $Q_{UV}$ whose correlation is less than the Hirschfeld-Gebelein-R\'enyi maximal correlation coefficient of $P_{XY}$ \cite{hirschfeld1935connection,gebelein1941statistische,renyi1959measures} can be produced at the distributed terminals. Furthermore, Witsenhausen showed that for any $\epsilon > 0$, Alice and Bob can simulate the Gaussian measure $Q_{UV}$ up to a desired variational distance $\epsilon$ with $d = \operatorname{poly}(|\mathcal{X}|, |\mathcal{Y}|, \log(1/\epsilon))$. Additionally, an explicit algorithm to construct $f_d(\cdot)$ and $g_d(\cdot)$ with run-time $\operatorname{poly}(d)$ was provided, thus answering the decidability, input complexity, and implementability questions and solving the Gaussian  NISS problem completely.
However, the discrete-output problem remains open. 

%An algorithm for constructing the corresponding mapping pair $(f(\cdot),g(\cdot))$ was provided, and it was shown that for jointly Gaussian target distributions,  
More recently, a set of impossibility results for discrete-output NISS were introduced in \cite{kamath2016non}, where 
hypercontractivity techniques were used to provide necessary conditions for the simulatability of $Q_{UV}$ for a given $P_{XY}$. These impossibility results were further improved upon in \cite{li2020boolean,shirani2023non}.
The decidability of NISS with finite alphabet outputs was studied in \cite{ghazi2016decidability,de2018non}. Particularly, it was shown that given $P_{XY}$ and $Q_{UV}$, the input complexity is $O(\exp \operatorname{poly}(\frac{1}{\epsilon}, \frac{1}{1-\rho_{XY}}))$, where $\rho_{XY}$ is the input maximal correlation. As a result, by using a brute-force search over the set of all simulating functions, a Turing machine can decide whether $Q_{UV}$ is simulatable for a distribution $P_{XY}$ with a run-time complexity\footnote{It should be noted that the main focus of \cite{ghazi2016decidability} was on proving decidability, and the brute-force algorithm is only discussed as an intermediate step in the decidability proof.}
\begin{align*}
O(\exp \exp \exp \operatorname{poly}(\frac{1}{\epsilon}, \frac{1}{1-\rho_{XY}}, \log \frac{1}{\alpha} )),    
\end{align*}
 where $\rho_{XY}$ is the maximal correlation of $P_{XY},$ and $\alpha$ is the minimum non-zero value of $P_{XY}$.  

In this work, we study several classes of NISS scenarios, and answer the input complexity and implementability questions for each class. In particular, we make the following contributions: 
\begin{itemize}[leftmargin=*]
    \item \textbf{Doubly-Symmetric Input, Binary-Output NISS:} For the scenario where the input is a doubly-symmetric  binary source  (i.e., $P_X(1)=P_Y(1)=\frac{1}{2}$) and the output is binary,  we show that the input complexity is $\Theta(\log{\frac{1}{\epsilon}})$, thus achieving a super-exponential improvement. Furthermore, we provide an explicit characterization of the simulating functions for a given input distribution $P_{XY}$ and target distribution $Q_{UV}$. (Theorem \ref{th:IC_BB_NISS})
    \item \textbf{Finite-input, Binary-Output NISS:} For finite-input, binary-output NISS (FB-NISS), we show that the implementability question can be reduced to the problem of finding the simulating functions achieving maximal correlation (Proposition \ref{prop:Sol_BB_NISS}).  Furthermore, we provide a primal and dual formulation of the maximal correlation problem (Theorems \ref{prop:primal} and  \ref{prop:dual}). The dual formulation is only applicable to inputs with uniform marginals, and yields a linear program which finds the pair of simulating functions for a given (simulatable) target distribution.
    As for the primal formulation, we introduce a Fourier-path following algorithm (F-PATH) to find the pair of simulating functions for an input distribution with non-uniform marginals and a simulatable target distribution (Algorithm \ref{alg:1}).
    \item \textbf{Finite-input, Finite-Output NISS:} In this general scenario, we show that the set of simulatable distributions can be decomposed into a union of star-convex sets. Consequently, we introduce the notion of directional maximal correlation, where direction is defined with respect to the center of the start-convex set, and show that the implementability question can be reduced to the problem of finding the simulating functions achieving  directional maximal correlation (Proposition \ref{prop:Sol_BF-NISS}). For inputs with uniform marginals, the dual formulation of Theorem \ref{prop:dual} can be used to find the simulating functions using a linear program. For non-uniform input marginals, the F-PATH algorithm can be used to find the simulating pair of functions (Algorithm \ref{alg:1}).   
    \item {\textbf{NISS Scenarios with Non-IID Inputs:}} The main focus of this work is to find solutions to the input complexity and implementability questions in NISS scenarios with IID inputs. However, the Fourier framework developed in the sequel is generally applicable to problems involving quantification of distributed correlation. To illustrate this, we provide several examples including an example with independent input sequences which are not identically distributed (Case Study 1 in Section \ref{subsec:3.4}), entanglement-assisted NISS (Case Study 2 in Section \ref{subsec:3.4}), and NISS scenarios with Markovian Sources (Case Study 3 in Section \ref{subsec:3.4}). As described in the subsequent sections, the application of the Fourier framework hinges on the orthonormality of the underlying function basis which is used to perform the Fourier expansion. In non-IID-input NISS scenarios, such orthonormal basis can be constructed by starting from the standard basis used for Fourier expansion in IID-input NISS, and then applying the Gram-Schmidt procedure, hence making the framework applicable to such scenarios. 
\end{itemize}

\noindent \textit{Other Related Works:} The NISS problem has been studied extensively under various scenarios. A comprehensive survey of relevant problems and their connections to NISS is given in \cite{yu2022common}. The decidability of NISS for finite output sets is presented in \cite{de2018non}. Further impossibility results for binary input and binary-output are introduced in \cite{yu2021non} that improves upon the hypercontractivity-based bounds in \cite{kamath2016non}. These derivations rely on the Fourier expansion of the functions $(f(\cdot),g(\cdot))$ over the Boolean cube. Other important variants of  the NISS problem are studied in G\'acs and K\"orner \cite{gacs1973common}, and Wyner \cite{wyner1975common}.  Decidability and solvability of secure NISS have also been studied in the literature \cite{Kiltz2023,Khorasgani2021,agarwal2022secure}.
The Fourier expansion on the Boolean cube  has been widely used in computational learning, analysis of Boolean functions,  and quantifying correlation among distributed functions \cite{mossel2004learning,courtade2014boolean,shirani2017correlation,shirani2019sub,yu2021non,ghazi2016decidability}.

\noindent \textit{Organization:} 
Section \ref{sec:prelim} provides the notation, problem formulation, and a brief background on the Fourier analysis techniques used throughout the paper. Section \ref{sec:Fourier_Tensor_Derand} explains the tools that we have developed, including extensions of known Fourier expansion, randomization, and derandomization techniques, to derive our main results. Section \ref{sec:MC} introduces  the notion of biased and directional maximal correlation and shows their sufficiency for solving NISS. Section \ref{sec:PD} presents our main results including the primal and dual optimizations for solving the NISS problem. Section \ref{sec:examples} provides several examples of NISS problems with non-IID input sequences to illustrate the general applicability of the Fourier framework. Section \ref{sec:FPATH} describes the F-PATH algorithm.  Section \ref{sec:IC_BB} derives a tight bound on the input complexity of BB-NISS with uniform input marginals. Section \ref{sec:sim} provides numerical simulations of the F-PATH algorithm. Section \ref{sec:conc} conlcudes the paper.

\section{Problem Formulation and Preliminaries}
\label{sec:prelim}
\noindent \textit{Notation.}
%The random variable $\mathbbm{1}_{\mathcal{E}}$ is the indicator function of the event $\mathcal{E}$.
 The set $\{1,2,\cdots, d\}$ is represented by $[d]$. For a given prime number $q\in \mathbb{N}$, the finite field of order $q$ is denoted by $\mathbb{F}_q$. 
The  vector $(x_1,x_2,\hdots, x_d)$ is written as $x^d$, and for a given $1\leq i\leq j\leq d$, the subvector $(x_i,x_{i+1},\cdots,x_j)$ is written as $x_{i}^j$. Sets are denoted by calligraphic letters such as $\mathcal{X}$. For the event $\mathcal{E}$, the variable $\mathbbm{1}(\mathcal{E})$ denotes the indicator of the event, i.e., $\mathbbm{1}(\mathcal{E})=1$ if $\mathcal{E}$ and $\mathbbm{1}(\mathcal{E})=-1$, otherwise. For a given alphabet $\mathcal{X}$, the notation $\Delta_{\mathcal{X}}$ represents the probability simplex on $\mathcal{X}$. $Conv(\cdot)$ represents the convex-hull.  Also, $\sigma_X, \sigma_Y$ are the standard deviation of $X$ and $Y$, respectively. By $P_{XY}^d$, denote the d-fold product distribution, i.e. $P^d_{X,Y}(x^d,y^d)=\prod_{i=1}^dP_{X Y}(x_i,y_i), x^d,y^d\in \mathcal{X}^d\times \mathcal{Y}^d$. 
Furthermore, we use the following functions and notations throughout the paper:
\begin{itemize}
    \item $d_{TV}(P,Q)$ represents the variational distance between distributions $P$ and $Q$ defined on a shared alphabet $\mathcal{X}$, i.e., $d_{TV}(P,Q)\triangleq \sum_{x\in \mathcal{X}}|P(x)-Q(x)|$.
    \item The Pearson correlation coefficient between a pair of random variables $X$ and $Y$  is defined as $\rho_{XY} \triangleq  \mathbb{E}\left[\frac{(X-\mu_X)(Y-\mu_Y)}{\sigma_X\sigma_Y}\right]$, where $\mu_X, \mu_Y$ are the expected value of $X$ and $Y$, respectively.
    \item  The notation $f\equiv h$  for any generic functions $f: \CX \rightarrow \mathcal{A}$ and $h:\CX \to \mathcal{B}$ means the two functions are equal up to a bijection. More precisely, there is an invertible map $\Gamma: \mathcal{A} \to\mathcal{B} $ such that $h(x) = \Gamma(f(x))$. 
\end{itemize}

\subsection{The Non-interactive Source Simulation Problem}
The NISS problem is formally defined below:
\begin{definition}[NISS] \label{def:NISS}
Let $P_{XY}$ be a probability measure on the finite set $\mathcal{X}\times\mathcal{Y}$  and  $Q_{UV}$ be the target (output) probability distribution over the finite set $\mathcal{U}\times\mathcal{V}$.  The distribution $Q_{UV}$ is said to be non-interactively simulated (simulatable) using $P_{XY}$ if there exists a sequence of functions $f_d:\mathcal{X}^d \to \mathcal{U}$ and $g_d:\mathcal{Y}^d\to \mathcal{V}$, with $d\in \NN$, such that 
\begin{align}\label{eq:NISS TV}
    \lim_{d\rightarrow \infty } d_{TV}(P_{U_dV_d}, Q_{UV})=0,
\end{align}
where  $U_d=f(X^d)$ and $V_d=g(Y^d)$ with $(X^d, Y^d)\sim P_{XY}^d$, and  $P_{U_dV_d}$ is the joint distribution of $(U_d, V_d)$.  Given, $P_{XY}$, the set of all non-interactively simulatable distributions $Q_{UV}$ is denoted by $\QNISS$.
\label{def:1}
\end{definition}
\begin{remark}
 One might consider a more general formulation involving non-deterministic mappings instead of deterministic $(f_d, g_d)$. However, any simulatable  target $Q_{UV}$ with stochastic mappings is also simulatable with deterministic functions $(f_d, g_d)$ and local randomness distilled via copies of the  samples $X$ and $Y$, respectively.  Hence, the restriction to determinsitic mappings does not loose generality.
\end{remark}

\begin{definition}[NISS with threshold $\epsilon$]
    Given $\epsilon>0$ and probability distributions  $P_{XY}$ over $\mathcal{X}\times\mathcal{Y}$  and  $Q_{UV}\in \mathcal{Q}(P_{XY},\mathcal{U},\mathcal{V})$, the NISS problem $(P_{XY},Q_{UV},\epsilon)$ is the problem of finding the parameter $d\in \mathbb{N}$ and pair of functions $f_d:\mathcal{X}^d \to \mathcal{U}$ and $g_d:\mathcal{Y}^d\to \mathcal{V}$ such that  $d_{TV}(P_{U_dV_d}, Q_{UV})\leq \epsilon,$ where  $U_d=f(X^d)$ and $V_d=g(Y^d)$ with $(X^d, Y^d)\sim P_{XY}^d$, and  $P_{U_dV_d}$ is the joint distribution of $(U_d, V_d)$.
\end{definition}
%The focus of this paper is on the  NISS scenarios with finite input and output alphabet ($|\CX|, |\CY|, |\CU|, |\CV|<\infty$). Our approach relies on the Fourier analysis tools described in the following. 

\subsection{Discrete Fourier Expansion}
\label{subsec:Fourier}
The uniform Fourier expansion on the Boolean cube has been well-studied and applied in various problems over the past decades (e.g., \cite{o2014analysis,Wolf2008}). It decomposes any real-valued function $f:\pmm^d\rightarrow \RR$ on the Boolean cube  as 
$f(\bfx)=\sum_{\mathcal{S}\subseteq [d]} \fSu \chi_S(x^d), x^d\in \{-1,1\}^d$ 
    where $\chi_S(x^d) \triangleq \prod_{j\in \CS} x_j$ for all $\CS \subseteq [d]$, and $\fSu \in \RR$ are called the Fourier coefficients of $f$ and are calculated as $\fSu \triangleq \frac{1}{2^d}\sum_{x^d\in \pmm^d} \fSu \chi_\CS(x^d)$. 
    
   This Fourier expansion can be extended to general product probability spaces. To elaborate, given a distribution $P_X$ over a finite field $\mathbb{F}_q$\footnote{For $q>2$, we consider the field $\mathbb{F}_q$ as the set $\{0,1,\cdots,q-1\}$ equipped with modulo addition and multiplication. For $q=2$, we consider the set $\{-1,1\}$ which reduces clutter and leads to more concise derivations.}, let us consider the vector space $\mathcal{L}^d_X$ of real-valued functions over $\mathbb{F}_q^d$. The inner-product between two functions  $f,g$ is defined as $\<f, g\> \deq \EE_{X^d\sim P_X^d}[f(X^d)g(X^d)]$. Then, the following defines the stochastic discrete Fourier extension over this probability space.

\begin{fact}[Stochastic Discrete Fourier Expansion \cite{o2014analysis}]\label{fact:fourier_2}
   Let $P^d_X$ be a product probability measure on the field $\mathbb{F}_{q}$, where $q$ is prime. The function $f:\mathbb{F}_{q}^d\rightarrow \RR$ decomposes as 
   % \begin{align*}
 $f_d(x^d)=\sum_{s^d\in \mathbb{F}^d_q}
 f_{s^d}~\phi_{s^d}(x^d), \quad \text{for all}~ x^d \in \mathbb{F}_q^d,$
 %\end{align*}
    where $\phi_{s^d}(x^d)$ is the parity associated with the vector $s^d \in \mathbb{F}_q^d$, defined as
$\phi_{s^d}(x^d)\triangleq \prod_{i\in [d]}\psi_{s_i}(x_i)
,  x^d \in \mathbb{F}_q^d$,
and $\psi_{s}(\cdot), s\in \mathbb{F}_q$ is an orthonormal basis for the space of functions $f:\mathbb{F}_q\to \mathbb{R}$ equipped with probability measure $P_X$. Moreover, $\fS$ are the Fourier coefficients of $f(\cdot)$ and are calculated as $f_{s^d}=\EE_{X^d\sim P_X^d}[f(X^d)\phi_{s^d}(X^d)]$ for all $s^d\in \mathbb{F}_q^d$.
\end{fact}

In particular, the Stochastic Boolean Fourier expansion can be further simplified by choosing the orthonormal basis elements $\psi_{-1}(x)= 1$ and $\psi_{1}(x)=\alpha x-\beta $ for appropriately chosen constants $\alpha,\beta$. This is explained below. 
\begin{fact}[Stochastic Boolean Fourier Expansion \cite{o2014analysis}]\label{fact:fourier}
   Let $P^d_X$ be a product probability measure on $\pmm^d$. The function $f:\pmm^d\rightarrow \RR$ decomposes as 
   % \begin{align*}
 $f_d(x^d)=\sum_{\mathcal{S}\subseteq [d]} \fS~\ps(x^d), \quad \text{for all}~ x^d \in \pmm^d,$
 %\end{align*}
    where $\pS(x^d)$ is the parity associated with a subset $\mathcal{S}\subseteq [d]$, defined as:
    \begin{align*}
\pS(x^d)\deq \prod_{i\in \mathcal{S}}\frac{x_i-\mu_X}{\sigma_X},\qquad  x^d \in \{-1,1\}^d,
\end{align*}
where $\mu_X$ and $\sigma_X$ are the expected value and standard deviation of $X\sim P_X$, respectively. Moreover, $\fS$ are the (biased) Fourier coefficients of $f(\cdot)$ and are calculated as $\fS=\EE_{X^d\sim P_X^d}[f(X^d)\pS(X^d)]$ for all $\CS\subseteq [d]$.
\end{fact}

Note that due to the orthonormality of $\psi_{s}(\cdot), s\in \mathbb{F}_q$ in Fact \ref{fact:fourier_2}, the parity functions $\phi_{s^d}, s^d\in \mathbb{F}_q^d$ form an orthonormal basis for this vector space $\mathcal{L}^d_X$ . Particularly,  $\<\phi_{s^d} ,\phi_{t^d}\> = \mathbbm{1}({s^d}={t^d})$, where ${s^d},{t^d}\in \mathbb{F}_q^d$.  The following facts summarize some basic properties of the Fourier expansion.
\begin{fact}[\cite{o2014analysis}]\label{fact:fourier1}
For any bounded pair of functions $f,g: \mathbb{F}_q^d\to \RR$, the following hold: 
\begin{itemize}
\item Plancherel Identity: $\EE[f(X^d)g(X^d)]=\sum_{s^d\in  \mathbb{F}_q^d} f_{s^d} g_{s^d} $. 
\item Parseval’s identity: $\norm{f}_2^2=\sum_{s^d\in  \mathbb{F}_q^d} f_{s^d}^2$.
\end{itemize}
\end{fact}

%\paragraph{Fourier Expansion Pairs.} To study NISS we consider joint pairs of functions $f_d(X^d)$ and $g_d(Y^d)$ and their Fourier expansion with respect to the marginal measures $P_X$ and $P_Y$, respectively. We elaborate on this expansion in Section \ref{sec:Fourier_Tensor_Derand}.

  % \begin{fact}\label{fact:disg prob binary}
  % \label{fact:4}
  % Given Boolean functions $f(\cdot),g(\cdot)$ and a joint probability measure $P_{XY}$, the following holds
  %     \begin{align}\label{eq:1}
  %       P(f(X^d)\neq g(Y^d))= \frac{1}{2}-\frac{1}{2} \sum_{\mathcal{S}\subseteq [d]} \fS \gS  \rho^{|\mathcal{S}|}.
  % \end{align}
  % \end{fact}
\section{A Fourier Framework for Quantifying Distributed Correlation}
\label{sec:Fourier_Tensor_Derand}
  \subsection{Functions as Overparametrized Vectors}\label{subsec:tensor}
As discussed in the prequel, our objective is to apply the Fourier expansion techniques of Section \ref{subsec:Fourier} to study pairs of distributed functions operating on correlated input sequences. 
In Section \ref{subsec:F2}, we develop Fourier analysis techniques for pairs of distributed functions in different probability spaces. These techniques are only directly applicable to binary-output functions. 
 To make them applicable to non-binary-output functions, we introduce an \textit{overparameterized vector representation} of such functions in terms of vectors of binary-output functions. Let $f:\CX^d\rightarrow \CU$ be a generic function with $\CU$ a finite set. Then, for any $u\in \CU$ define  the indicators $f_{u}(x^d)\triangleq 2 \11(f(x^d)=u)-1$. Note that $f_{u}(x^d)= 1$ if $f(x^d)=u$; otherwise  $f_{u}(x^d)= -1$. The indicator functions are sufficient to determine the output of $f(\cdot)$. We write $f\equiv  (f_u)_{u\in \mathcal{U}}$ to denote the vector of the indicator functions. 
Note that $f_u, u\in \CU$ satisfy the following conditions: 
\begin{align}
    &\label{eq:c1:0}
    \text{Condition (1):}\qquad &f_u(x^d)\in \{-1,1\},~~ \forall x^d\in \CX^d \\
    &\label{eq:c2:0}\text{Condition (2):}\qquad  &\sum_{u}f_u(x^d) =2-|\CU|, ~~ \forall x^d\in \CX^d.
\end{align}
Consequently, in the sequel, we often focus on binary-output functions, and apply the resulting solutions
to general finite-output alphabets using the above overparametrized representation.

\subsection{Fourier Expansion and Disagreement Probability}
\label{subsec:F2}
 The Fourier expansion is closely related to the probability of disagreement between the outputs of pairs of functions as shown in the following.

% \begin{lemma}
% \label{Lem:6}
%      Let $P_{U|X^d}:\CX^d \rightarrow \Delta_{|\CU|}$ and $P_{V|Y^d}:\CY^d \rightarrow \Delta_{|\CV|}$ be  two channels with $\CX=\CY=\CU=\CV=\pmm$. Then 
% %    \begin{align*}
%         \[\EE(UV) =\sum_{\mathcal{S}\subseteq [d]} u_\CS v_\CS  \rho^{|\mathcal{S}|},\]
%    % \end{align*}
%     where $u_\CS=\EE[U\pS(X^d)]$ and  $v_\CS=\EE[V\psi_\CS(Y^d)]$, and $\pS=\prod_{i\in \CS}\frac{X_i-\mu_X}{\sigma_X}$ and $\psi_\CS=\prod_{i\in \CS}\frac{Y_i-\mu_Y}{\sigma_Y}$ are the corresponding parity functions defined for the marginal spaces. 
% \end{lemma}

\begin{lemma}\label{cor:Fourier functions}
    Let $P_{XY}$ be a probability measure over $\mathbb{F}_q\times \mathbb{F}_q$. For any bounded pair of functions $f,g: \mathbb{F}_q\mapsto \RR$, the following holds 
\begin{align}\label{eq: Efg}
     & \mathbb{E}_{(X^d,Y^d)}[f(X^d)g(Y^d)]= \sum_{s^d,t^d\in \mathbb{F}_q^d} f_{s^d}g_{t^d}\prod_{s,t\in \mathbb{F}_q} \rho_{s,t}^{n(s,t|s^d,t^d)},
  \end{align}
  where $\rho_{s,t}= \mathbb{E}_{X,Y}[\psi_{s}(X)\psi'_t(Y)], s,t\in \mathbb{F}_q$ is the correlation coefficient between $\psi_{s}(X)$ and $\psi'_t(Y)$ under $P_{XY}$, $(\psi_s, s\in \mathbb{F}_q)$ and $(\psi'_t, t\in \mathbb{F}_q)$ are orthogonal basis for the space of functions $\mathcal{L}_X$ and $\mathcal{L}_Y$, respectively, 
   $(f_{s^d},g_{s^d}), s^d\in \mathbb{F}_q^d$ are the Fourier coefficient of $f$ with respect to $P_X$ and $g$ with respect to $P_Y$, respectively, and $n(s,t|s^d,t^d)\triangleq \sum_{i\in [d]}\mathbbm{1}(s_i=s,t_i=t)$.
\end{lemma}
The proof of Lemma \ref{cor:Fourier functions} follows by applying linearity of expectation, the fact that the probability space is a product probability space, and using the definition of Fourier expansion for finite input alphabets given in Fact \ref{fact:fourier_2}.

Note that for Stochastic Boolean Fourier expansion, we have $\psi_{-1}(X)=1$ and $\psi_1(X)=\frac{X-\mu_X}{\sigma_X}$. Consequently,
\[ \forall s,t\!\in \{-1,1\}\!:\! \rho_{s,t} =
\begin{cases}
    1 \quad\! & s=t=-1,\\
    \rho_{XY} & s=t=1,\\
    0& \text{otherwise}
\end{cases}\quad \!\!
\Rightarrow  \forall s^d,t^d\in \{-1,1\}^d\!:\!\rho_{s^d,t^d}= \rho_{XY}^{w_H(s^d)} \mathbbm{1}(s^d=t^d),\] where $w_H(\cdot,\cdot)$ denotes the Hamming weight. Consequently, for Boolean functions, the following simplified result can be derived.

\begin{corollary}\label{cor:bn_Fourier functions}
    Let $P_{XY}$ be a probability measure over $\pmm^d\times \pmm^d$. For any bounded pair of functions $f,g: \pmm^d\mapsto \RR$, the following holds 
\begin{align}\label{eq: bn_Efg}
     & \mathbb{E}_{(X^d,Y^d) \sim P_{XY}^d}[f(X^d)g(Y^d)]= \sum_{\mathcal{S}\subseteq [d]} \fS \gS  \rho_{XY}^{|\mathcal{S}|},
  \end{align}
  where $\rho_{XY}$ is the correlation coefficient defined for $P_{XY}$,  $\fS$ is the Fourier coefficient of $f$ with respect to $P_X$, and $\gS$ is that of $g$ with respect to $P_Y$.
\end{corollary}

The following lemma provides a relation between $\mathbb{E}(f_{u}(X^d)g_{v}(Y^d))$ and $\PP(U_d=u, V_d=v)$. In the subsequent sections, we write $\mathbb{E}(f_{u}(X^d)g_{v}(Y^d))$ in terms of the Fourier coefficients of $f_u$ and $g_v$, which allows us to forumalte the NISS problem in terms of an optimization over the Fourier coefficients. The proof is provided in Appendix \ref{App:Lem:4}. 
\begin{lemma}\label{lem:Exp f_ug_v}
  Consider any pair of functions $f:\CX^d\rightarrow \CU$ and $g:\CY^d\rightarrow \CV$. Let $f\equiv (f_u)_{u\in \mathcal{U}}$ and $g\equiv (g_v)_{v\in \mathcal{V}}$ be the overparametrized vector representation of $f$ and $g$, respectively. Let $U_d=f(X^d)$ and $V_d=g(Y^d)$. Then,   
  \begin{equation*}
      \mathbb{E}(f_{u}(X^d)g_{v}(Y^d)) = 4\PP(U_d=u, V_d=v)-2\PP(U_d=u)-2\PP(V_d=v)+1, \quad u,v \in \mathcal{U}\times \mathcal{V}.
  \end{equation*}
\end{lemma}

 % This paper extends the above expression to  general but finite output alphabets $\CU\times \CV$. For that, we use binary indication functions. For any function $f:\{-1,1\}^d\to \mathcal{U}$, define its indications as 
 % \begin{equation}
 %     f_{u}(x^d) =\begin{cases}
 %                  1 & \text{if}~ f(x^d)=u\\
 %                  -1 & o.w.
 %                \end{cases}
 % \end{equation}
 % for all $u\in \mathcal{U}$. Similarly, define the indicators $g_v, v\in \CV$ of a given function $g:\{-1,1\}^d\to \mathcal{V}$.  

 % Let $U_d=f(X^d)$ and $V_d=g(Y^d)$. The indication functions are sufficient to characterize the output distribution $P_{U_d, V_d}$. 

 Note that the functions $f_u,g_v$ are binary-output functions. As a result, the joint distribution $\PP(f_{u}(X^d)=a,  g_{v}(Y^d)=b), a,b\in \set{-1,1}$ has three degrees of freedom and is completely characterized by the marginals $\prob{f_u(X^d)=1},\prob{g_v(Y^d)=1}$ and $\prob{f_u(X^d)=1,g_v(Y^d)=1}$. These quantities are in turn uniquely determined by   $\EE[f_{u}(X^d)]=2P_{U_d}(u)-1, \EE[g_{v}(Y^d)]=2P_{V_d}(v)-1$ and  $\EE[f_{u}(X^d) g_v(Y^d)]$. This is formalized in the following proposition.
 
\begin{proposition}
\label{prop:5}
   Let $U_d=f(X^d)$ and $V_d=g(Y^d)$, where $(X^d, Y^d)\sim P_{XY}^d, d\in \NN$. The vector $(\EE[f_{u}(X^d) g_v(Y^d)], u\in \mathcal{U}, v\in \mathcal{V})$ and marginals $P_{U_d}, P_{V_d}$ uniquely characterize the output distribution $P_{U_d, V_d}$.
\end{proposition}
% \begin{proof}
%     Let us fix $u,v\in \mathcal{U}\times \mathcal{V}$. Note that the functions $f_u,g_v$ are binary-output functions. As a result, the joint distribution $\PP(f_{u}(X^d)=a,  g_{v}(Y^d)=b), a,b\in \set{-1,1}$ has three degrees of freedom and is completely characterized by the marginals $P_{f_u(X^d)},P_{f_v(Y^d)}$ and $\PP(f_u(X^d)=1,f_v(Y^d)=1)$. These quantities are in turn uniquely determined by   $\EE[f_{u}(X^d)]=P_{U_d}(u), \EE[g_{v}(Y^d)]=P_{V_d}(v)$ and  $\EE[f_{u}(X^d) g_v(Y^d)]$ (Lemma \ref{lem:Exp f_ug_v}).
%     %From Fact \ref{fact:4}, $\EE[f_{u}(X^d) g_v(Y^d)]$ can be written as a function of the Fourier coefficients  of  $f_{u}$ and $g_v$.
%  %    On the other hand,  
%  % \begin{align*}
%  %     P_{U_d, V_d}(u,v) &= \EE[\11(U_d=u, V_d=v)]  = \prob{ f_{u}(X^d)=1,  g_{v}(Y^d)=1} = \EE[f_u(X^d)g_v(Y^d)]\\
%  %     &=\sum_{\CS\in[d]} \fSset{u, \CS}\gSset{v, \CS} \rho^{|\CS|}.
%  % \end{align*}
% This completes the proof. 
% \end{proof}
 As a result of Proposition \ref{prop:5}, characterizing the set of simulatable distributions $Q_{UV}$ as in Definition \ref{def:NISS} is equivalent to characterizing the set of feasible values for the vector ($\mathbb{E}(f_{u}(X^d)g_{v}(Y^d)), u\in \mathcal{U}, v\in \mathcal{V})$.  
% \begin{corollary}\label{lem:Fourier channel}
%     Let $P_{U|X^d}:\CX^d \rightarrow \Delta_{|\CU|}$ and $P_{V|Y^d}:\CY^d \rightarrow \Delta_{|\CV|}$ be  two channels with $\CX=\CY=\CU=\CV=\pmm$. The disagreement probability between the output variables $U$ and $V$ decomposes as 
%     \begin{align*}
%         \prob{U\neq V} = \frac{1}{2}-\frac{1}{2} \sum_{\mathcal{S}\subseteq [d]} u_\CS v_\CS  \rho^{|\mathcal{S}|},
%     \end{align*}
%     where $u_\CS=\EE[U\pS(X^d)]$ and  $v_\CS=\EE[V\pS(Y^d)]$
% \end{corollary}
% The proof of this corollary follows from the fact that for any pair of binary random variables $UV$ over $\pmm$, the identity $P(U\neq V)=\frac{1}{2}-\frac{1}{2}\EE(UV)$ holds.

\subsection{Non-Convexity of $\QNISS$ and Decomposition into Star-Convex Sets}\label{sec:SC}
\label{sec:stc}
The set of simulatable distributions $\QNISS$ is not a convex set in general. To see this, note that the extreme points of the probability simplex $\Delta_{\mathcal{U}\times \mathcal{V}}$ are always in $\QNISS$. To elaborate, for any given $(u^*,v^*)\in \mathcal{U}\times \mathcal{V}$, 
 Alice and Bob can produce deterministic outputs $U,V$ such that $P_{UV}(u,v)= \mathbbm{1}(u=u^*,v=v^*), u,v\in \mathcal{U}\times \mathcal{V}$. 
 Therefore, the convex hull of $\QNISS$ equals the set of all probability distributions on $\mathcal{U}\times \mathcal{V}$. For instance, for binary-output NISS with $\mathcal{U}=\mathcal{V}=\{-1,1\}$,
 the convex hull of  $\QNISS$  includes $Q_{U,V}(u, v)=\frac{1}{2}\11(u=v)$. However, this distribution is simulatable if and only if the correlation coefficient of $(X,Y)$ satisfies $|\rho_{XY}|=1$ \cite{witsenhausen1975sequences}. Hence, in general, when $|\rho_{XY}|\neq 1$, the set $\QNISS$ is not convex.

 The fact that  $\QNISS$ is not a convex set makes its characterization challenging. Consequently, we decompose $\QNISS$ into a union of star-convexed sets, which is more amiable to analysis. To elaborate, let us define
 \begin{equation}\label{eq:Q NISS marginals}
    \mathcal{P}(P_{XY},Q_U,Q_V)\triangleq \{Q'_{UV}\in \QNISS| Q'_U=Q_U, Q'_V=Q_V\}.
\end{equation}
The collection $\mathcal{P}(P_{XY},Q_U,Q_V), Q_U\in \Delta_{\mathcal{U}}, Q_V\in \Delta_{\mathcal{V}}$ partitions the set
$\QNISS$. That is, 
\begin{align*}
  & \QNISS= \bigcup_{Q_U,Q_V}  \mathcal{P}(P_{XY},Q_U,Q_V),
   \\&\mathcal{P}(P_{XY},Q_U,Q_V)\cap \mathcal{P}(P_{XY},Q'_U,Q'_V)=\phi, \quad  \forall (Q_U,Q_V)\neq (Q'_U,Q'_V)
\end{align*}

\begin{lemma}[\textbf{Star-Convexity of $\mathcal{P}(P_{XY},Q_U,Q_V)$
\footnote{This result and the accompanying proof was partially presented in the conference version \cite{shirani2023non}.}
}]
\label{lem:8}
Given finite alphabets $\mathcal{X},\mathcal{Y},\mathcal{U}$ and $\mathcal{V}$, and distributions $P_{XY}, Q_U,Q_V$ defined on $\mathcal{X}\times \mathcal{Y}$, $\mathcal{U}$, and $\mathcal{V}$, respectively, the set     $\mathcal{P}(P_{XY},Q_U,Q_V)$ defined in Equation \eqref{eq:Q NISS marginals} is star-convex. 
\end{lemma}
The proof is provided in Appendix \ref{App:lem:8}.

\subsection{Extended Search Space and the Randomization/Derandomization Procedure}
\label{subsec:3.4}
So far, we have established that in order to study the NISS problem, one can study the extreme points of the star-convex sets $\Qpartition$ in each direction from the center (Lemma \ref{lem:8}). Furthermore, 
in Section \ref{subsec:tensor}, we introduced the overparameterized representation of discrete functions, and showed that the distributions corresponding to these extreme points can be parameterized by  the probability of disagreement between the corresponding indicator functions, which in turn is bijectively related to the vector of their inner products (Proposition \ref{prop:5}). Furthermore, the inner products can be expressed in terms of the Fourier coefficients, as demonstrated in Lemma \ref{cor:Fourier functions}. Consequently, the problem of finding the extreme points can be posed as an optimization problem over the Fourier coefficients. However, this optimization is over a discrete search space of functions with discrete outputs. To enable the use of widely used optimization methods over convex spaces, we expand the search space to continuous valued functions in the following and show that this expansion does not affect the output of the optimization problem. 

Building upon the randomization approach of \cite{ghazi2016decidability,shirani2023non}, in the sequel, we often expand our search from the space of discrete functions with output alphabet $\mathcal{U}$, which are captured by vectors of indicator functions satisfying Conditions (1) and (2) in Equations \eqref{eq:c1:0} and \eqref{eq:c2:0}, to a search over the space of continuous-valued functions
$h_u: \CX^d \to [-1,1]$ for $u\in \CU$ Conditions (3) and (4) below:
\begin{align}
    &\text{Condition (3):}\qquad &\tilde{f}_u(x^d)\in [-1,1],~~ \forall x^d\in \CX^d, u\in\mathcal{U} \\
    \label{eq:conditions f_u}
    &\text{Condition (4):}\qquad  &\sum_{u\in \mathcal{U}}\tilde{f}_u(x^d) = 2-|\mathcal{U}|, ~~ \forall x^d\in \CX^d.
\end{align}

In the following, we provide a description of the randomization$/$derandomization (RD) approach which enables us to transform a pair of functions $\tilde{f}_u: \mathcal{X}^d\to [-1,1],\tilde{g}_v:\mathcal{Y}^d\to [-1,1], u,v\in \mathcal{U}\times \mathcal{V}$, satisfying conditions (3) and (4) into functions ${f}_u: \mathcal{X}^d\to \{-1,1\}, g_v:\mathcal{Y}^d\to \{-1,1\}, u,v\in \mathcal{U}\times \mathcal{V}$ satisfying Conditions (1) and (2) in a correlation preserving and marginal preserving manner, i.e., such that $\mathbb{E}(\tilde{f}_u(X^d))=\mathbb{E}({f}_u(X^d)) $, $\mathbb{E}(\tilde{g}_v(Y^d))=\mathbb{E}({g}_v(Y^d))$, and $\mathbb{E}(\tilde{f}_u(X^d)\tilde{g}_v(Y^d))=\mathbb{E}({f}_u(X^d)g_v(Y^d))$. We first provide the description of the RD procedure for the binary-output scenario, and then for the general case.
\subsubsection{Binary-Output Randomization and Derandomization} Alice is given input $X^d$ and continuous-valued functions $\tilde{f}_0(\cdot)$ and $\tilde{f}_1(\cdot)$ satisfying Conditions (3) and (4). Note that Condition (4) implies that $\tilde{f}_0(\cdot)=-\tilde{f}_1(\cdot)$. 
Alice produces a coin $C\in \{-1,1\}$ with bias $\frac{1+\tilde{f}_1(X^d)}{2}$, i.e., $P(C=1) = \frac{1+\tilde{f}_1(X^d)}{2}$,  and defines the $f_1(X^d)$ as the output of this coin\footnote{Note Alice can generate the coin locally using unused input samples, e.g. \cite{VonNeumann1951}.}. Furthermore, define $f_0(x^d)=-f_1(x^d), x^d\in \mathcal{X}^d$. Then, $f_{0}(\cdot),f_1(\cdot)$ satisfy Conditions (1) and (2). Bob constructs $g_0(\cdot),g_1(\cdot)$ from $\tilde{g}_0,\tilde{g}_1$ following a similar procedure. We have:
\begin{align*}
   & \mathbb{E}(f_0(X^d))= \PP(f_0(X^d)=1)-\PP(f_0(X^d)=-1)= 
    \frac{1+\mathbb{E}(\tilde{f}_0(X^d))}{2}- 
     \frac{1-\mathbb{E}(\tilde{f}_0(X^d))}{2}
     \\ & = \mathbb{E}(\tilde{f}_0(X^d)). 
\end{align*}
More generally, following the above line of arugment one can show that:
\begin{align}
    & \label{eq:RD:1}\mathbb{E}(f_u(X^d))=  \mathbb{E}(\tilde{f}_u(X^d)),\quad 
     \mathbb{E}(g_v(Y^d))= \mathbb{E}(g_v(Y^d)),
     \\&\label{eq:RD:2}
     \mathbb{E}(f_u(X^d)g_v(Y^d))= \mathbb{E}(\tilde{f}_u(X^d) \tilde{g}_v(Y^d)), \quad 
      \forall u,v\in \mathcal{U}\times \mathcal{V}. 
\end{align}

\subsubsection{Finite-Alphabet-Output Randomization and Derandomization} 
Define $\mathcal{U}_\phi= \mathcal{U}-\{0\}$ and $\mathcal{V}_{\phi}=\mathcal{V}-\{0\}$. Alice generates independent coins $C_u, u\in \mathcal{U}_{\phi}$ with biases $\frac{\tilde{f}_u(X^d)+1}{3-u- \sum_{u'<u}\tilde{f}_{u'}(X^d)}$, respectively, and Bob produces independent coins $C'_v, v\in \mathcal{V}_{\phi}$ with biases   $\frac{\tilde{g}_v(Y^d)+1}{3-v- \sum_{v'<v}\tilde{g}_{v'}(Y^d)}$, respectively. Note that Condition (4) ensures that $\sum_{u\in \mathcal{A}}\tilde{f}_u(x^d)\leq 2-|\mathcal{A}|, x^d\in \mathcal{X}^d$ for all $\mathcal{A}\subseteq \mathcal{U}_{\phi}$, which in turn, guarantees that the biases are in $[0,1]$ and the coin is well-defined. 
If $C_u=-1, u\in \mathcal{U}_\phi$, then Alice sets $f_0(X^d)=1$ and $f_u(X^d)=-1, u\in \mathcal{U}_{\phi}$. Otherwise, she sets $f_{u^*}(X^d)=1$ for $u^*=\arg\min_{u\in \mathcal{U}_\phi} \{u|C_u=1\}$ and $f_u(X^d)=-1, u\neq u^*$. Note that by construction for any $u\in \mathcal{U}_{\phi}$, we have:
\begin{align*}
    &\PP(f_u(X^d)=1|X^d=x^d)= \PP(C_u=1,C_{u'}=-1, u'<u)
    = \PP(C_u=1)\prod_{u'<u}(1-\PP(C_{u'}=1))
    \\&=\frac{1+\tilde{f}_u(x^d)}{2}.
\end{align*}
Similarly, for any $v\in \mathcal{V}_\phi$, we have $\PP(g_v(Y^d)=1|Y^d=y^d)=\frac{1+\tilde{g}_v(y^d)}{2}.$ Consequently, $\mathbb{E}(f_u(X^d))= \mathbb{E}(\tilde{f}_u(X^d))$ and  $\mathbb{E}(g_v(Y^d))= \mathbb{E}(\tilde{g}_v(Y^d))$, similar to the binary case. Furthermore, 
\begin{align*}
    &\mathbb{E}(f_u(X^d)g_v(Y^d))= \PP(f_u(X^d)=g_v(Y^d))-P(f_u(X^d)\neq g_v(Y^d))
    \\&= \mathbb{E}(\frac{1+\tilde{f}_u(X^d)}{2}\frac{1+\tilde{g}_v(Y^d)}{2})+\mathbb{E}(\frac{1-\tilde{f}_u(X^d)}{2}\frac{1-\tilde{g}_v(Y^d)}{2})
    \\&-\mathbb{E}(\frac{1-\mathbb{E}(\tilde{f}_u(X^d)}{2}\frac{1+\tilde{g}_v(Y^d)}{2})-
  \mathbb{E}(  \frac{1+\tilde{f}_u(X^d)}{2}\frac{1-\mathbb{E}(\tilde{g}_v(Y^d)}{2})
    \\& = \mathbb{E}(\tilde{f}_u(X^d)\tilde{g}_v(Y^d)),
\end{align*}
where we have used the fact that the local coins are generated using independent samples of the input sources. Consequently, Equation \eqref{eq:RD:1} and \eqref{eq:RD:2} hold for finite-output scenarios as well. 

\subsubsection{Expanded Search Space and A Bijective Mapping}
So, far we have introduced a procedure to transform continuous-valued functions $(\tilde{f}_{u},\tilde{g}_v)_{u,v \in \mathcal{U}\times \mathcal{V}}$ to discrete-valued functions $({f}_{u},{g}_v)_{u,v \in \mathcal{U}\times \mathcal{V}}$. We call this procedure the RD procedure, the functions $(\tilde{f}_{u},\tilde{g}_v)_{u,v \in \mathcal{U}\times \mathcal{V}}$ the randomized functions, and $({f}_{u},{g}_v)_{u,v \in \mathcal{U}\times \mathcal{V}}$ the derandomized functions.

\begin{definition}[Binary-Output Randomized Simulating Functions]
   Consider the sequence of function pairs $(\tilde{f}_d, \tilde{g}_d)_{d\in \NN}$ with $\tilde{f}_d, \tilde{g}_d: \pmm^d\rightarrow [-1,1]$. Let $\mu_d=\mathbb{E}(\tilde{f}_{d}(X^d)), \nu_d =\mathbb{E}(\tilde{g}_{d}(Y^d)), \eta_d = \mathbb{E}(\tilde{f}_{d}(X^d)\tilde{g}_{d}(Y^d)),$ for all $d\in \NN$.  Given $(\mu, \nu)\in [-1,1]^2$, define the set of randomized simulating functions  $\FBoolQUV(\mu,\nu)$ as the set of sequences of function-pairs $(\tilde{f}_d, \tilde{g}_d)_{d\in \NN}$ for which $ \lim_{d\to \infty} \mu_d= \mu$, $\lim_{d\to \infty} \nu_d= \nu$, and $\lim_{d\to \infty} \eta_d$ exists. 
\end{definition}

\begin{definition}[Finite-Output Randomized Simulating Functions]
Let  $\mu_u,\nu_v\in [-1,1]^2, u,v,\in \mathcal{U}\times \mathcal{V}$ such that $\sum_{u} \mu_u= 2-|\mathcal{U}|$ and $\sum_v \nu_v=2-|\mathcal{V}|$.
  The function tuples $(\tilde{f}_{u,d}, \tilde{g}_{v,d})_{u,v\in \mathcal{U}\times \mathcal{V}, d\in \NN}$ with $\tilde{f}_{u,d}, \tilde{g}_{v,d}: \pmm^d\rightarrow [-1,1]$ are called randomized simulating functions with parameters  $(\mu_u,\nu_v)_{u,v,\in \mathcal{U}\times \mathcal{V}}$ if they
  satisfy conditions (3) and (4), and $(\tilde{f}_{u,d},\tilde{g}_{v,d})_{d\in \mathbb{N}}\in \mathcal{F}_{XY}(\mu_u,\nu_v), u,v\in \mathcal{U}\times \mathcal{V}$.
   The set of randomized simulating functions parametrized by $(\mu_u,\nu_v)_{u,v,\in \mathcal{U}\times \mathcal{V}}$ is denoted by $\mathcal{F}_{XY}( Q_U,Q_V)$, where $Q_U(u)\triangleq \frac{1+\mu_u}{2}$ and $Q_V(v)\triangleq \frac{1+\nu_v}{2}, u,v\in \mathcal{U}\times\mathcal{V}$. 
    \end{definition}

Let us define the set $\mathcal{E}(P_{X,Y},Q_U,Q_V)$ as follows:
\begin{align*}
    \ENISS \triangleq \bigg\{(e_{u,v})_{u\in \mathcal{U}, v \in \mathcal{V}}\big|~& \exists (\tilde{f}_{u,d},\tilde{g}_{v,d})_{u,v \in \mathcal{U}\times \mathcal{V}, d\in \mathbb{N}}\in   \FBoolQUV(Q_U,Q_V):\\\numberthis \label{eq:ENISS}
 & e_{u,v}=\lim_{d\to \infty} \mathbb{E}(\tilde{f}_{d,u}(X^d)\tilde{g}_{d,v}(Y^d))\bigg\}.
\end{align*}
 The randomization procedure in the prequel implies that there is a bijection between   $\ENISS$ and $\mathcal{P}(P_{XY},Q_U,Q_V)$. Hence, it suffices to characterize $\ENISS$ in order to characterize  $\mathcal{P}(P_{XY},Q_U,Q_V)$. This is formally stated below. 

\begin{lemma}
Given marginals $Q_U$ and $Q_V$ and $P_{XY}$, there exists bijective mapping $\Psi$ from $\mathcal{P}(P_{XY},Q_U,Q_V)$ to $\ENISS$. 
\label{prop:4}
% We require the simulating function to be $f: \CX^d \rightarrow \CU$ and $g: \CY^d \rightarrow \CV$. Since this space is, discrete conventional optimization techniques for continuous settings are prohibitive. Therefore, we extend the search space to a continuous space of stochastic mappings. To see the idea consider generic finite sets $\mathcal{Z}$ and $\CU$ and  let $h:\mathcal{Z} \rightarrow \CU$ be a generic function. The vector of $|\CU|$ indicator functions  $\11(h(z) = u)$ can represent this function. These vectors belong to a subspace of $\RR^{|\CU|}$. That subspace is $\Delta_m$, the set of all vectors  $(p_1, p_2, ..., p_m) \in [0, 1]^{m}$ that $\sum_{j=1}^m p_j = 1$. With this definition, in NISS problem, we extend our search space to stochastic mappings over $\CX\times \CY$, that is, functions $P_f:\CX \rightarrow \Delta_{|\CU|}$ and $P_g:\CX \rightarrow \Delta_{|\CV|}$.
%Once we find the desired stochastic mappings $P_f, P_g$, we use randomized rounding to generate the function with outputs in $\CU$ and $\CV$.       
%To prove Lemma \ref{prop:4}, we explicitly construct a bijection between $\QNISS$ and $\ENISS$.  Define the mapping $\Psi:\QNISS \to \ENISS$ such that 
\begin{equation}\label{eq:bijection}
    \Psi(Q_{U,V}) = (4Q_{U,V}(u,v)-2(Q_U(u)+Q_V(v))+1)_{u\in \mathcal{U},v\in \mathcal{V}}.
\end{equation} 
\end{lemma}
The proof is provided in Appendix \ref{App:prop:4}.

\section{The Maximal Correlation Problem  and Solvability of NISS}
\label{sec:MC}
In this section, we first consider FB-NISS problems with uniform outputs and reproduce Witsenhausen's (\cite{witsenhausen1975sequences}) result which shows that any output distribution can be simulated if and only if its Hirschfeld-Gebelein-R\'enyi 
 maximal correlation is less than that of the input distribution. Next, we provide a simulating protocol for this scenario (Corollary \ref{Cor:Sol_BB_NISS}). Furthermore, we generalize the notion of maximal correlation by defining the \textit{biased maximal correlation} and prove that solving the FB-NISS for general non-uniform outputs is equivalent to characterizing the biased maximal correlation and the simulating protocol to simulate the joint distribution that achieves it. Next, we consider the general finite-input, finite-output NISS problem and further extend the notion of maximal correlation by defining the \textit{directional maximal correlation}. We show that solving the FB-NISS problem for a given $Q_{UV}$ and $P_{XY}$ is equivalent to finding the directional maximal correlation in a specific direction.
 
The maximal correlation coefficient was first introduced by  Hirschfeld-Gebelein and then R\'enyi  \cite{hirschfeld1935connection,gebelein1941statistische,renyi1959measures}. It is formally defined as follows:
\begin{definition}[\textbf{Maximal Correlation}]
\label{Def:MC}
Given alphabets $\mathcal{X}$ and $\mathcal{Y}$, and joint distribution $P_{XY}$, the maximal correlation $\rho(\mathcal{X},\mathcal{Y},P_{XY})$ is defined as:
\begin{align*}
\rho(\mathcal{X},\mathcal{Y},P_{XY})= \max_{\substack{f:\mathcal{X}\to \mathbb{R}, g:\mathcal{Y}\to \mathbb{R}, \\ \mathbb{E}(f)=\mathbb{E}(g)=0,\\
Var(f)=Var(g)=1}}\mathbb{E}_{X,Y}(f(X)g(Y)),
\end{align*}
where $\EE(f), \EE(g),  Var(f), Var(g)$ are shorthand for the expectation and variance of $f(X)$ and $g(Y)$, respectively.
\end{definition}

Witsenhausen \cite{witsenhausen1975sequences} showed that the maximal correlation coefficient $\rho(\mathcal{X},\mathcal{Y},P_{XY})$ tensorizes and satisfies the data processing inequality. This is summarized below:

\begin{proposition}[\cite{witsenhausen1975sequences}]
\label{prop:3}
Given alphabets $\mathcal{X}$ and $\mathcal{Y}$, and joint distribution $P_{XY}$, the maximal correlation $\rho(\mathcal{X},\mathcal{Y},P_{XY})$ satisfies the following:
\begin{itemize}[leftmargin=*]
    \item \textbf{Tensorization:} Let $P^{d}_{X,Y}(x^d,y^d)=\prod_{i=1}^dP_{XY}(x_i,y_i), x^d\in  \mathcal{X}^d,y^d\in \mathcal{Y}^d$ be the n-letter product distribution associated with $P_{XY}$. Then, $\rho(\mathcal{X},\mathcal{Y},P_{XY})= \rho(\mathcal{X}^d,\mathcal{Y}^d,P^{d}_{XY})$.
    \item \textbf{Data Processing:} Given alphabets $\mathcal{U}$ and $\mathcal{V}$, for any $f:\mathcal{X}\to \mathcal{U}$ and $g:\mathcal{Y}\to \mathcal{V}$, let $U=f(X)$ and $V=g(Y)$. Then, $\rho(\mathcal{X},\mathcal{Y},P_{XY})\geq \rho(\mathcal{U},\mathcal{V},P_{UV})$. 
\end{itemize}
\end{proposition}
Proposition \ref{prop:3} has an important consequence in solving a special instance of the  FB-NISS problem $(P_{XY},Q_{UV},\epsilon)$. Specifically, one can derive the solution for the case where $U$ and $V$ are uniformly distributed, i.e., $P_U(1)=P_V(1)=\frac{1}{2}$. This is formally stated in the following proposition which is proved in Appendix \ref{App:Cor:15}.

\begin{proposition}[\textbf{Solution to Uniform-Output FB-NISS}]
    \label{Cor:Sol_BB_NISS}
     Consider the NISS problem $(P_{XY},Q_{UV},\epsilon)$, where $X,Y$ and finite alphabet sources defined on $\mathbb{F}_q$, $U,V$ are binary variables, and $Q_U(1)=Q_V(1)=\frac{1}{2}$. The distribution $Q_{UV}$ is simulatable up to $\epsilon$ total variation distance if and only if there exists $Q'_{UV}\in \mathcal{S}(P_{XY},Q_U,Q_V)$ such that $d_{TV}(Q_{UV},Q'_{UV})< \epsilon$,  where:
    \begin{align*}
         \mathcal{S}(P_{XY},Q_U,Q_V)= \{P_{UV}| P_U(1)=P_V(1)=\frac{1}{2}, |\rho(\mathcal{U},\mathcal{V},P_{UV})|\leq|  \rho(\mathcal{X},\mathcal{Y},P_{XY})|\}.
     \end{align*}
     Furthermore, the solution is achieved by the pair of functions 
     \begin{align}
         U=\begin{cases}
        C_X(p_1)\qquad & \text{ if } C_X(\lambda)=1\\
        C_X(\frac{1}{2})\qquad & \text{ if } C_X(\lambda)=-1
         \end{cases},\quad 
              V=\begin{cases}
        C_Y(p_2)\qquad & \text{ if } C_Y(\lambda)=1\\
        C_Y(\frac{1}{2})\qquad & \text{ if } C_Y(\lambda)=-1
         \end{cases},
         \label{eq:Sym_BB_NISS}
     \end{align}
     where $C_X$ and $C_Y$ are independent coins\footnote{The biased coins $C_X(\lambda)$, $C_Y(\lambda) ,C_X(p_1),C_Y(p_2),C_X(\frac{1}{2})$ and $C_Y(\frac{1}{2})$ are assumed to be generated using non-overlaping samples of $X^d$ and $Y^d$, so that they are independent of each other given $p_1$ and $p_2$.}, $p_1\triangleq\frac{1+\tilde{f}(X^d)}{2}$, $p_2\triangleq\frac{1+\tilde{g}(Y^d)}{2}$, $\lambda\triangleq\sqrt{\frac{|\rho(\mathcal{U},\mathcal{V},Q'_{UV})|}{|\rho(\mathcal{X},\mathcal{Y},P_{XY})|}}$, and $\tilde{f}(X^d)=\sum_{i\in [q-1]} \tilde{f}^*_i\phi_i(X_1)$, $\tilde{g}(Y^d)=\sum_{i\in [q-1]} \tilde{g}^*_i\phi'_i(Y_1)$, where $\phi_i,\phi'_i, i\in \mathbb{F}_{q}$ are the orthonormal basis of functions defined on probability spaces $\mathcal{L}_X$ and $\mathcal{L}_Y$, respectively;  we have defined the first basis element as the identity functions, i.e., $\phi_0(X)=\phi'_0(Y)=1$; and the Fourier coefficients $\tilde{f}^*_i, \tilde{g}^*_i, i\in [q-1]$ are the outputs of the following optimization:
     \begin{align*}
         &(\tilde{f}^*_i, \tilde{g}^*_i)_{i,j\in [q-1]}=\arg \max_{(\tilde{f}_i,\tilde{g}_i)_{i\in [q-1]}\in \mathcal{A}} \sum_{i\in [q-1]}\tilde{f}_i\tilde{g}_j\rho_{i,j},
         \\&\mathcal{A}\triangleq \{(\tilde{f}_i,\tilde{g}_i)_{i\in [q-1]}| \sum_{i\in [q-1]}\tilde{f}_i\phi_i(x)\in [-1,1], \sum_{i,j\in [q-1]}\tilde{g}_i\phi'_i(y)\in [-,1,1], x,y\in \mathcal{X}\times \mathcal{Y}\},
     \end{align*}
     and $\rho_{i,j}\triangleq \mathbb{E}(\phi_i(X)\phi'_j(Y)), i,j \in [q-1]$. Particularly, for BB-NISS, we have:
    \[\tilde{f}(X^d)=\frac{X_1-\mathbb{E}(X)}{{1+|\mathbb{E}(X)|}},\qquad  \tilde{g}(Y^d)=\frac{Y_1-\mathbb{E}(Y)}{1+|\mathbb{E}(Y)|}.\]
\end{proposition}

\begin{remark}
Note that the simulating protocol in Proposition \ref{Cor:Sol_BB_NISS} requires one sample of $X^d$ and $Y^d$ to generate the distribution achieving the maximal correlation. The biased coins can be generated using other independent samples of the distributed sources (e.g., \cite{VonNeumann1951}).
\end{remark}

As shown in Proposition \ref{Cor:Sol_BB_NISS}, the characterization of the maximal correlation coefficient defined in Definition \ref{Def:MC}, along with the introduction of the pair of functions 
$\tilde{f}(X^d)$ and $\tilde{g}(Y^d)$ achieving this maximal correlation completely solves the uniform-output FB-NISS problem. However, the results do not naturally extend to non-uniform output FB-NISS and FF-NISS scenarios. We will show in the next sections, that a key step in solving the general binary-output NISS problem for given distributions $P_{XY},Q_{UV}$ is to characterize the \textit{biased maximal correlation coefficient} of $P_{XY}$ with respect to $Q_{UV}$ which is defined below.

\begin{definition}[\textbf{Biased Maximal Correlation}]
\label{def:biased_MC}
Let $\mathcal{X}=\mathcal{Y}=\mathbb{F}_q$ and  $\mathcal{U}$ and $\mathcal{V}$ be binary alphabets, and consider the joint distribution $P_{XY}$ and marginal distributions $Q_{U}$, and $Q_V$, the biased maximal correlation $\rho_b(P_{XY},Q_{U},Q_{V})$ is defined as:
\begin{align*}
\rho_b(P_{XY},Q_{U},Q_{V})= 
\sup_{d\in \mathbb{N}}\max_{\substack{f:\mathcal{X}^d\to [-1,1], g:\mathcal{Y}^d\to [-1,1], \\ \mathbb{E}(f)=2Q_U(1)-1, \mathbb{E}(g)=2Q_V(1)-1}}\mathbb{E}_{X^d,Y^d}(f(X^d)g(Y^d)).
\end{align*}
\end{definition}

\begin{proposition}[\textbf{Solution to the General FB-NISS}]
\label{prop:Sol_BB_NISS}
     Consider the NISS problem $(P_{XY},Q_{UV},\epsilon)$, where $X,Y$ are defined on $\mathbb{F}_q$ and $U,V$ are binary variables. The problem has a solution if and only if there exists $P_{UV}\in \mathcal{S}(P_{XY},Q_U,Q_V)$ such that $d_{TV}(Q_{UV},P_{UV})\leq \epsilon$, where:
    \begin{align*}
        & \mathcal{S}(P_{XY},Q_U,Q_V)=
        \\&\{P_{UV}| P_U(1)=Q_U(1), P_V(1)=Q_V(1), |1-2\PP(U\neq V)|\leq|  \rho_b(P_{XY},Q_U,Q_V)|\}.
     \end{align*}
     Furthermore, the solution is achieved by the pair of functions 
     \begin{align}
         U=\begin{cases}
        C_X(p_1)\qquad & \text{ if } C_X(\lambda)=1\\
        C_X(Q_U(1))\qquad & \text{ if } C_X(\lambda)=-1
         \end{cases},\quad 
              V=\begin{cases}
        C_Y(p_2)\qquad & \text{ if } C_Y(\lambda)=1\\
        C_Y(Q_V(1))\qquad & \text{ if } C_Y(\lambda)=-1
         \end{cases},
         \label{eq:BB_NISS}
     \end{align}
     where $p_1\triangleq\frac{1+f_{\delta}(X^d)}{2}$, $p_2\triangleq\frac{1+g_{\delta}(Y^d)}{2}$, $\lambda\triangleq\sqrt{\frac{|1-2\PP(U\neq V)|}{\rho_b(P_{XY},Q_U,Q_V)}}$, and $f_{\delta}:\mathcal{X}^d\to [-1,1]$ and $g_{\delta}:\mathcal{Y}^d\to [-1,1]$ are a pair of functions for which $\rho_b(P_{XY},Q_U,Q_V)-\mathbb{E}(f_{\delta}(X^d),g_{\delta}(Y^d))\leq \delta$ and $\delta\triangleq \epsilon- d_{TV}(Q_{UV},P_{UV})$.
\end{proposition}
The proof follows by similar arguments as that of Proposition \ref{Cor:Sol_BB_NISS} and is omitted for brevity. To solve the more general FF-NISS problem, we extend the notion of biased maximal correlation in the following definition.

\begin{definition}[\textbf{Directional Maximal Correlation}]
\label{def:dir_cor}
Let $\mathcal{X},\mathcal{Y}=\mathbb{F}_q$, and $\mathcal{U}=\{0,1,\cdots,|\mathcal{U}|\}$ and $\mathcal{V}=\{0,1,\cdots,|\mathcal{V}|\}$ finite alphabets. Define $\mathcal{U}_{\phi}\triangleq\mathcal{U}-\{0\}$ and $\mathcal{V}_{\phi}\triangleq\mathcal{V}-\{0\}$.
Consider the joint distribution $P_{XY}$, marginal distributions $Q_{U}$,and $Q_V$, and the direction vector $\bm{\alpha}=(\alpha_{u,v})_{u\in \mathcal{U}_\phi,v\in \mathcal{V}_\phi}$, where $\alpha_{u,v}\in\mathbb{R}$ and $\sum_{u\in \mathcal{U}_\phi,v\in \mathcal{V}_\phi}\alpha^2_{u,v}=1$.   
The directional maximal correlation $\rho_d(P_{XY},Q_{U},Q_{V})$ is defined as:
\begin{align*}
\rho_d(P_{XY},Q_{U},Q_{V},\bm{\alpha})= 
\sup_{d\in \mathbb{N}}\max_{t\in \mathcal{T}_d}
|t|,
\end{align*}
where $t\in \mathcal{T}_d$ if an only if there exists $\tilde{f}_{d,u}:\mathcal{X}^d\to [-1,1], u\in \mathcal{U}_\phi$ and $\tilde{g}_{d,v}:\mathcal{Y}^d\to [-1,1],v\in \mathcal{V}_\phi$ such that $\mathbb{E}(\tilde{f}_{d,u}(X^d))=2Q_U(u)-1, u\in \mathcal{U}_\phi$, $\mathbb{E}(\tilde{g}_{d,v}(Y^d))=2Q_V(v)-1, v\in \mathcal{V}_\phi$, and
\begin{align*}
    t=\frac{\mathbb{E}(\tilde{f}_{d,u}(X^d)\tilde{g}_{d,v}(Y^d))-(2Q_U(u)-1)(2Q_V(v)-1)}{\alpha_{u,v}}, \forall u,v\in \mathcal{U}_{\phi}\times \mathcal{V}_\phi.
\end{align*}
\end{definition}
\begin{remark}
Note that the set $\mathcal{T}_d$ is always non-empty since the product distribution $Q_UQ_V$ can always be generated via local biased coins by Alice and Bob. So, the directional maximal correlation is always defined. 
\end{remark}
\begin{remark}
 Note that for binary output NISS, the direction vector can only take one direction characterized by $\alpha_{1,1}=1$. It can be verified that in this case, $\rho_d(P_{XY},Q_U,Q_V,1)=\rho_b(P_{XY},Q_U,Q_V)-2(Q_U(u)-1)(2Q_V(v)-1)$. So, finding the directional maximal correlation in this case is equivalent to finding the biased maximal correlation, and the simulating protocol achieving the directional maximal correlation is the same as the one achieving the biased maximal correlation.  
\end{remark}
Next, we generalize Proposition \ref{prop:Sol_BB_NISS} to the FF-NISS problem by leveraging the concept of directional maximal correlation. To this end, let us consider  finite alphabets $\mathcal{U}=\{0,1,\cdots,|\mathcal{U}|\}$ and $\mathcal{V}=\{0,1,\cdots,|\mathcal{V}|\}$ and a given joint probability distribution $Q_{UV}$ on $\mathcal{U}\times \mathcal{V}$. We define the direction vector  $\bm{\alpha}(Q_{UV})$ associated with $Q_{UV}$ as follows:
\begin{align*}
    &\beta_{u,v}(Q_{UV})\triangleq 4(Q_{UV}(u,v)-Q_U(u)Q_V(v)),
    \\& \alpha_{u,v}(Q_{UV})\triangleq  \frac{\beta_{u,v}(Q_{UV})}{\sum_{u'\in \mathcal{U}_\phi,v'\in \mathcal{V}_\phi}\beta_{u',v'}^2(Q_{UV})}, u\in \mathcal{U}_\phi, v\in \mathcal{V}_\phi,
    %4Q_{UV}-2Q_U(u)-2Q_V(v)+1-(2Q_U(u)-1)(2Q_V(v)-1)
\end{align*}
where $\mathcal{U}_{\phi}\triangleq\mathcal{U}-\{0\}$ and $\mathcal{V}_{\phi}\triangleq\mathcal{V}-\{0\}$. Furthermore, we define the directional correlation associated with $Q_{UV}$ as: 
\[t(Q_{UV})\triangleq \sum_{u\in \mathcal{U}_\phi,v\in \mathcal{V}_\phi}\beta_{u,v}^2(Q_{UV}).\]

Note that by construction, for any given $Q_{UV}$ and associated directional maximal correlation 
$\rho_d( P_{XY},$ $ Q_U, Q_V, \bm{\alpha} (Q_{UV}))$, there is a unique distribution $Q'_{UV}$ such that $\bm{\alpha}(Q'_{UV})=\bm{\alpha}(Q_{UV})$ and 
$t(Q'(U,V))= \rho_d(P_{XY},Q_U,Q_V,\bm{\alpha}(Q_{UV}))$. Furthermore, by definition of maximal directional correlation, for any $\delta>0$, there exist $d\in \mathbb{N}$ and  functions $\tilde{f}_{d,u}:\mathcal{X}^d\to [-1,1], u\in \mathcal{U}_\phi$ and $\tilde{g}_{d,v}:\mathcal{Y}^d\to [-1,1], v\in \mathcal{V}_\phi$ such that 
\begin{align*}
    \left|t(Q'(U,V))-\frac{\mathbb{E}(\tilde{f}_{d,u}(X^d)\tilde{g}_{d,v}(Y^d))-(2Q_U(u)-1)(2Q_V(v)-1)}{\alpha_{u,v}}\right|\leq \delta, \forall u,v\in \mathcal{U}_{\phi}\times \mathcal{V}_\phi.
\end{align*}
Hence, using the bijection described in the proof of Lemma \ref{prop:4}, the distribution $Q'_{UV}$ can be simulated using $P_{XY}$. We denote by $C_{X,\delta}(Q'_{UV})$ and  $C_{Y,\delta}(Q'_{UV})$  the functions $f:\mathcal{X}^d\to \mathcal{U}$ and $g:\mathcal{Y}^d\to \mathcal{V}$ which simulate $Q'_{UV}$ with variational distance smaller than $\delta$. The following proposition follows by similar arguments as that of Proposition \ref{Cor:Sol_BB_NISS}.

\begin{proposition}[\textbf{Solution to FF-NISS}]
\label{prop:Sol_BF-NISS}
   Consider the NISS problem $(P_{XY},Q_{UV},\epsilon)$, where $X,Y,U,V$ are finite random variables.
   The problem has a solution if and only if there exists $Q'_{UV}$ such that $d_{TV}(Q_{UV},Q'_{UV})< \epsilon$ and 
   \[|t(Q'_{UV})|\leq \rho_d(P_{XY},Q_U,Q_V,\bm{\alpha}(Q'_{UV})).\] 
     Furthermore, the solution is achieved by the pair of functions 
     \begin{align}
         U=\begin{cases}
        C_{X,\delta}(Q'_{UV})\qquad & \text{ if } C_X(\lambda)=1\\
        C_{X,\delta}(Q'_UQ'_V)\qquad & \text{ if } C_X(\lambda)=-1
         \end{cases},\quad 
              V=\begin{cases}
        C_{Y,\delta}(Q'_{UV})\qquad & \text{ if } C_Y(\lambda)=1\\
        C_{X,\delta}(Q'_UQ'_V)\qquad & \text{ if } C_Y(\lambda)=-1
         \end{cases},
         \label{eq:BF_NISS}
     \end{align}
     where  $\lambda\triangleq\sqrt{\frac{t(Q'_{UV})}{\rho_d(\mathcal{U},\mathcal{V},Q'_{UV},\bm{\alpha}(Q'_{UV}))}}$, and $\delta\triangleq \epsilon-d_{TV}(Q_{UV}-Q'_{UV})$.
\end{proposition}

\section{Primal and Dual Formulations of the Maximal Correlation Problem}
\label{sec:PD}
In the previous section, we showed that the FB-NISS problem can be reduced to the problem of finding the biased maximal correlation and the FF-NISS problem to the problem of finding the directional maximal correlation, and the associated generating functions. 
In the following, we reformulate the problem of characterizing the biased maximal correlation and its associated functions as a norm optimization problem over a convex polytope and its dual. We show that in the case of uniform input distributions, the dual problem can be solved as a linear programming problem. For non-uniform input scenarios, in the next sections, we develop the low computational complexity optimization algorithms to find approximate solutions to the primal problem. 

In the following proposition, we formulate the biased maximal correlation defined in Definition \ref{def:biased_MC}, and by
 considering the extended search space (as in Section \ref{sec:stc}) and the corresponding overparametrized Fourier. 
 
\begin{theorem}[\textbf{Biased Maximal Correlation - Primal Form}]
\label{prop:primal}
Let $\mathcal{X},\mathcal{Y}=\mathbb{F}_q$ and $\mathcal{U}$ and $\mathcal{V}$ be binary alphabets, and consider the joint distribution $P_{XY}$ and marginal distributions $Q_{U}$,and $Q_V$, the biased maximal correlation $\rho_b(P_{XY},Q_{U},Q_{V})$ can be computed by solving the following optimization:
\begin{align}
\label{eq:primal}
 \rho_b(P_{XY},Q_{U},Q_{V})=\sup_{d\in \mathbb{N}}\sup_{\substack{(f_{s^d} , s^d\in \mathbb{F}_q^d)\in\mathcal{F}(Q_U)\\(g_{t^d} ,  t^d\in \mathbb{F}_q^d)\in\mathcal{G}(Q_V)}}\sum_{s^d,t^d\in \mathbb{F}_q^d} \tilde{f}_{s^d} \tilde{g}_{{t}^d} \prod_{s,t\in\mathbb{F}_q}\rho_{s,t}^{n(s,t|s^d,t^d)}, 
\end{align}
where $\rho_{s,t}$ is defined in Lemma \ref{cor:Fourier functions}, and  
\begin{align}
\label{eq:conv_R1}
&\mathcal{F}(Q_U)\triangleq \{(\tilde{f}_{s^d} ,  s^d\in \mathbb{F}_q^d)\big| |\sum_{ s^d\in \mathbb{F}_q^d}\tilde{f}_{s^d} \prod_{i\in [d]}\phi_{s_i}(x^d)|\leq 1, \forall x^d\in \mathbb{F}_q^d, \tilde{f}_{\phi}=2Q_U(1)-1\},
\\&\label{eq:conv_R2}
\mathcal{G}(Q_V)\triangleq \{(\tilde{g}_{t^d} ,  t^d\in \mathbb{G}_q^d)\big| |\sum_{ t^d\in \mathbb{G}_q^d}\tilde{g}_{t^d} \prod_{i\in [d]}\phi'_{t_i}(y^d)|\leq 1, \forall y^d\in \mathbf{G}_q^d, \tilde{g}_{\phi}=2Q_V(1)-1\},
\end{align}
where $\phi_{s}, s\in \mathbb{F}_q$ and $\phi'_t, t\in \mathbb{F}_q$ are the orthonormal basis for functions in $\mathcal{L}_X$ and $\mathcal{L}_Y$, respectively. Particularly, for BB-NISS, we have:
\begin{align}
\label{eq:bn_primal}
 \rho_b(P_{XY},Q_{U},Q_{V})=\sup_{d\in \mathbb{N}}\sup_{\substack{(\tilde{f}_S , \mathcal{S}\subseteq [d])\in\mathcal{F}(Q_U)\\(\tilde{g}_S , \mathcal{S}\subseteq [d])\in\mathcal{G}(Q_V)}}\sum_{\mathcal{S}\subseteq[d]} \tilde{f}_S  \tilde{g}_S \rho^{|\mathcal{S}|}, 
\end{align}
where 
\begin{align*}
&\mathcal{F}(Q_U)\triangleq \{(\tilde{f}_S  , \mathcal{S}\subseteq [d])\big| |\sum_{\mathcal{S}\subseteq [d]}\tilde{f}_S 
 \phi_{\mathcal{S}}(x^d)|\leq 1, \forall x^d\in \{-1,1\}^d, \tilde{f}_{\phi}=2Q_U(1)-1\},
\\&\mathcal{G}(Q_V)\triangleq \{(\tilde{g}_S , \mathcal{S}\subseteq [d])\big| |\sum_{\mathcal{S}\subseteq [d]}\tilde{g}_S \psi_{\mathcal{S}}(y^d)|\leq 1, \forall y^d\in \{-1,1\}^d, \tilde{g}_{\phi}=2Q_V(1)-1\},
\end{align*}
and $\phi_{\mathcal{S}}(x^d)\triangleq \prod_{i\in \mathcal{S}}\frac{x_i-\mathbb{E}(X)}{\sqrt{Var(X)}}$ and $\psi_{\mathcal{S}}(y^d)\triangleq \prod_{i\in \mathcal{S}}\frac{y_i-\mathbb{E}(Y)}{\sqrt{Var(Y)}}$  are the corresponding parity functions, and $\rho\triangleq \rho(\mathcal{X},\mathcal{Y},P_{XY})$.
%\\Furthermore, the maximum correlation is achieved by the functions $f^*(X^d)= \sum_{\mathcal{S}\subseteq [d]}f^*_{\mathcal{S}}\phi_{\mathcal{S}}(x^d)$ and $g^*(Y^d)= \sum_{\mathcal{S}\subseteq [d]}g^*_{\mathcal{S}}\psi_{\mathcal{S}}$, where $(f^*_{\mathcal{S}},\mathcal{S}\subseteq [d])$ and $(g^*_{\mathcal{S}},\mathcal{S}\subseteq [d])$
\end{theorem}

 Note that the optimization in Equation \eqref{eq:primal} is an inner-product optimization over two intersecting polytopes. In particular, the following corollary shows that for binary-output NISS with $Q_U(1)=Q_V(1)$, the problem reduces to a norm optimization over a convex polytope. The proof is provided in Appendix \ref{App:Cor:28}.

% \begin{corollary}[Equi-Biased Maximal Correlation]
% \label{cor:28}
%    In the setting of Theorem \ref{prop:primal}, if $Q_U(1)=Q_V(1)$, then
%    \begin{align}
% \label{eq:primal_2}
%  \rho_b(P_{X,Y},Q_{U},Q_{V})=\sup_{d\in \mathbb{N}}\sup_{\substack{(f_{s^d} , s^d\in \mathbb{F}_q^d)\in\mathcal{F}(Q_U)}}\sum_{s^d,t^d\in \mathbb{F}_q^d} \tilde{f}_{s^d} \tilde{f}_{{t}^d} \prod_{s,t\in\mathbb{F}_q}\rho_{s,t}^{n(s,t|s^d,t^d)}, 
% \end{align}
% where $\mathcal{F}(Q_U)$ is defined in Theorem \ref{prop:primal}. 
% \end{corollary}

\begin{corollary}[Equi-Biased Maximal Correlation]
\label{cor:28}
   In the setting of Theorem \ref{prop:primal}, if $Q_U(1)=Q_V(1)$ and $\mathcal{X}=\mathcal{Y}=\{-1,1\}$, then
   \begin{align}
\label{eq:primal_2}
 \rho_b(P_{XY},Q_{U},Q_{V})=\sup_{d\in \mathbb{N}}\sup_{\substack{(\tilde{f}_\mathcal{S}  , \mathcal{S}\subseteq [d])\subseteq \mathcal{F}(Q_U))}}\sum_{\mathcal{S}\subseteq[d]} \tilde{f}^2_{\mathcal{S}}\rho^{|\mathcal{S}|}, 
\end{align}
where $\mathcal{F}(Q_U)\triangleq \{(\tilde{f}_\mathcal{S}  , \mathcal{S}\subseteq [d])\big| |\sum_{\mathcal{S}\subseteq [d]}\tilde{f}_\mathcal{S} \phi_{\mathcal{S}}(x^d)|\leq 1, \forall x^d\in \{-1,1\}^d, f_{\phi}=2Q_U(1)-1\}$. 
\end{corollary}

\begin{remark}
    It can be observed from proof of Corollary \ref{cor:28} that the result can be extended to non-binary input alphabets as long as a positive-definiteness condition holds. To elaborate, consider the vector space $\mathcal{R}= \{(v_{s^d}, s^d\in \mathbb{F}_q^d)| v_{s^d}\in [-1,1], s^d\in \mathbb{F}_q^d\}$ and define the inner-product $\langle\mathbf{v},\mathbf{w}\rangle\triangleq \sum_{s^d,t^d} v_{s^d}w_{t^d}\prod_{s,t\in [d]}\rho_{s,t}^{n(s,t|s^d,t^d)}$, where $\rho_{s,t}$ is defined in Lemma \ref{cor:Fourier functions}. If $P_{XY}$ is such that the inner-product is positive-definite, then the Cauchy-Schwarz inequality used in the proof of Corollary \ref{cor:28} holds, and the result can be extended to non-binary input alphabets. 
\end{remark}

The optimization in Equation \eqref{eq:primal}  is a quadratic program which in general is NP-hard \cite{pardalos1988checking}. In the next sections, we provide a convex-concave optimization problem to estimate the optimal solution. 

In the special case when the input marginal distributions are uniform (i.e., $\mathcal{X}=\mathcal{Y}=\mathbb{F}_q$ and $P_X(x)=P_Y(y)=\frac{1}{q},x,y\in \mathbb{F}_q$), one can use the 
dual to the optimization in Equation \eqref{eq:primal}, which is given in the following proposition and can be solved by linear programming. We derive the dual formulation by considering the Karush–Kuhn–Tucker (KKT) conditions. Note that there are $4\times q^d$ constraints in the primal formulation, and hence they are assigned $4\times q^d$ coefficients in the KKT conditions. To elaborate, for each set of constraints $f(x^d)\leq 1, x^d\in \mathbb{F}_q$, $f(x^d)\geq -1, x^d\in \mathbb{F}_q$, $g(y^d)\leq 1, y^d\in \mathbb{F}_q$, $g(y^d)\geq -1, y^d\in \mathbb{F}_q$, we define the KKT coefficients $\lambda^+_f(x^d), \lambda^-_f(x^d), \lambda^+_g(y^d)$ and $\lambda^-_g(y^d)$. To prove the following result, we apply a uniform Fourier expansion on the KKT conditions $\lambda^+_f(x^d), \lambda^-_f(x^d), \lambda^+_g(y^d)$ and $\lambda^-_g(y^d)$.
 The dual objective function and constraints are then expressed in terms of the  resulting Fourier coefficients. The complete proof is provided in Appendix \ref{app:th:dual}.
\begin{theorem}[\textbf{Biased Maximal Correlation - Dual Form}]
\label{prop:dual}
Let $\mathcal{X}=\mathcal{Y}=\mathbb{F}_q$ and  $\mathcal{U}$ and $\mathcal{V}$ be binary alphabets, and consider the joint distributions $P_{XY},Q_{UV}$ such that $P_X(x)=P_Y(y)=\frac{1}{q},x,y\in \mathbb{F}_q$. The biased maximal correlation $\rho_b(P_{XY},Q_{UV})$ can be computed by solving the following optimization:
\begin{align}
\label{eq:dual}
 &\rho_b(P_{XY},Q_{U},Q_{V})=(2Q_U(1)-1)(2Q_V(1)-1)
 \\&\nonumber  \qquad\qquad +\sup_{d\in \mathbb{N}} \sup_{\substack{(\lambda^+_{f,s^d},\lambda^-_{f,s^d} s^d\in \mathbb{F}_q^d)\in \Lambda(Q_U)\\(\lambda^+_{g,s^d},\lambda^-_{g,s^d} s^d\in \mathbb{F}_q^d)\in \Lambda(Q_V)}}
 Q_U(0)\lambda^+_{f,\mathbf{0}}+Q_U(1)\lambda^-_{f,\mathbf{0}}+Q_V(0)\lambda^+_{g,\mathbf{0}}+Q_V(1)\lambda^-_{g,\mathbf{0}}, 
\end{align}
where $(\lambda^+_{f,s^d},\lambda^-_{f,s^d}, s^d\in \mathbb{F}_q^d)\in \Lambda(Q_U)$ if and only if: 
\begin{align*}
    &|2Q_V(1)-1 +\sum_{s^d\in \mathbb{F}_q^d, s^d\neq \mathbf{0}}\bar{f}_{s^d}(\lambda)\chi_{s^d}(x^d)|\leq 1, \quad \forall x^d\in \mathbb{F}_q^d,\\
&\sum_{\mathcal{S}\subseteq [d]}\lambda^+_{f,s^d}\chi_{s^d}(x^d)\geq 0,\quad \sum_{s^d\in \mathbb{F}_q^d}\lambda^-_{f,s^d}\chi_{s^d}(x^d)\geq 0,\quad  \forall x^d\in\mathbb{F}_q^d,
\end{align*}
where $\bar{f}_{s^d}(\lambda_f), s^d\in \mathbb{F}_q^d$ is the solution to the linear system $\mathbf{P}\mathbf{f}= \mathbf{L}_f$, $\mathbf{P}\triangleq [\prod_{s,t}\rho_{s,t}^{n(s,t|s^d,t^d)}]_{s^d,t^d\in \mathbb{F}_q^d}$, $\mathbf{f}\in \mathbb{R}^{q^d}$, $\mathbf{L}_f\triangleq [\lambda^+_{f,s^d}-\lambda^-_{f,s^d}]_{s^d\in \mathbb{F}_q^d}$.
Similarly, $(\lambda^+_{g,s^d},\lambda^-_{g,s^d}, s^d\in \mathbb{F}_q^d)\in \Lambda(Q_V)$ if and only if: 

\begin{align*}
    &|2Q_U(1)-1 +\sum_{s^d\in \mathbb{F}_q^d, s^d\neq \mathbf{0}}\bar{g}_{s^d}(\lambda)\chi_{s^d}(x^d)|\leq 1, \quad \forall x^d\in \mathbb{F}_q^d,\\
&\sum_{s^d\in\mathbb{F}_q^d}\lambda^+_{g,s^d}\chi_{s^d}(y^d)\geq 0,\quad  \sum_{s^d\in \mathbb{F}_q^d}\lambda^-_{g,s^d}\chi_{s^d}(y^d)\geq 0,\quad  \forall y^d\in\mathbb{F}_q^d,
\end{align*}
where $\bar{g}_{s^d}(\lambda_g), s^d\in \mathbb{F}_q^d$ is the solution to the linear system $\mathbf{P}\mathbf{g}= \mathbf{L}_g$, $\mathbf{g}\in \mathbb{R}^{q^d}$, $\mathbf{L}_g\triangleq [\lambda^+_{g,s^d}-\lambda^-_{g,s^d}]_{s^d\in \mathbb{F}_q^d}$.
\end{theorem}

Theorem \ref{prop:primal} can be generalized to the BF-NISS scenario as given below.

\begin{theorem}[\textbf{Directional Maximal Correlation - Primal Form}]
\label{prop:primal_2}
Let $\mathcal{X},\mathcal{Y}$, $\mathcal{U}$ and $\mathcal{V}$ finite alphabets. 
Consider the joint distribution $P_{XY}$, marginal distributions $Q_{U}$,and $Q_V$, and direction vector $\bm{\alpha}$. The directional maximal correlation $\rho_d(P_{XY},Q_{U},Q_{V},\bm{\alpha})$ can be computed by solving the following optimization:
\begin{align}
\label{eq:dir_primal}
& \rho_d(P_{XY},Q_{U},Q_{V},\bm{\alpha})= \notag
\\&\sup_{d\in \mathbb{N}}\sup_{(\tilde{f}_{s^d} ,\tilde{g}_{s^d}, s^d\in \mathbb{F}_q^d)\in\mathcal{C}(Q_U,Q_V,\bm{\alpha})} \hspace{-40pt}\frac{\sum_{s^d,t^d\in \mathbb{F}_q^d}\tilde{f}_{s^d} \tilde{g}_{t^d} \prod_{s,t}\rho_{s,t}^{n(s,t|s^d,t^d)}-(2Q_U(1)-1)(2Q_V(1)-1)}{\alpha_{1,1}}, 
\end{align}
where $\mathcal{C}(Q_U,Q_V,\bm{\alpha})$ is the set of all $(\tilde{f}_{s^d} ,\tilde{g}_{s^d} , s^d\in \mathbb{F}_q^d)$ for which there exists $\tilde{f}_{u,s^d},\tilde{g}_{v,s^d}, u\in \mathcal{U}_\phi, v\in \mathcal{V}_{\phi}, s^d\in \mathbb{F}_q^d$ such that:
\begin{itemize}[leftmargin=*]
    \item Directionality constraint:
    \begin{align*}
& \frac{\sum_{s^d,t^d\in \mathbb{F}_q^d}f_{s^d} g_{t^d}\prod_{s,t}\rho_{s,t}^{n(s,t|s^d,t^d)}-(2Q_U(1)-1)(2Q_V(1)-1)}{\alpha_{1,1}}= \\&\qquad \qquad  \frac{\sum_{s^d,t^d\in \mathbb{F}_q^d}\tilde{f}_{u,s^d} \tilde{g}_{v,t^d} \prod_{s,t}^{n(s,t|s^d,t^d)}-(2Q_U(u)-1)(2Q_V(v)-1)}{\alpha_{u,v}}, \forall u\in \mathcal{U}_\phi, v\in \mathcal{V}_{\phi}
\end{align*}
\item Bias constraints:
\begin{align*}
    \tilde{f}_{u,\mathbf{0}}=2Q_U(u)-1, \quad \tilde{g}_{v,\mathbf{0}}=2Q_V(v)-1,\quad \forall u\in \mathcal{U}_\phi, v\in \mathcal{V}_\phi
\end{align*}
\item Valid distribution constraints:
\begin{align*}
  &  |\sum_{s^d\in \mathbb{F}_q^d} f_{u,s^d}\phi_{s^d}(x^d)|\leq 1,\quad  |\sum_{s^d\in \mathbb{F}_q^d} g_{v,s^d}\phi'_{s^d}(x^d)|\leq 1, \quad \forall x^d,y^d\in \{-1,1\}^d
  \\& \sum_{u\in \mathcal{U}}\sum_{s^d\in \mathbb{F}_q^d} f_{u,s^d}\phi_{s^d}(x^d)\leq 2-|\mathcal{U}|,\quad \sum_{v\in \mathcal{V}}\sum_{s^d\in \mathbb{F}_q^d} g_{v,s^d}\phi'_{s^d}(x^d)\leq 2-|\mathcal{V}|.
\end{align*}
\end{itemize}

%and $\phi_{s^d}(x^d)\triangleq \prod_{i\in \mathcal{S}}\frac{x_i-\mathbb{E}(X)}{\sqrt{Var(X)}}$, $\psi_{\mathcal{S}}(y^d)\triangleq \prod_{i\in \mathcal{S}}\frac{y_i-\mathbb{E}(Y)}{\sqrt{Var(Y)}}$, and $\rho\triangleq \rho(\mathcal{X},\mathcal{Y},P_{X,Y})$.
%\\Furthermore, the maximum correlation is achieved by the functions $f^*(X^d)= \sum_{\mathcal{S}\subseteq [d]}f^*_{\mathcal{S}}\phi_{\mathcal{S}}(x^d)$ and $g^*(Y^d)= \sum_{\mathcal{S}\subseteq [d]}g^*_{\mathcal{S}}\psi_{\mathcal{S}}$, where $(f^*_{\mathcal{S}},\mathcal{S}\subseteq [d])$ and $(g^*_{\mathcal{S}},\mathcal{S}\subseteq [d])$
\end{theorem}
\textit{Proof Outline:} It suffices to argue that the Fourier coefficients associated with any pair of functions admissible in the optimization in Definition \ref{def:dir_cor} satisfy the constraints in the proposition statement. The directionality constraints must be satisfied by the definition of directional maximal correlation. The bias constraints enforce that the marginal distributions produced by the pair of functions are equal to $(Q_U,Q_V)$. Furthermore, as argued in Equation \eqref{eq:conditions f_u} in Section \ref{subsec:tensor}, the valid distribution constraints are satisfied by the corresponding Fourier coefficients of any $|\mathcal{U}|$-ary and $|\mathcal{V}|$-ary pair of functions. Consequently, optimizing the objective function of directional maximal correlation over the set of Fourier coefficients satisfying these constraints yeilds the maximal correlation in direction $\bm{\alpha}$ as desired.

\section{Case Studies in Classical and Quantum Distributed Correlation Quantification}
\label{sec:examples}
In the previous sections, we have introduced  a general framework for quantifying the correlation produced among distributed terminals in source simulation scenarios. Our focus in this work is  to study the NISS scenario with independent and identically distributed pairs of input samples (Figure \ref{fig:1}) by applying the Fourier framework. However, it should be mentioned that the framework is general, and it is potentially applicable in various other problems of interest which require quantification of distributed correlation. To provide examples of these general scenarios, and clarify some of the  notation introduced in the previous sections, in this subsection we consider three source simulation scenarios, focusing on i) classical sequences with independent but not identically distributed elements, ii) classical sources generated by sequential measurements of entangled qubits, and iii) classical sources with Markovian structure.

\noindent \textbf{Case Study 1: Non-IID Classical Source with Common Randomness}

 \begin{figure}[!t]
\centering 
\includegraphics[width=0.8\textwidth]{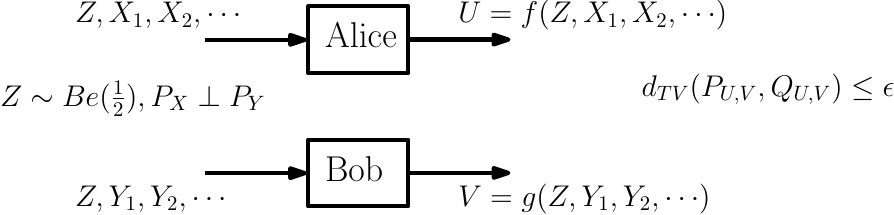}
\caption{An NISS scenario with one-bit common randomness and unlimited local randomness available to the agents.
%The source sequences $X^d$ and $Y^d$ are IID with distributions $P_X$ and $P_Y$, respectively. . 
}
\label{fig:2n}
\end{figure}
In this scenario (Figure \ref{fig:2n}), the two agents are provided with one bit of common randomness, $Z\sim Be(\frac{1}{2})$, and an unlimited number of bits of local randomness $X_i, i \in \mathbb{N}$ and $Y_i, i\in \mathbb{N}$, respectively. The agents wish to simulate a pair of binary outputs. 
Let us denote the set of simulatable joint distributions by $\mathcal{Q}_1$. 
 To find $\mathcal{Q}_1$, we first characterize the set of simulatable distributions with marginals $Q_U$ and $Q_V$, denoted by $\mathcal{P}(Q_U,Q_V)$, where $Q_U(1)=a$ and $Q_V(1)=b$ for some $a,b\in [0,1]$. Then, $\mathcal{Q}_1$ is found by taking the union of  $\mathcal{P}(Q_U,Q_V)$ over all $a,b \in [0,1]$. 
 Let $U,V$ be produced by following the RD procedure using functions $\tilde{f}(Z,X^d)$ and $\tilde{g}(Z,Y^d)$, satisfying Conditions (3) and (4) in Section \ref{subsec:3.4}, for some $d\in \mathbb{N}$. Let $\tilde{f}_{s^{d+1}}, \tilde{g}_{s^{d+1}}, s^{d+1}\in \{-1,1\}^{d+1}$ be the Fourier coefficients resulting from decomposing $\tilde{f}$ and $\tilde{g}$, respectively, where the decomposition is performed with the following orthonormal basis:
 \begin{align*}
     \phi_{s^{d+1}}(Z,X^d)\triangleq  
     \begin{cases}
         Z\prod_{i:s_i=1, i>1} X_{i-1} \qquad & \text{ if } s_1=1\\
         \prod_{i:s_i=1, i>1} X_{i-1} & \text{ if } s_1=-1
     \end{cases}, \qquad s^{d+1}\in \{-1,1\}^{d+1}.
 \end{align*}
We have:
\begin{align}
&\label{ex:1:1} \tilde{f}_{\phi} = \mathbb{E}[\tilde{f}] = 2a - 1, \quad \tilde{g}_{\phi} = \mathbb{E}[\tilde{g}] = 2b - 1, \quad 
\mathbb{E}[\tilde{f}\tilde{g}] = (2a - 1)(2b - 1) + \tilde{f}_{1}\tilde{g}_1,
\end{align}
where in the last equality we have used the fact that $X^d$ and $Y^d$ are independent local randomness to conclude that $\rho_{s^{d+1}}=0$ for all $s^{d+1}$ such that there exists $s_i=1, i>1$. Furthermore, from Conditions (3)  in Section \ref{subsec:3.4}, we have: 
\begin{align}
\label{eq:MCQ}
\begin{cases}
    &-1 \leq \tilde{f}_{\phi} + \tilde{f}_1 \leq 1
    \\& -1 \leq \tilde{f}_{\phi} - \tilde{f}_1 \leq 1
    \\& -1 \leq \tilde{g}_{\phi} + \tilde{g}_1 \leq 1
    \\& -1 \leq \tilde{g}_{\phi} - \tilde{g}_1 \leq 1    
\end{cases}
&\quad \quad \quad 
\Rightarrow \quad \quad \quad 
\begin{cases}
    & -2a \leq \tilde{f}_1 \leq 2(1 - a)
    \\& -2(1 - a) \leq \tilde{f}_1 \leq 2a
    \\& -2b \leq \tilde{g}_1 \leq 2(1 - b)
    \\& -2(1 - b) \leq \tilde{g}_1 \leq 2b
\end{cases}
\end{align}
As a result, 
\begin{align}
    \begin{cases}
       &\max\{-2a, -2(1 - a)\} \leq \tilde{f}_1 \leq \min\{2(1 - a), 2a\}
    \\& \max\{-2b, -2(1 - b)\} \leq \tilde{g}_1 \leq \min\{2(1 - b), 2b\} 
    \end{cases}
    \label{ex:1:2}
\end{align}
Hence, using Lemma \eqref{lem:Exp f_ug_v} and Equations $\eqref{ex:1:1}$ and \eqref{ex:1:2},
the set $\mathcal{P}(Q_U,Q_V)$ is given by:
\begin{align*}
 \mathcal{P}(Q_U,Q_V)=  \{Q_{UV}|   Q_U(1)=a,Q_V(1)=b,\frac{2-\zeta_{a,b} - \beta_{ab}}{2} \leq Q(U= V) \leq \frac{2-\zeta_{a,b} + \beta_{ab}}{2}\},
\end{align*}
where $\zeta_{a,b}\triangleq 1-(2a - 1)(2b - 1)$ and $\beta_{a,b} \triangleq  \min\{2a, 2(1 - a)\} \min\{2b, 2(1 - b)\}
$. Consequently, $\mathcal{Q}_1= \bigcup_{a,b\in [0,1]}\mathcal{P}(Q_U,Q_V)$. 
 \begin{figure}[!b]
\centering 
\includegraphics[width=0.8\textwidth]{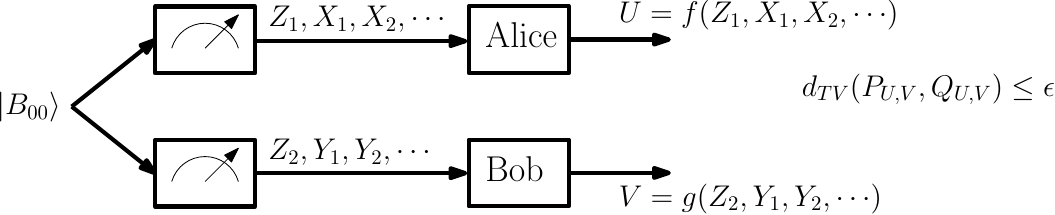}
\caption{A quantum NISS scenario with a Bell state shared among two agents along with unlimited local randomness available to the agents.
%The source sequences $X^d$ and $Y^d$ are IID with distributions $P_X$ and $P_Y$, respectively. . 
}
\label{fig:3n}
\end{figure}

\vspace{.1in}
\noindent\textbf{Case Study 2: Sequential Measurements of Entangled Qubits}

The quantification of classical correlations generated by quantum measurements has been studied extensively, and is closely related to the celebrated Bell's theorem and the  CHSH inequality \cite{clauser1969proposed,cirel1980quantum,nielsen2001quantum}. There has been significant interest and progress in this area in recent years, e.g.,  \cite{slofstra2019set,ji2021mip}. We consider a simple example of classical correlation generation via quantum measurements to illustrate the applicability of the Fourier framework to such problems. 
In this scenario (Figure \ref{fig:3n}), the two agents share an entangled pair of qubits in the
Bell state, $\ket{\beta}_{00}=\frac{\ket{00}+\ket{11}}{2}$ and can perform sequential measurements --- restricted to a specific set of measurements as discussed below --- of their corresponding qubits. Note that in addition to generating correlated measurement outcomes from measuring the entangled qubits, the agents can produce unlimited amounts of local classical randomness $X_i, i \in \mathbb{N}$ and $Y_i, i\in \mathbb{N}$, respectively, by performing subsequent measurements on the (unentangled) output qubits of the first measurement. Each agent performs a  two-outcome measurement on its qubit, and uses the classical measurement output for non-interactive source simulation along with the available local randomness. Let the set of simulatable distributions be denoted by $\mathcal{Q}_2$. We show that $\mathcal{Q}_1=\mathcal{Q}_2$. That is, in this specific scenario, classical randomness generated by quantum measurements does not allow for simulating joint distributions other that those which are simulatable using classical common randomness achieved in Case Study 1.

We restrict the measurements $M_1$ and $M_2$, performed by Alice and Bob, respectively, to have two possible outcomes each, corresponding to $\ket{\phi_i}= \alpha_{0,i}\ket{0}+\alpha_{1,i}\ket{1}, i=\{-1,1\}$ for $M_1$ and  $\ket{\psi_i}= \beta_{0,i}\ket{0}+\beta_{1,i}\ket{1}, i=\{-1,1\}$ for $M_2$, satisfying the completeness relation. We denote the classical output of $M_i$ by $Z_i\in \{-1,1\}$ for $i\in \{-1,1\}$. Then,
\[
P_{Z_1,Z_2}(i,j) = \left|\langle B_{00}|\phi_i \otimes \psi_j\rangle \right|^2=\frac{1}{2}(\alpha_{0,i} \beta_{0,j}+\alpha_{1,i}\beta_{1,j})^2
\]
The completeness relation requires the following:
\begin{align*}
 &\alpha_{0,-1}^2+\alpha_{0,1}^2=1, \quad  \alpha_{1,-1}^2+\alpha_{1,1}^2=1,\quad \alpha_{0,-1}\alpha_{1,-1}+\alpha_{0,1}\alpha_{1,1}=0
 \\& \beta_{0,-1}^2+\beta_{0,1}^2=1, \quad  \beta_{1,-1}^2+\beta_{1,1}^2=1,\quad \beta_{0,-1}\beta_{1,-1}+\beta_{0,1}\beta_{1,1}=0
\end{align*}
Parametrizing with $\theta,\theta'\in [0,2\pi]$, we have:
\begin{align*}
  &\alpha_{0,-1}= sin(\theta), \quad  \alpha_{0,1}= cos(\theta), \quad \alpha_{1,-1}=-cos(\theta), \quad \alpha_{1,1}= sin(\theta) 
  \\&  \beta_{0,-1}= sin(\theta'), \quad 
  \beta_{0,1}= cos(\theta'), \quad \beta_{1,-1}=-cos(\theta'), \quad \beta_{1,1}= sin(\theta').
\end{align*}
Consequently:
\begin{align*}
    &P_{Z_1,Z_2}(-1,-1)=\frac{cos^2(\theta'-\theta)}{2}, \quad \quad P_{Z_1,Z_2}(-1,1)= \frac{sin^2(\theta-\theta')}{2},\\
    &P_{Z_1,Z_2}(1,-1)= \frac{sin^2(\theta-\theta')}{2}, \quad P_{Z_1,Z_2}(1,1)= \frac{cos^2(\theta-\theta')}{2}.
\end{align*}
Note that $Z_1$ and $Z_2$ have symmetric marginal distributions and hence following the arguments in the previous scenario, the constrains in Equation \eqref{eq:MCQ} hold, consequently, 
$\mathcal{Q}_2\subseteq \mathcal{Q}_1$. On the other hand, if $\theta=\theta'$, we achieve one bit of classical common randomness, hence $\mathcal{Q}_1\subseteq \mathcal{Q}_2$. Consequently, $\mathcal{Q}_1=\mathcal{Q}_2$. 

\noindent \textbf{Case Study 3: First Order Markov Sources}

\begin{figure}[!t]
\centering 
\includegraphics[width=0.8\textwidth]{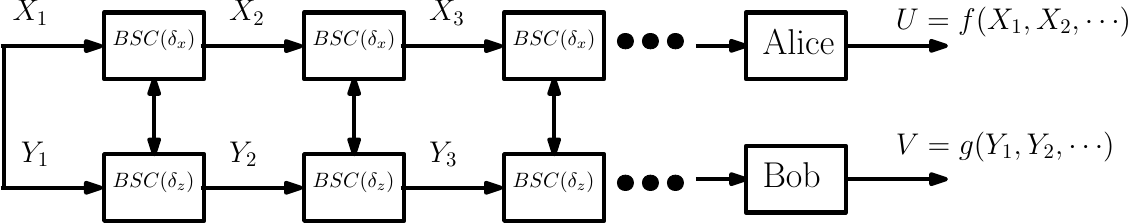}
\caption{An NISS scenario with first-order Markov sources. We have defined $\delta_z\triangleq \delta_x\ast\delta_y$. 
%The source sequences $X^d$ and $Y^d$ are IID with distributions $P_X$ and $P_Y$, respectively. . 
}
\vspace{-.2in}
\label{fig:4n}
\end{figure}

 In this scenario (Figure \ref{fig:4n}) we consider a Markov source $(X^d,Y^d)$, where $X_1=Y_1$ is a uniform common random bit, i.e., $X_1\sim Be(\frac{1}{2})$.  The elements $X_i,Y_i, i>1$ are produced by passing $X_{i-1},Y_{i-1}$ through correlated binary symmetric channels \footnote{For an input $X\in \{-1,1\}$ and bias $\delta\in [0,1]$, a $BSC(\delta)$ produces output $Y\in \{-1,1\}$ with bit-flip probability $\delta$.}. The relation between $(X_i,Y_i)$ and $(X_{i-1},Y_{i-1})$ is given below.  
\begin{align*}
& P(X_i \neq X_{i-1}, Y_i \neq Y_{i-1}) = \delta_x(1-\delta_y),\qquad 
P(X_i \neq X_{i-1}, Y_i = Y_{i-1}) = \delta_x \delta_y \\
&P(X_i = X_{i-1}, Y_i \neq Y_{i-1}) = (1-\delta_x)\delta_y,  \qquad
P(X_i = X_{i-1}, Y_i = Y_{i-1}) = (1-\delta_x)(1-\delta_y),
\end{align*}
 where $\delta_x,\delta_y\in [0,1]$. Under this formulation both sequences $X_i,i \in \mathbb{N}$ and $Y_i, i\in \mathbb{N}$ are first order Markov processes.   

Note that $X_i$ and $Y_i$ have symmetric marginals by construction. Let us define \footnote{In this example, an orthonormal basis is constructed directly. In general, one can start with the basis associated with uniform IID sequences and perform the Gram-Schmidt procedure to derive an orthonormal basis.}
\begin{align*}
   & \psi_1(X^d)\triangleq X_1,\quad \psi_i(X^d)\triangleq \frac{1}{\sigma_1}(X_{i-1}X_i-  2\delta_x+1), \quad i\in \{2,\cdots,d\}
    \\&
    \psi'_1(Y^d)\triangleq Y_1,\quad \psi'_i(Y^d)\triangleq \frac{1}{\sigma_2}(Y_{i-1}Y_i- 2\delta_x
\ast \delta_y+1),\quad  i\in \{2,\cdots,d\}
\\& \phi_{\mathcal{S}}(X^d)= \prod_{i\in \mathcal{S}}\psi_i(X^d), \quad \phi'_{\mathcal{S}}(Y^d)= \prod_{i\in \mathcal{S}}\psi'_i(Y^d), \qquad \mathcal{S}\subseteq [d],
\end{align*}
where $p\ast q\triangleq p(1-q)+q(1-p), p,q\in [0,1]$, $\sigma_1\triangleq 2\sqrt{\delta_x(1-\delta_x)}$ and $\sigma_2\triangleq 2\sqrt{\delta_x\ast \delta_y(1-\delta_x\ast \delta_y)}$. 
It is straightforward to check that $\phi_{\mathcal{S}}(X^d), \mathcal{S}\subseteq [d]$ and  $\phi'_{\mathcal{S}}(Y^d), \mathcal{S}\subseteq [d]$ form an orthonormal basis for the space of functions operating on $X^d$ and $Y^d$, respectively. Furthermore, $\mathbb{E}(\phi_{\mathcal{S}}(X^d)\phi'_{\mathcal{S}'}(Y^d))= \mathbbm{1}(\mathcal{S}=\mathcal{S}') \prod_{i\in \mathcal{S}}\rho_i$, where $\rho_1=1$ and $\rho_i=\rho'\triangleq 4\delta_x(1-\delta_x)(1-2\delta_y)
, i>1$. Consequently, we have:
\begin{align*}
    \mathbb{E}(f(X^d)g(Y^d))= \sum_{\mathcal{S}} f_{\mathcal{S}}g_{\mathcal{S}}{\rho'}^{|\mathcal{S}|-\mathbbm{1}(1\in \mathcal{S})}.
\end{align*}
Given a set of Fourier coefficients $f_{\mathcal{S}},g_{\mathcal{S}},\mathcal{S}\subseteq [d]$, we can find $Q(U=V)$ using the above equation and Lemma \ref{lem:Exp f_ug_v}. This along with the marginals of the target distribution uniquely characterize $Q_{UV}$. In the subsequent sections, we provide an optimization technique over the Fourier coefficients to find the set of simulatable distributions using this characterization. 

\section{Convex-Concave Relaxation and the F-Path Algorithm}
\label{sec:FPATH}

In the previous sections, we have formulated the NISS problem as a quadratic program by writing the primal optimization in Theorems \ref{prop:primal} and \ref{prop:primal_2}. This optimization is NP-hard in general \cite{pardalos1988checking}. In this section, we introduce a path-following method to find approximate solutions to the primal formulation. To elaborate, we relax the problem to a quadratic concave maximization  problem. We then obtain a solution path of a convex-concave problem obtained by linear interpolation of the convex and concave problems, starting from the concave relaxation, to iteratively converge to the global maximum of the convex relaxation. Similar convex-concave optimization approaches have been used in the literature to address quadratic programming optimization problems such as in graph matching \cite{zaslavskiy2008path}, general quadratic assignment problems \cite{anstreicher2001new,blake1987visual}. The relaxation is based on the observation that the optimal point in a convex maximization problem, such as norm optimization over a convex polytope, is located on the boundaries of the optimization region. One can manipulate the original objective function $\mathcal{L}_1$ (e.g., the norm function) into a new objective function $\mathcal{L}_0$ such that its value is unchanged on the boundaries of the optimization region, but it becomes a concave function inside the optimization region. Then, concave maximization over the function $\mathcal{L}_0$ will yield an optimal point $\mathbf{P}_0$ which lies inside the optimization region. This point is used as initial point to optimize the function $\mathcal{L}_\lambda\triangleq \lambda \mathcal{L}_1+(1-\lambda) \mathcal{L}_0$, where $\lambda$ is a small positive number. The resulting optimal point $\mathbf{P}_\lambda$ is used as initial point for the optimization 
of $\mathcal{L}_{\lambda'}\triangleq \lambda' \mathcal{L}_1+(1-\lambda') \mathcal{L}_0$, where $\lambda'=\lambda+d\lambda$ and $d\lambda$ is a small positive value. This process is continued until the value of $\lambda$ reaches one at which point the original objective function $\mathcal{L}_1$ is optimized. For a comprehensive discussion on the choice of $\lambda$ and related optimization methods please refer to \cite{zaslavskiy2008path}.

%n the previous sections, we have formulated the NISS problem as a quadratic program by writing the primal optimization in Theorems \ref{prop:primal} and \ref{prop:primal_2}. This optimization is NP-hard in general \cite{pardalos1988checking}. In this section, we present a path-following method to find approximate solutions to the primal formulation. For this purpose, first we transform the problem into a quadratic convex maximization problem by adding terms that are constant on the boundary of convex region. 

The original objective function is given as
\begin{align*}
    \mathcal{L}(P_{XY},Q_U,Q_V)= \sup_{\substack{(f_{s^d} , s^d\in \mathbb{F}_q^d)\in\mathcal{F}(Q_U)\\(g_{t^d} ,  t^d\in \mathbb{F}_q^d)\in\mathcal{G}(Q_V)}}\sum_{s^d,t^d\in \mathbb{F}_q^d} \left(\tilde{f}_{s^d} \tilde{g}_{{t}^d} \prod_{s,t\in\mathbb{F}_q}\rho_{s,t}^{n(s,t|s^d,t^d)}\right).
\end{align*}
This objective function is neither concave nor convex. From previous sections, we know that the optimal value lies on the boundary of the search space function. The reason is that we started with an optimization on the boundary (over discrete-valued functions) and expanded the search space to the continuous-valued functions using the RD procedure, and proved in Section \ref{sec:Fourier_Tensor_Derand} that this relaxation does not change the optimization solution. We note that for any pair of binary-output  (indicator) functions $f:\mathcal{X}^d\to \{-1,1\}$ and $g:\mathcal{Y}^d\to \{-1,1\}$, we have that $\mathbb{E}[f^2(X^d)]=\sum_{s^d\in  \mathbb{F}_q^d} f^2_{s^d}=1$, and $\mathbb{E}[g^2(Y^d)]=\sum_{s^d\in  \mathbb{F}_q^d} g^2_{s^d}=1$, and for any continuous-valued $\tilde{f}:\mathcal{X}^d\to [-1,1]$ and $\tilde{g}:\mathcal{Y}^d\to [-1,1]$, we have that 
$\mathbb{E}[f^2(X^d)]=\sum_{s^d\in  \mathbb{F}_q^d} f^2_{s^d}\leq 1$, and $\mathbb{E}[g^2(Y^d)]=\sum_{s^d\in  \mathbb{F}_q^d} g^2_{s^d}\leq 1$. Consequently, the following optimization yeilds the same solution as the original optimization for any $\alpha_1,\beta_1>0$: 
\begin{align}
 &\nonumber\mathcal{L}_{1,\alpha_1,\beta_1}(P_{XY},Q_{U},Q_{V})
 \\&\label{eq:convex} \qquad \qquad =\sup_{\substack{(f_{s^d} , s^d\in \mathbb{F}_q^d)\in\mathcal{F}(Q_U)\\(g_{t^d} ,  t^d\in \mathbb{F}_q^d)\in\mathcal{G}(Q_V)}}\sum_{s^d,t^d\in \mathbb{F}_q^d} \left(\tilde{f}_{s^d} \tilde{g}_{{t}^d} \prod_{s,t\in\mathbb{F}_q}\rho_{s,t}^{n(s,t|s^d,t^d)}
 + \alpha_1 \tilde{f}^2_{s^d}
 +\beta_1 \tilde{g}^2_{t^d}\right)-\alpha_1-\beta_1.
\end{align}
We choose the values of $\alpha_1,\beta_1$ large enough so that the objective function $\mathcal{L}_{1,\alpha_1,\beta_1}(P_{XY},Q_{U},Q_{V})$ is convex. Similarly, we define

%where by choosing $\alpha_1$ and $\beta_1$ big enough, the convexity of Equation \ref{eq:convex} is guaranteed. The maximization of a convex function over a relaxed convex domain results in the solution lying on the boundary, which corresponds to the optimal solution of the initial problem.
%In the next step, we identify a concave objective function equivalent to the original objective function on the boundary. However, it should be noted that if we relax the problem into a concave maximization over a convex set, there is no guarantee that the solution will lie on the boundary. We can manipulate the original objective function $\mathcal{L}_1$ (e.g., the norm function) into a new objective function $\mathcal{L}_0$ such that its value is unchanged on the boundaries of the optimization region, but it becomes a concave function inside the optimization region.
\begin{align}
&\nonumber\mathcal{L}_{0,\alpha_0,\beta_0}(P_{XY},Q_{U},Q_{V})=
\\&\label{eq:concave}\qquad \qquad \sup_{\substack{(f_{s^d} , s^d\in \mathbb{F}_q^d)\in\mathcal{F}(Q_U)\\(g_{t^d} ,  t^d\in \mathbb{F}_q^d)\in\mathcal{G}(Q_V)}}\sum_{s^d,t^d\in \mathbb{F}_q^d} \left(\tilde{f}_{s^d} \tilde{g}_{{t}^d} \prod_{s,t\in\mathbb{F}_q}\rho_{s,t}^{n(s,t|s^d,t^d)}
 - \alpha_0 \tilde{f}^2_{s^d}
 -\beta_0 \tilde{g}^2_{t^d}\right)+\alpha_0+\beta_0,
\end{align}
and choose $\alpha_0,\beta_0>0$ large enough such that  $\mathcal{L}_{0,\alpha_0,\beta_0}(P_{XY},Q_{U},Q_{V})$ is concave. 
%where by choosing $\alpha_2$ and $\beta_2$ big enough, the convexity of Equation \ref{eq:concave} is guaranteed.
We then derive a solution path for a convex-concave problem by constructing a new objective function using linear interpolation between the convex and concave problems.
\[
\mathcal{L}_\lambda\triangleq \lambda \mathcal{L}_{1,\alpha_1,\beta_1}+(1-\lambda) \mathcal{L}_{0,\alpha_0,\beta_0},
\quad \lambda \in [0,1].
\]
The resulting Algorithm is described in Algorithm \ref{alg:1}. As seen in the description, the algorithm is initialized with $\lambda=0$ and the optimal point of the objective function $\mathcal{L}_\lambda$ is found by a method such as the Franke-Wolfe algorithm \cite{frank1956algorithm,levitin1966constrained}. Then, we gradually increase $\lambda$ by $d\lambda$.  The choice of constant $\epsilon_{\lambda}$ controls the tradeoff between computational complexity and accuracy in converging to the global optimum \cite{bertsekas1997nonlinear}.  
\\\textbf{Computational Complexity:} The number of iterations needed for the algorithm to reach $\lambda=1$ depends on the choice of  $\epsilon_{\lambda}$. Since by construction we have $\mathcal{L}_{0,\alpha_0,\beta_0}\in [-1,1+\alpha_0+\beta_0]$, the number of iterations in the algorithm is upper-bounded by $\frac{2+\alpha_0+\beta_0}{\epsilon_{\lambda}}$.
The Franke-Wolfe optimization step involves a gradient optimization which has $O(n^3)$ complexity, where $n$ is the number of optimization parameters \cite{frank1956algorithm,levitin1966constrained}.  Note that the number of optimization parameters is $O(\operatorname{exp}(d))$. So the Franke-Wolfe optimization step has complexity $O(\operatorname{exp}(3d))$, where $d$ is the input dimension. In Section \ref{sec:IC_BB}, we show that for BB-NISS scenarios with uniform input marginals, the sample complexity is $d=\Theta(\log{\frac{1}{\epsilon}})$. Characterizing tight bounds on the sample complexity in the general case is an ongoing research direction. 
\\\textbf{Sufficient Conditions and Theoretical Guarantees.} 
Sufficient conditions for global optimality of path-following algorithms were derived in the context of graph matching \cite{zaslavskiy2008path}.
Following the arguments in \cite{zaslavskiy2008path} we show the following proposition which provides sufficient guarantees for optimality of the solution of F-PATH. 
%Subsequently, by iteratively increasing the weight of the convex function in the objective function and decreasing the weight of the concave function—and utilizing the solution from the preceding step as the starting point for each iteration—we approach the optimal solution. 
%Similar convex-concave optimization approaches have been used in the literature to address quadratic programming optimization problems such as in graph matching \cite{zaslavskiy2008path}, general quadratic assignment problems \cite{anstreicher2001new,blake1987visual}. 
%For a comprehensive discussion on the choice of $\lambda$ and related optimization methods please refer to \cite{zaslavskiy2008path}. It is proved that if for some $\lambda^*$ the optimal value of $\mathcal{L}_{\lambda^*}$ acieved on the boundary of the convex region and the function $\mathcal{L}_{\lambda}$ be concave for $\lambda^*$, then that solution is the optimal solution. 

\begin{proposition}[\textbf{Sufficient Conditions for Optimallity of F-PATH}]
    \label{prop:opt}
    The solution of Algorithm \ref{alg:1} is globally optimal if $\exists \lambda^* \in (0, 1)$ such that:
\\i) the function pair $(\tilde{f}^*_{\lambda^*},\tilde{g}^*_{\lambda^*})$ optimizing $\mathcal{L}_{\lambda^*}$ is on the boundary region, i.e., $\sum_{s^d}({\tilde{f}^*_{\lambda^*,s^d}})^2=1$ and $\sum_{t^d}({{{\tilde{g}}^*}_{\lambda^*,t^d}})^2=1$, and
\\ii) the function $\mathcal{L}_{\lambda^*}$ is concave.
\end{proposition}
The proof is provided in Appendix \ref{App:prop:opt}.

\begin{algorithm}[H]
\textbf{1) Initialization:}
\\\text{\hspace{0.2in}}
 $\lambda\triangleq0$
\\\text{\hspace{0.2in}}
 $\bm{P}(0)=\text{argmax }\mathcal{L}_0$ --- concave maximization problem, global maximum is found by Franke-Wolfe algorithm
\\
\textbf{2) Cycle over $\lambda$:}
\\
\text{\hspace{0.2in}}\text{while }{$\lambda<1$}
\\\text{\hspace{0.3in}}$\lambda_\text{new}\triangleq\lambda+d\lambda$
\\\text{\hspace{0.3in}}
\text{if }{$|\mathcal{L}_{\lambda_\text{new}}(\bm{P}(\lambda)) - \mathcal{L}_\lambda(\bm{P}(\lambda))| < \epsilon_\lambda$}
\\\text{\hspace{0.4in}}
$\lambda\triangleq\lambda_\text{new}$
\\
\text{\hspace{0.3in}}
\text{else} $\bm{P}(\lambda_\text{new})=\text{argmin }
\mathcal{L}_{\lambda_\text{new}}$ is found by Frank-Wolfe starting from $\bm{P}(\lambda)$
\\\text{\hspace{0.4in}} $\lambda\triangleq\lambda_\text{new}$
\\
\textbf{Output:}
 $\bm{P}^*=\bm{P}(1)$
\\
\caption{The F-Path Algorithm}
\label{alg:1}
\end{algorithm}

%Using the above representation the optimal solution of $\mathcal{L}_\lambda$ is denoted by $\bm{P}_{\lambda}$ which is optimal vector $(\tilde{f}_{s^d},\tilde{g}_{t^d}, s^d, t^d \in \mathbb{F}_q^d)$.
%The Fourier path-following algorithm (F-Path) is then given in Algorithm \ref{alg:1}. 

\section{Input Complexity for BB-NISS with Uniform Input Marginals}
\label{sec:IC_BB}
\begin{figure}[!b]
\centering 
\includegraphics[scale=0.4]{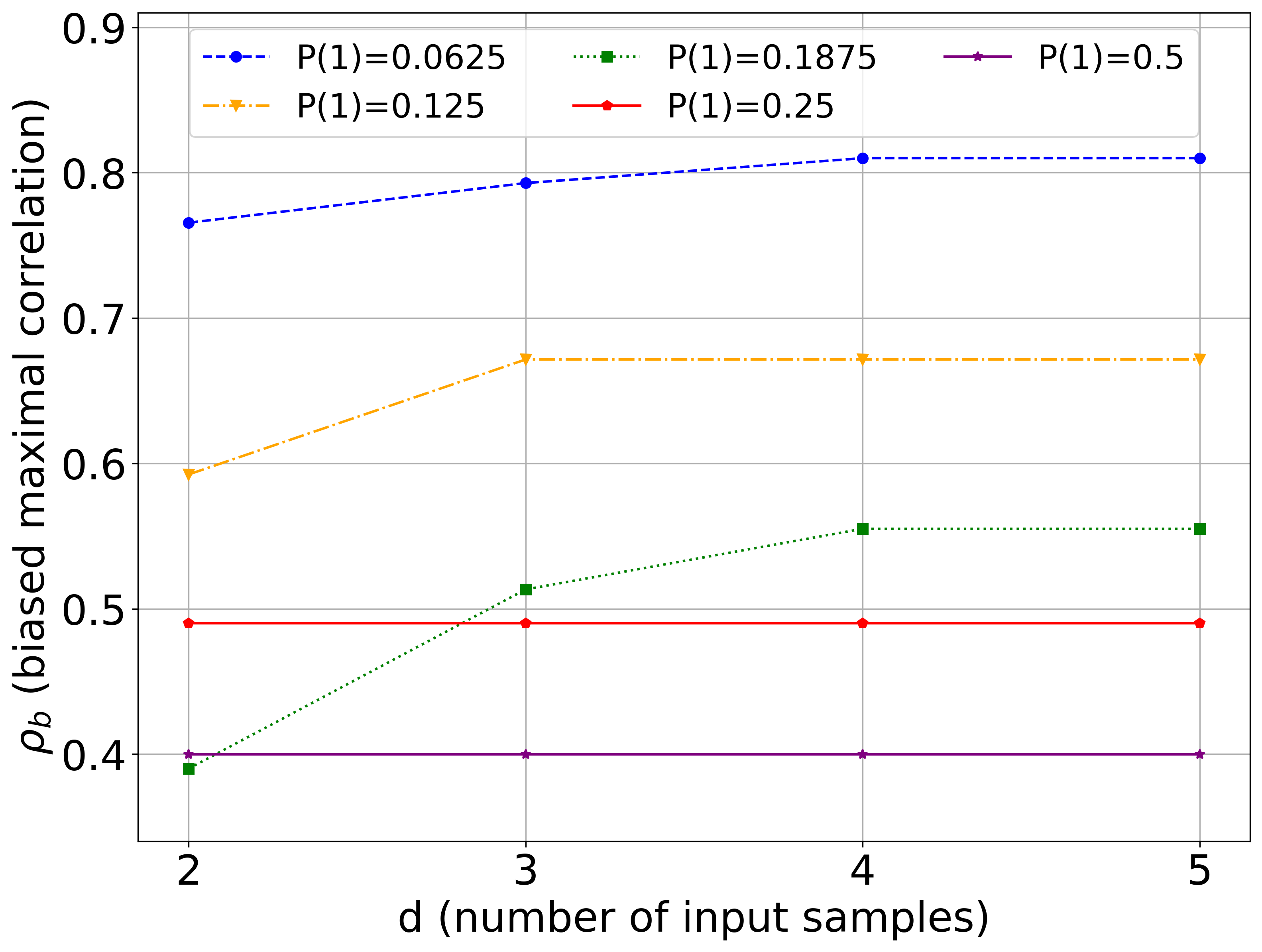}
\caption{Achievable biased maximal correlation $\rho_b$ for the symmetric-output BB-NISS using F-PATH as a function of number of input samples $d$ and output marginals $P_U(1)=P_V(1)=P(1)$. 
}
\label{fig:2}
\end{figure}

So far, we have provided two optimization problems, the dual and primal formulations, for solving the implementability question for NISS.  In the primal(dual) formulation, for a given input length $d\in \mathbb{N}$, a quadratic(linear) program is given for finding the simulating functions achieving maximal correlation. Furthermore, the F-PATH algorithm  provides an efficient path-following algorithm for finding the solution to the quadratic program for a given input length $d\in \mathbb{N}$. To apply these algorithms, one needs to first determine the value of $d$  given a desired total variation distance $\epsilon$. The previous best-known bound on input complexity was $O(\exp \operatorname{poly}(\frac{1}{\epsilon}))$. Our numerical simulations of the F-PATH algorithm and the linear program in the dual formulation suggest that this bound on input complexity is loose (e.g., Figure \ref{fig:2} in Section 
\ref{sec:sim}). In this section, we focus on the BB-NISS problems with uniform input marginals, and show that the input complexity is $\Theta(\log{\frac{1}{\epsilon}})$, hence achieving a super-exponential improvement in this scenario.  

Using Proposition \ref{prop:Sol_BB_NISS}, for a given fixed output marginal pair $Q_U,Q_V$ and $\epsilon>0$, it suffices to characterize the input complexity needed to achieve correlation within $\epsilon$ distance of the biased maximal correlation. We derive this characterization by time-domain analysis in contrast to the Fourier-domain analysis in the previous sections. 
The following lemma provides a time-domain characterization of output correlation in terms of the simulating functions and input correlation coefficient.

\begin{lemma}[\textbf{Time-Domain Characterization of Output Correlation}]
\label{lem:td}
Given binary alphabets $\mathcal{X},\mathcal{Y},\mathcal{U},\mathcal{V}$, symmetric input marginal $P_X(1)=P_Y(1)=\frac{1}{2}$, output marginals $Q_U,Q_V$, and simulating functions $f,g:\{-1,1\}^d\to \{-1,1\}$, the following holds:
\begin{align}
\label{eq:cor:time}
    \mathbb{E}(f(X^d)g(Y^d))=\left(\frac{1+\rho}{4}\right)^d \Big(2\sum_{\substack{x^d,y^d\in \{-1,1\}^d
    \\ f(x^d)=g(y^d)=1}} \beta(x^d,y^d)-2^d\left(1+Q_V(1)+Q_U(1)\right)C_{\rho}\Big),
\end{align}
where $\beta(x^d,y^d)\triangleq \left(\frac{1-\rho}{1+\rho}\right)^{d_H(x^d,y^d)}$, $d_H(x^d,y^d)\triangleq \sum_{i=1}^d \mathbbm{1}(x_i\neq y_i)$, $\rho$ is the correlation coefficient between $X$ and $Y$, $C_{\rho}\triangleq \sum_{\substack{x^d\in \{-1,1\}^d}}\beta(x^d,-\mathbf{i})$, and $\mathbf{i}= (1,1,\cdots,1)$ is the all-ones vector of length $d$. 
\end{lemma}
The proof is provided in Appendix \ref{App:lem:td}. 

The following definition introduces the notion of distance spectrum of a pair of functions which is an effective measure to quantify the correlation between their outputs.
\begin{definition}[\textbf{Distance Spectrum and Dominating Spectrum}]
\label{def:domin}
Given a pair of simulating functions $f_d,g_d:\{-1,1\}^d\to \{-1,1\}$, their distance spectrum is defined as the vector $S(f_d,g_d)\triangleq (n_0,n_1,\cdots,n_d)$, where $n_i\triangleq  |\{(x^d,y^d)| d_H(x^d,y^d)=i, f(x^d)=g(y^d)=1\}|, i\in \{0\}\cup [d]$.  The pair of  simulating functions $(f'_d,g'_d)$ is said to dominate the pair $(f_d,g_d)$ in spectrum if:
\begin{align*}
    \sum_{k=0}^{\ell} n_k\leq  \sum_{k=0}^{\ell} n'_k, \quad \forall \ell\in [d],
\end{align*}
where $S(f',g')=(n'_0,n'_1,n'_2,\cdots,n'_d)$ and $S(f,g)=(n_0,n_1,n_2,\cdots,n_d)$. In  this case, we write $S(f,g)\preceq S(f',g')$\footnote{The relation $\preceq$ resembles the Lorenz ordering relations and majorization which are widely used in various fields such as economics, computability theory, and quantum information theory \cite{arnold2012majorization}.  However, it differs from a Lorenz ordering relation since it does not pre-sort the sequences prior to comparison.}. 
\end{definition}
\begin{proposition}[\textbf{Correlation and Distance Spectrum}]
\label{prop:domin}
Given a joint distribution $P_{XY}$ on binary variables $X,Y$ with uniform marginals, and two simulating pair of functions $(f,g)$ and $(f',g')$, where $f,g,f',g':\{-1,1\}^d\to \{-1,1\}$, the following statements are equivalent:
\\i) $\mathbb{E}(f(X^d)g(Y^d))\leq \mathbb{E}(f'(X^d)g'(Y^d)$.
\\i) $S(f,g)\preceq S(f',g')$. 
\end{proposition}
The proof follows by noting that Equation \eqref{eq:cor:time} can be rewritten as follows: 
\begin{align*}
    \mathbb{E}(f(X^d)g(Y^d))=\left(\frac{1+\rho}{4}\right)^d \Big(2\sum_{k=1}^d n_k\left(\frac{1-\rho}{1+\rho}\right)^{k}-2^d\left(Q_V(1)+Q_U(1)\right)C_{\rho}\Big),
\end{align*}
where $S(f,g)=(n_0,n_1,n_2,\cdots,n_d)$.

In the following, we characterize several classes of correlation-preserving operations on pairs of simulating functions $(f_d,g_d)_{d\in \mathbb{N}}$, which are then used to derive a tight bound on input complexity. A correlation-preserving operator is formally defined below.

\begin{definition}[\textbf{Correlation-Preserving Operator}]
 An operator $\Gamma: \mathcal{L}_{X^d}\to \mathcal{L}_{X^d}$ is called correlation-preserving if  \begin{align*}
    &i) \quad \mathbb{E}(f(X^d)g(Y^d))\leq \mathbb{E}(\Gamma(f(X^d))\Gamma(g(Y^d))
\\    &ii) \quad \mathbb{E}(f(X^d))= \mathbb{E}(\Gamma(f(X^d))),
\end{align*}    
for all input distributions $P_{XY}$ with uniform marginals, and simulating functions $f_d,g_d:\{-1,1\}^d\to \{-1,1\}$.
\end{definition}
We will show that the following classes of operators are correlation-preserving.
\begin{definition}[\textbf{Projection Operator}]
 Given $d\in \mathbb{N}$, the projector operator $\Xi_k:\mathcal{L}^d_X\to \mathcal{L}^d_X, k\in [d]$ is defined as:   
 \begin{align}
 \label{eq:proj_rule}
     \Xi_k(f(x^d))=
     \begin{cases}
         1\quad & \text{ if } x_k=-1 \text{ and } f(x^d)\lor f(\xi_k(x^d))=1\\
         1 & \text{ if } x_k=1 \text{ and }
         f(x^d)\land f(\xi_k(x^d))=1\\
         -1& \text{otherwise}
     \end{cases},
 \end{align}
for all $x^d\in \{-1,1\}^d$ and  $f:\{-1,1\}^d\to \{-1,1\}$, where $\xi_k(x^d)$ is the $k$th bit-flip operator, i.e.,
 $y^d=\xi_k(x^d)$ if and only if $y_i=x_i, i\neq k$ and $y_k\neq x_k$. 
\end{definition}
\begin{definition}[\textbf{Shuffling Operator}]
Given $d\in \mathbb{N}$, and a permutation mapping $\pi:[d]\to [d]$, the shuffling operator $\Pi_{\pi}:\mathcal{L}^d_X\to \mathcal{L}^d_X$ is defined as:
\begin{align*}
    \Pi_{\pi}(f(x^d))= f(\pi(x^d)),\quad  \forall x^d\in \{-1,1\}^d, f:\{-1,1\}^d\to \{-1,1\}.
\end{align*}
\end{definition}
\begin{lemma}[Correlation-Preservation of Projection and Shuffling Operators]
\label{lem:cor_pres}
Given $d\in \mathbb{N}$, the following hold:
\\i) For any given $k\in [d]$, the projection operator $\Xi_k(\cdot)$ is correlation-preserving.
\\ii) For any given bijective mapping $\pi:[d]\to [d]$, the shuffling operator $\Pi_{\pi}$ is correlation-preserving. 
\end{lemma}
The proof is provided in Appendix \ref{App:lem:cor_pres}.

In addition to the correlation-preserving operators described above, we use the following recursive representation of the Hamming distance matrix to characterize the simulating functions achieving biased maximal correlation.
\begin{lemma}[Recursive Representation of the Hamming Distance Matrix]
\label{lem:dist_mat}
Let the Hamming distance matrix for sequences of length $d$ be defined as
$D_d=[d_H(\mathbf{j},\mathbf{k})]_{\mathbf{j},\mathbf{k}\in \{-1,1\}^d}$, where $d_{H}(\cdot,\cdot)$ is the binary Hamming distance measure. Then,
\begin{align}
\label{eq:dist_1}
   D_d = 
 \mathbb{I}_1 \otimes D_{d-1} + D_1 \otimes \mathbb{I}_{d-1},
\end{align}
where, $\mathbb{I}_k=[1]_{i,j\in [2^k]}$ represents the $2^k\times 2^k$ all-ones matrix for $k\in \mathbb{N}$ and $\otimes$ denotes the Kronecker product. Alternatively, we have:
\[
D_d = \sum_{k=0}^{d-1} \mathbb{I}_{d-k-1} \otimes D_1 \otimes \mathbb{I}_k.
\]
where $D_1 = \begin{bmatrix} 0 & 1 \\ 1 & 0 \end{bmatrix}
$.
\end{lemma}
The proof follows by induction and is omitted for brevity. The following presents the main result of this section.
\begin{theorem}[Input Complexity of BB-NISS with Symmetric Input Marginals]
\label{th:IC_BB_NISS}
Consider a BB-NISS scenario with uniform input marginals, i.e., $P_X(1)=P_Y(1)=\frac{1}{2}$. Let $Q^*_{U,V}$ be the simulatable distribution achieving the biased maximal correlation for a desired pair of output marginals $Q_U,Q_V$.   
Define the sequence of lexicographical functions $(f_{L,d},g_{L,d})_{d\in \mathbb{N}}$ associated with $Q_U,Q_V$ as follows:
\begin{align*}
    f_{L,d}(x^d)\triangleq \mathbbm{1}(x^d\sqsubset \mathbf{x}_c(d,Q_U)), \quad    g_{L,d}(y^d)\triangleq \mathbbm{1}(y^d\sqsubset \mathbf{y}_c(d,Q_V)),
\end{align*}
where $\mathbf{x}_c(d,Q_U)$ and $\mathbf{y}_c(d,Q_V)$ are the binary representations of $\lceil2^dQ_U(1)\rceil$ and
 $\lceil2^dQ_V(1)\rceil$, respectively, and $\sqsubset$ denotes the lexicographical ordering relation in the binary vector space.
 \\The following hold:
 \\i) The sequence of lexicographical functions  associated with $Q_U,Q_V$ are a simulating sequence of functions for $Q^*_{U,V}$.
 \\ii) For a given $d\in \mathbb{N}$, let $U_{L}=f_{L,d}(X^d),V_{L}=g_{L,d}(Y^d)$, then:
 \begin{align*}
     d_{TV}(P_{U_L,V_L},Q_{UV})\leq c2^{-d},
 \end{align*}
 for a universal constant $c>0$. 
 \\iii) The input complexity of the BB-NISS problem with uniform output marginals is $\Theta(\log{\frac{1}{\epsilon}})$, where $\epsilon$ denotes an upper-bound on the total variation distance. 
\end{theorem}
The proof is provided in Appendix \ref{App:th:IC_BB_NISS}. 

% \begin{figure}[ht]
% \centering 
% \includegraphics[scale=0.3]{New_Figs/rho_Variable.png}
% \caption{Achievable biased maximal correlation $\rho_b$ for the symmetric input asymmetric output BB-NISS using F-PATH as a function of the number of input samples $d$ and the input correlation $\rho$. Here $P_X(1)=0.5, P_Y(1)=0.5, P_U(1)=0.25, P_V(1)=0.125$. 
% }
% \vspace{-.25in}
% \label{fig:3}
% \end{figure}
\section{Numerical Simulations}
\label{sec:sim}
In this section, we provide simulation results by implementing the F-PATH algorithm. In our simulations, we have $\alpha_0=\beta_0=1, \alpha_1=\beta_1=1.1$ in Equations \eqref{eq:concave} and \eqref{eq:convex}, and we have set $d\lambda = 2\times 10^{-5}$ and $\epsilon_{\lambda}= 0.04$ in Algorithm \ref{alg:1}.

\begin{figure}[!h]
\centering 
\includegraphics[scale=0.35]{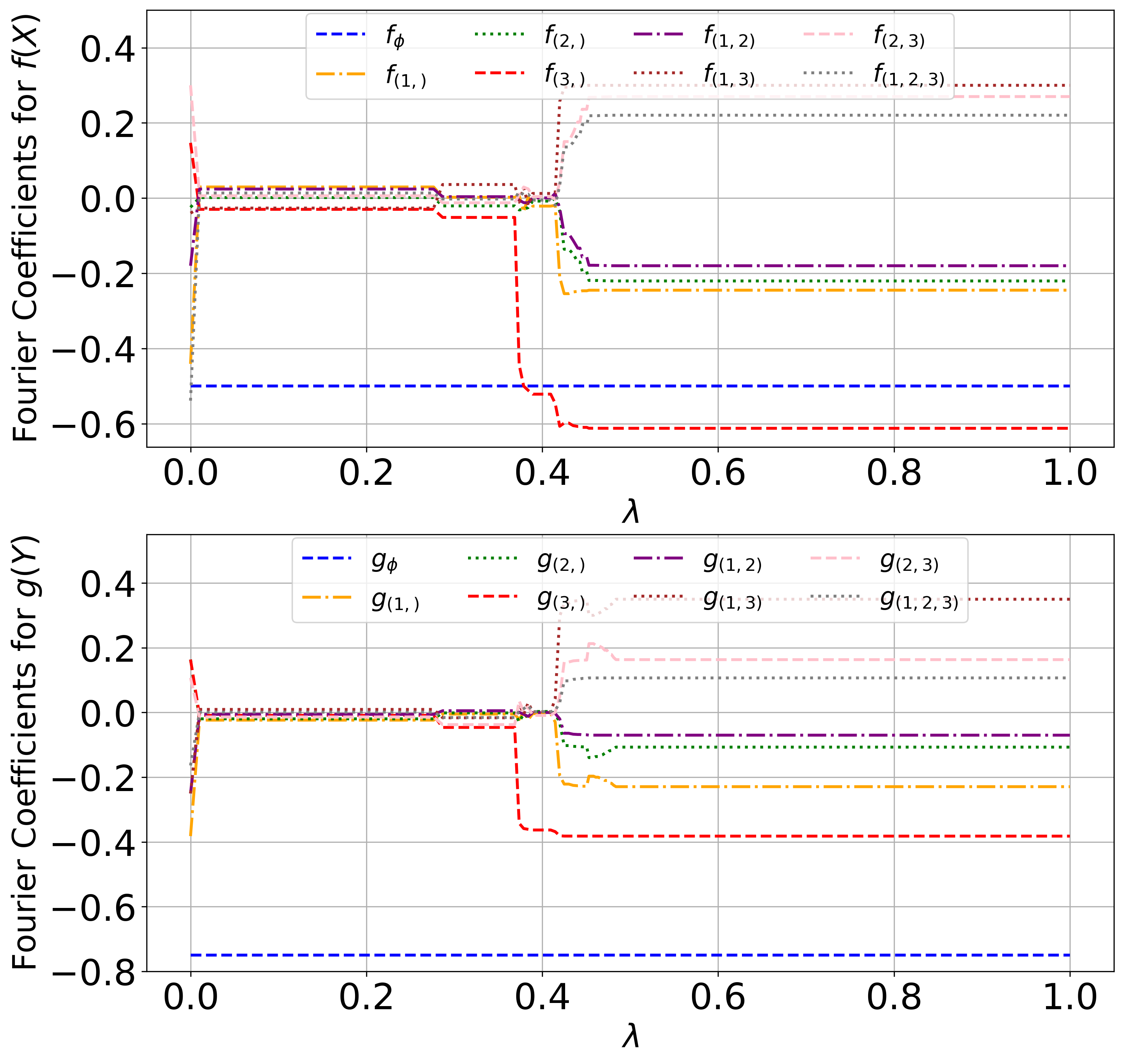}
\caption{Evolution of the Fourier coefficients as a function of $\lambda$ in the F-PATH algorithm. Here $P_X(1)=0.6, P_Y(1)=0.7, P_U(1)=0.25, P_V(1) = 0.125$ and $\rho_{XY}=0.4$.
}
\vspace{-.2in}
\label{fig:5}
\end{figure}

\begin{figure}[!h]
\centering 
\includegraphics[scale=0.3]{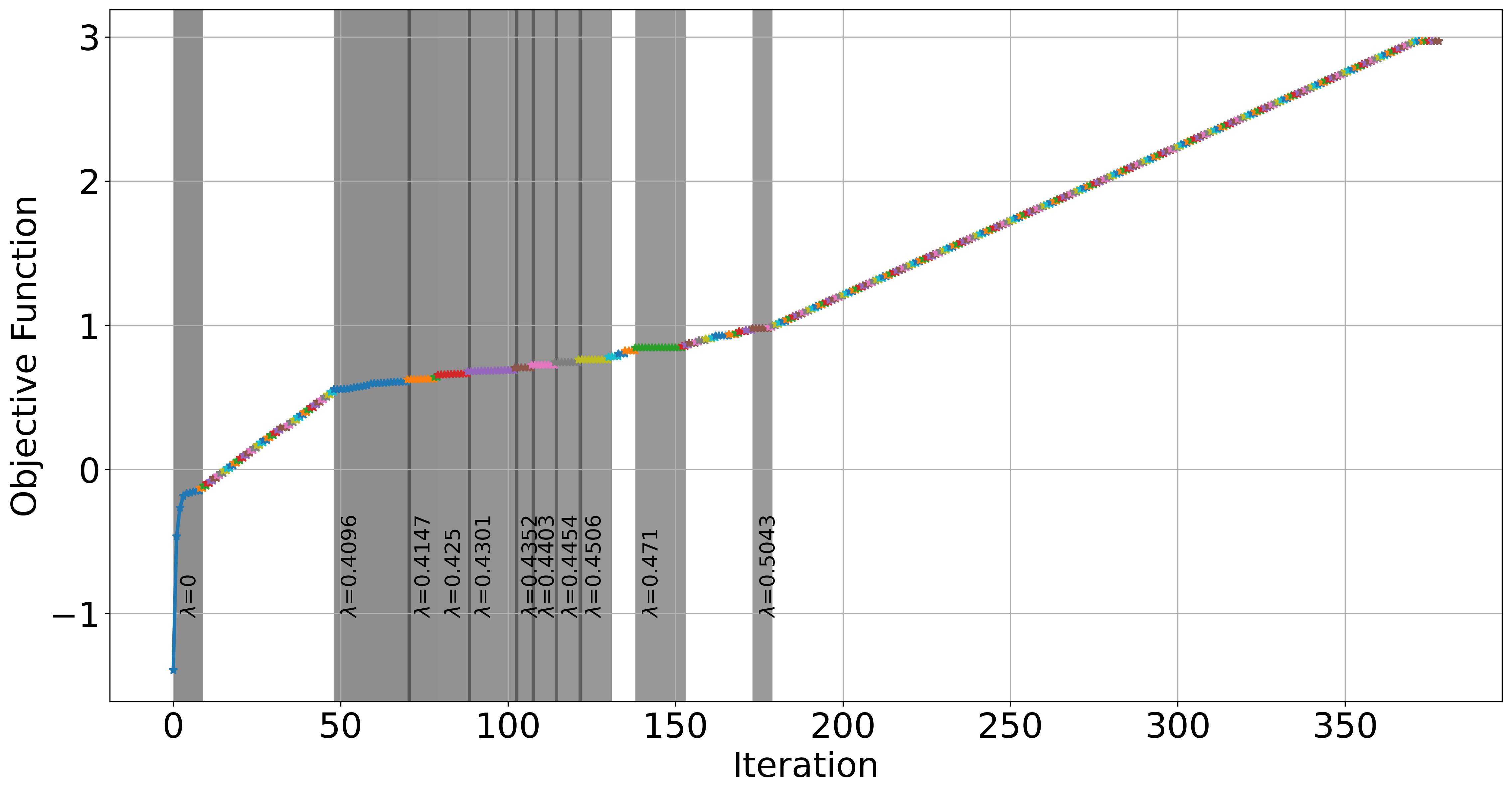}
\vspace{-.1in}
\caption{Evolution of the (shifted) objective function $\mathcal{L}_\lambda+\lambda(\alpha_1+\beta_1)-(1-\lambda)(\alpha_0+\beta_0)$ in the F-PATH algorithm. Here $P_X(1)=0.6, P_Y(1)=0.7, P_U(1)=0.25, P_V(1) = 0.125$ and $\rho_{XY}=0.4$. 
}
\label{fig:6}
\end{figure}

\begin{figure}[!h]
\centering 
\includegraphics[scale=0.55]{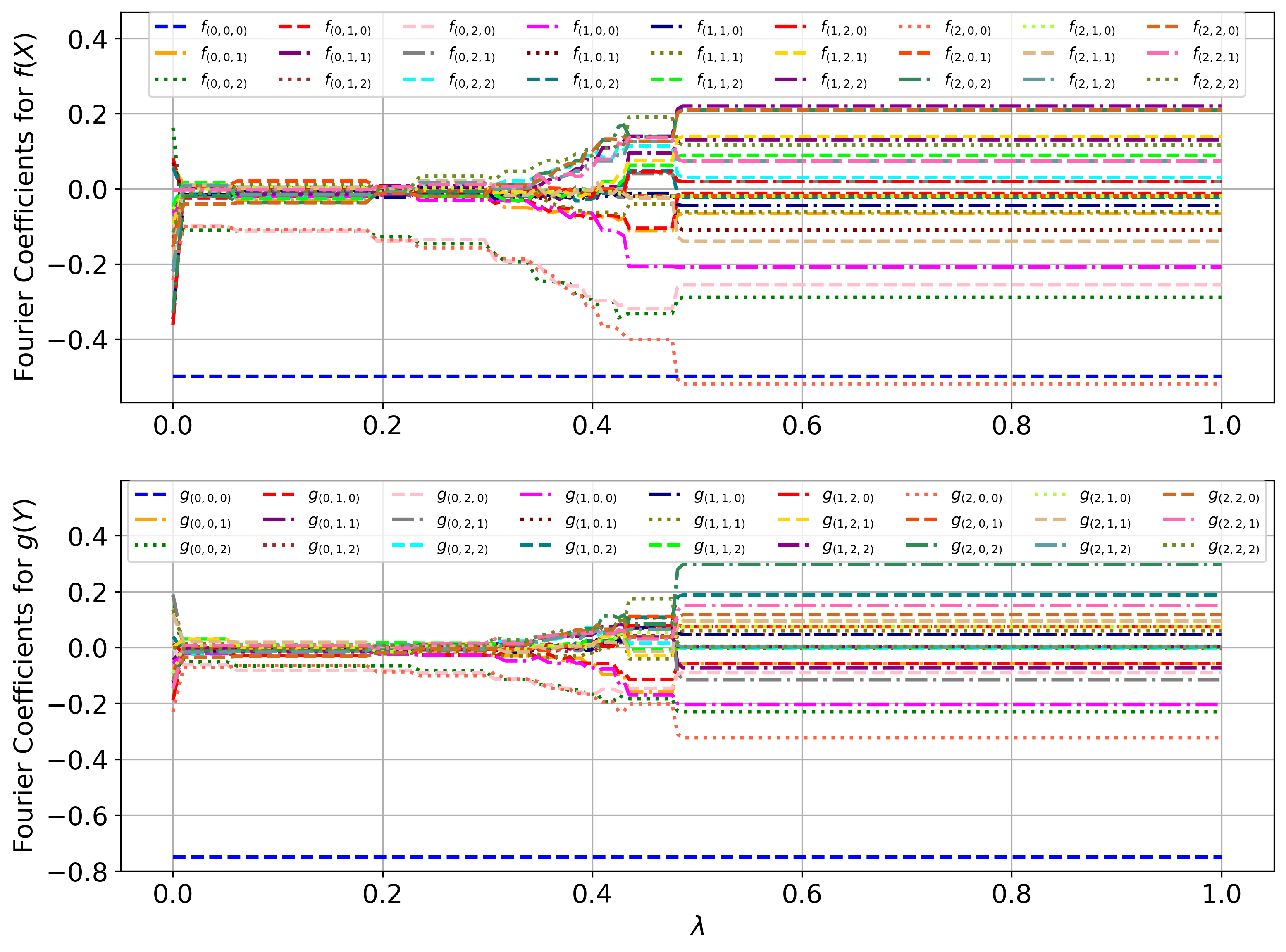}
\caption{Evolution of the Fourier coefficients as a function of $\lambda$ in the F-PATH algorithm. 
}
\label{fig:7}
\end{figure}
We consider the following NISS scenarios:

\paragraph{BB-NISS with Uniform Input Marginals:}
In Section \ref{sec:IC_BB}, we have studied the BB-NISS scenario with uniform input marginals, and explicitly characterized the simulating functions achieving maximal correlation, and the associated input complexity. In Figure \ref{fig:2}, we have numerically simulated the F-PATH algorithm for this scenario, where $\rho_{XY}=0.4$ and $P_U(1)=P_V(1)=P(1) \in\{\frac{1}{2^{-4}}| k=1,2,3,4,8\}$. It can be observed from the proof of Theorem \ref{th:IC_BB_NISS} that when $P(1)=\frac{k}{2^{-4}}$, the optimal simulating functions are produced using $k$ samples. As observed in the simulations of Figure \ref{fig:2}, the output correlation does not increase by increasing the number of samples after this point, confirming the predictions of Theorem \ref{th:IC_BB_NISS}. 
%Figure \ref{fig:3} shows the biased maximal correlation when $P_U(1)= 0.25, P_V(1)=0.125$ and $\rho_{X,Y}\in\{0.1,0.2,0.3,0.4,0.5,0.6,0.7,0.8\}$.

\paragraph{BB-NISS with Non-Uniform Marginals:} We simulate a BB-NISS scenario with non-uniform input marginals, where $P_X(1)=0.6$ and $P_Y(1)=0.7$. 
Figures \ref{fig:5} and \ref{fig:6} present the evolution of the objective function and Fourier coefficients in Algorithm \ref{alg:1}, as  a function of $\lambda$, for $Q_U(1)=Q_V(1)=0.4$ and $\rho=0.3$. It can be observed that the solution falls on the boundary around $\lambda=0.5$. Note that Proposition \ref{prop:opt} shows that if the solution of the F-PATH algorithm falls on the boundary for some $\lambda<1$, and the function $\mathcal{L}_{\lambda}$ is concave for the value of $\lambda$, then the solution is optimal for the original convex objective function $\mathcal{L}_1$.

\paragraph{Numerical Simulations of FB-NISS} We have numerically simulated a ternary-input, binary-output NISS scenario 
with input distribution 
\begin{align*}
&P_{XY}(0,0)=0.133,\quad   P_{XY}(0,1)=0.133, \quad  P_{XY}(0,2)=0.133,
\\& P_{XY}(1,0)=0.2,\quad  P_{XY}(1,1)=0.1, \quad P_{XY}(1,2)=0,
\\&P_{XY}(2,0)=0.066,\quad  P_{XY}(2,1)=0.066,\quad   P_{XY}(2,2)=0.166.
\end{align*}

Recall that in this case, we need to first construct  an orthonormal Fourier basis for the space of functions on $X$ and $Y$ as discussed in Section \ref{sec:Fourier_Tensor_Derand}.  To this end,  we have used the Gram-Schmidt procedure to find the following orthonormal basis functions for Fourier decomposition:
\begin{align*}
    &\psi_0(0)=1, \quad \psi_0(1)=1,\quad  \psi_0(2)=1\\
    &\psi_1(0)=1.225, \quad  \psi_1(1)=-0.816,\quad  \psi_1(2)=-0.816\\
    &\psi_2(0)=0, \quad \psi_2(1)=1.290,\quad  \psi_2(2)=-1.290.
\end{align*}
We run the F-PATH algorithm for $d=3$. Note that in this case, we have $3^d=27$ Fourier coefficients for each agent's simulating function. 
The evaluation of  Fourier coefficients when appying the F-PATH algorithm is shown in Figure \ref{fig:7}. 
The resulting biased maximal correlation is $\rho_b= 0.6454$.

\section{Conclusion and Future Directions}
\label{sec:conc}
A Fourier analysis framework along with a path-following algorithm for solving the non-interactive source simulation (NISS) problem was presented. The input complexity and implementability questions were answered for several classes of NISS scenarios. For binary-output NISS scenarios with doubly-symmetric binary inputs, it was shown that the input complexity is $\Theta(\log{\frac{1}{\epsilon}})$. Furthermore, an explicit characterization of the simulating pair of functions was provided. For general finite-input scenarios, a constructive algorithm, F-PATH, was introduced that explicitly finds the  simulating functions $(f_d(X^d),g_d(Y^d))$.  
Various numerical simulations of NISS scenarios with IID inputs were provided to illustrate the application of the F-PATH algorithm. Furthermore, to illustrate the general applicability of the Fourier framework, several examples of NISS scenarios with non-IID inputs, including entanglement-assisted NISS and NISS with Markovian inputs, were provided and the input complexity and implementability questions were investigated for each scenario. There are several future avenues of research. One direction is to evaluate the input complexity for the general finite-alphabet NISS scenarios, and investigate whether the super-exponential gains observed in input complexity in BB-NISS scenarios with uniform marginals, compared to the previous best-known upper-bounds, can be replicated. Another possible direction of future work is to develop a general framework for non-IID NISS scenarios by designing an efficient algorithm for finding the orthonormal Fourier basis for sources with memory such as Markovian sources. A third direction is to evaluate the set of classical correlations which can be generated in entanglement assisted scenarios. A question of interest in this area is whether the availability of the entanglement resource enlarges the set of simulatable distributions compared to classical resources such as limited common randomness, and to quantify such gains in different NISS scenarios. 
\appendices
\section{Proof of Lemma \ref{lem:Exp f_ug_v}}
\label{App:Lem:4}
  Since, $f_u(x^d),g_v(y^d)\in \pmm$, we have 
    \begin{align*}
 \mathbb{E}(f_{u}(X^d)g_{v}(Y^d))&= \prob{f_u(X^d)=g_v(Y^d)}-\prob{f_u(X^d)\neq g_v(Y^d)}
 \\&= 2 \prob{f_{u}(X^d) = g_{v}(Y^d)}-1.
\end{align*}
Note that 
\begin{align*}
    &\prob{f_{u}(X^d) = g_{v}(Y^d)} = \prob{f_{u}(X^d) = g_{v}(Y^d)=1}+ \prob{f_{u}(X^d) =  g_{v}(Y^d)=-1}
    \\& = \prob{f_{u}(X^d) = g_{v}(Y^d)=1}+\prob{f_u(X^d)=-1}- \prob{f_u(X^d)=-1, g_v(Y^d)= 1}
    \\& = \prob{f_{u}(X^d) = g_{v}(Y^d)=1}+\prob{f_u(X^d)=-1}
    - \prob{g_v(Y^d)=1}+ \prob{f_u(X^d)=1, g_v(Y^d)= 1}
        \\& =2 \prob{f_{u}(X^d) = g_{v}(Y^d)=1}-\prob{f_u(X^d)=1} - \prob{g_v(Y^d)=1}+ 1
\end{align*}
This completes the proof. 
$\qed$
\section{Proof of Lemma \ref{lem:8}}
\label{App:lem:8}
    We show that  $\mathcal{P}(P_{XY},Q_U,Q_V)$ is star-convex with the independent distribution $Q_UQ_V$ as its center. To show this, choose an arbitrary $Q_{UV}\in \mathcal{P}(P_{XY},Q_U,Q_V)$ and let $\epsilon>0$. Since $Q_{UV}$ is simulatable, we conclude that there exist $d\in \mathbb{N}$ and simulating pair of functions $(f_d,g_d)$ such that $d_{TV}(P_{U_dV_d},Q_{UV})\leq \epsilon$, where $U_d\triangleq f_d(X^d), V_d\triangleq g_d(Y^d)$. Furthermore, using standard results from single-source simulation it is known that Alice and Bob can each produce a source with marginal $Q_U$ ad $Q_V$, respectively \cite{VonNeumann1951,Elias1972,Knuth1976,Cover}. Let $f'_{d'}$ and $g'_{d''}$ be such that $d_{TV}(P_{U'_{d'}},Q_U)\leq \epsilon'$ and  $d_{TV}(P_{V'_{d''}},Q_V)\leq \epsilon'$, where $U'_{d'}\triangleq f'_{d'}(X_{d+1}^{d+d'})$ and $V'_{d''}\triangleq g'_{d''}(Y_{d+d'+1}^{d+d'+d''})$, and the value of $\epsilon'$ will be determined later. We have:
    \begin{align}
&  \label{eq:lem8:1}     d_{TV}(P_{U'_{d'}},Q_U)\leq \epsilon'\rightarrow \sum_{u}|P_{U'_{d'}}(u)-Q_U(u)|\leq \epsilon' \\
& \label{eq:lem8:2}  d_{TV}(P_{V'_{d''}},Q_V)\leq \epsilon' \rightarrow \sum_{v}|P_{V'_{d''}}(v)-Q_V(v)|\leq \epsilon'
    \end{align}
From Equations \eqref{eq:lem8:1} and \eqref{eq:lem8:2}, we get:
\begin{align}
 \label{eq:lem8:2.5}
    \sum_{u,v} |P_{U'_{d'}}(u)P_{V'_{d''}}(v)-Q_U(u)P_{V'_{d''}}(v)-P_{U'_{d'}}(u)Q_V(v)+Q_U(u)Q_V(v)|< {\epsilon'}^2.
\end{align}
Furthermore, from Equations \eqref{eq:lem8:1} and \eqref{eq:lem8:2}, we get:
    \begin{align}
&   \label{eq:lem8:3}    \sum_{u,v}P_V(v)|P_{U'_{d'}}(u)-Q_U(u)|\leq \epsilon' \\
&  \label{eq:lem8:4} \sum_{u,v}P_U(u)|P_{V'_{d''}}(v)-Q_V(v)|\leq \epsilon'
    \end{align}
Adding Equations \eqref{eq:lem8:2.5}-\eqref{eq:lem8:4} and using the triangle inequality, we have:
\begin{align*}
    d_{TV}(P_{U'_{d'}}P_{V'_{d''}}, Q_{U}Q_V)\leq 2\epsilon'+{\epsilon'}^2.
\end{align*}
We choose $\epsilon'$ such that $2\epsilon'+{\epsilon'}^2=\epsilon$. Next, Alice and Bob each produce local coins $C_{p_1}$ and $C'_{p_1}$ with bias $p_1\in [0,1]$, respectively. That is, $\PP(C_{p_1}=1)=\PP(C'_{p_1}=1)=p_1$ and $\PP(C_{p_1}=-1)=\PP(C'_{p_1}=-1)=1-p_1$. Alice sets $U''= U_d$ if $C_{p_1}=1$ and $U''= U'_{d'}$, otherwise. Similarly, Bob sets $V''= V_d$ if $C'_{p_1}=1$ and $V''=V'_{d''}$, otherwise. We have:
\begin{align*}
    &P_{U'',V''}= p_1^2 P_{U_dV_d}+ (1-p_1)p_1 P_{U'_{d'}}P_{V_d}+ p_1(1-p_1)P_{U_d}P_{V_{d''}}+(1-p_1)^2P_{U'_{d'}}P_{V'_{d''}}
    \\&\Rightarrow |P_{U'',V''}- (p_1^2 P_{U_dV_d}+(1-p_1^2) P_{U_d}P_{V_d})|\leq \epsilon,
\end{align*}
where in the last equality, we have used the fact that all distributions in $\mathcal{P}(P_{XY},Q_U,Q_V)$ have the same marginal. Consequently, by choosing value of $p_1$ in the unit interval, we conclude that $\lambda P_{U_dV_d}+(1-\lambda) P_{U_d}P_{V_d} $ is simulatable for all $\lambda\in [0,1]$, and the set $\mathcal{P}(P_{XY},Q_U,Q_V)$ is star-convex with $Q_UQ_V$ as its center. 
$\qed$

% \section{Proof of Lemma \ref{Lem:6}}
% \label{App:Lem:6}
% Starting from the law of total expectations, we have that
%     \begin{align*}
%         \EE(UV) = \EE(\EE(UV|X^d, Y^d)) = \EE(\EE(U|X^d) \EE(V | Y^d) ) = \EE(P_{U|X^d}(1 | X^d) P_{V|Y^d}(1 | Y^d)).
%     \end{align*}
%     Note that the Fourier expansion of $P_{U|X^d}(1 | X^d)$ is $\sum_{\CS \subseteq [d]} u_{\CS} \pset{\CS}(x^d)$. Similarly, for $P_{V|Y^d}(1 | Y^d)$ we can write the Fourier expansion as $\sum_{\CS \subseteq [d]} u_{\CS} \psi_\CS(y^d)$ where $\psi_\CS$ are the parities with respect to the marginal $P_Y$. As a result, the term above equals to 
%     \begin{align*}
%         \EE(P_{U|X^d}(1 | X^d) P_{V|Y^d}(1 | Y^d)) = \sum_{\CS, \CT \subseteq [d]} u_\CS v_\CT \EE(\pS(X^d) \psi_\CT(Y^d)). 
%     \end{align*}
%     Note that for $\CS\neq \CT$
%     \begin{align*}
%         \EE(\pS(X^d) \psi_\CT(Y^d)) = \EE\Big(\prod_{j\in \CS}\prod_{i\in \CT} \frac{X_j-\mu_X}{\sigma_X}\frac{Y_i-\mu_Y}{\sigma_Y}\Big)=0,
%     \end{align*}
%     where we used the independentness of the samples across time. When $\CS=\CT$, the above expression equals to $\prod_{i\in \CS} \EE(\frac{X_i-\mu_X}{\sigma_X}\frac{Y_i-\mu_Y}{\sigma_Y}) = \rho^{|\CS|},$ where $\rho$ is the correlation coefficient of $(X,Y)$. Therefore, the proof is completed by noting that: 
%     \begin{align*}
%          \EE(UV) = \sum_{\CS, \CT \subseteq [d]} u_\CS v_\CT \rho^{|\CS|}.
%     \end{align*}

\section{Proof of Lemma \ref{prop:4}}
\label{App:prop:4}
Clearly, $\Psi(\cdot)$ is an injective mapping since it is an affine transformation of $Q_{UV}$. 
We show that  i) $ \Psi(Q_{UV})\in \ENISS, \forall Q_{UV}\in \mathcal{P}(P_{XU},Q_U,Q_V)$, and ii)  $ \Psi(Q_{UV}) $ is surjective on  $\ENISS$. To prove (i), note that since $Q_{UV}$ is a simulatable target distribution, from Definition \ref{def:NISS}, there exists a sequence of functions $(f_d, g_d)_{d\in \NN}$ such that \eqref{eq:NISS TV} holds. For each $d\in \NN$, consider the components $f_d$ and $g_d$ in the overparametrized representation of $f_d$ and $g_d$, that is $f_d\equiv (f_{d,u})_{u\in \mathcal{U}}$ and $g_d\equiv (g_{d,v})_{v\in \mathcal{V}}$.  Then, from Lemma \ref{lem:Exp f_ug_v}, for any $(u,v)\in \CU\times \CV$ we have that  \begin{align*}
\mathbb{E}(f_{d,u}(X^d)g_{d,v}(Y^d)) &= 4P_{U_d V_d}(u,v)-2(P_{U_d}(u)+P_{V_d}(v))+1,
\end{align*}
where $U_d=f_d(X^d)$ and $V_d = g_d(Y^d)$. As a result, from \eqref{eq:NISS TV}, by taking the limit, we conclude:
 \begin{align}
\lim_{d\to \infty} \mathbb{E}(f_{d,u}(X^d)g_{d,v}(Y^d)) &= 4Q_{UV}(u,v)-2(Q_u(u)+Q_V(v))+1 = \Psi(Q_{UV}).
\label{eq:5}
\end{align}
Consequently, $\Psi(Q_{UV})\in \mathcal{P}(P_{XU},Q_U,Q_V)$. 
To show (ii), we argue that  for any $(e_{u, v})_{u,v\in \mathcal{U}\times \mathcal{V}}\in \ENISS$ there exist a $Q_{UV} \in \QNISS$ such that $\Psi(Q_{U V}) = (e_{u, v})_{u,v\in \mathcal{U}\times \mathcal{V}}$. To see this, let  $(\tilde{f}_{u,d}, \tilde{g}_{v,d})_{u,v\in \mathcal{U}\times \mathcal{V}, d\in \NN} \in \FBoolQUV(Q_U,Q_V)$ be the sequence of functions that generate $(e_{u, v})_{u,v}$ as in \eqref{eq:ENISS}. Then, using the RD procedure described in the prequel, these simulation functions generate a target distribution $Q_{UV} \in \mathcal{P}(P_{XY},Q_U,Q_V)$ since the limit $\lim_{d\to \infty} \mathbb{E}({f}_{d,u}(X^d)g_{d,v}(Y^d))$ exists for the derandomizad functions $(f_{d,u},g_{d,v})_{u,v\in \mathcal{U}\times \mathcal{V},d\in \mathbb{N}}$. As a result,  $\Psi(Q_{UV}) = (e_{u, v})_{u,v}$. 
$\qed$

\section{Proof of Proposition \ref{Cor:Sol_BB_NISS}}
\label{App:Cor:15}
The fact that the NISS problem is not solvable for any distribution for which there does not exist $Q'_{U,V}$ such that $d_{TV}(Q_{UV},Q'_{UV})\leq \epsilon$ and $Q'_{U,V}\in \mathcal{S}(P_{XY},Q_U,Q_V)$ follows from the data processing property in Proposition \ref{prop:3} (e.g., see \cite{ghazi2016decidability}). We show that if such  $Q'_{U,V}$ exists, then the NISS problem is solvable. The proof method forms the foundation for several of the proofs in the rest of the paper. 
To this end, let us consider the $U$ and $V$ defined in  \eqref{eq:Sym_BB_NISS} and denote their joint distribution by $P_{UV}$. It suffices to show that $P_{UV}=Q_{UV}$. We have:
\begin{align*}
    P(U=1)= \lambda \mathbb{E}(p_1)+(1-\lambda)(\frac{1}{2})= \frac{1}{2},\quad 
     P(V=1)= \lambda \mathbb{E}(p_2)+(1-\lambda)(\frac{1}{2})= \frac{1}{2},
\end{align*}
\vspace{-.3in}

where we have used the fact that $\mathbb{E}(f(X^d))=\mathbb{E}(g(Y^d))=0$ to conclude that $\mathbb{E}(p_1)=\mathbb{E}(p_2)=\frac{1}{2}$. Furthermore, 
\begin{align*}
    P(U\neq V)&= \frac{1-\mathbb{E}(UV)}{2}
= \frac{1-\lambda^2\mathbb{E}(C_X(p_1)C_Y(p_2))}{2} 
\end{align*}
Note that given $p_1, p_2$, the coins $C_X(p_1), C_Y(p_2)$ are independently generated as non-overlapping samples are used. However, $p_1$ and $p_2$ themselves are correlated. Therefore, as the coins take values from $\pmm$, we have that
\begin{align*}
    \mathbb{E}(C_X(p_1) C_Y(p_2)) &=\prob{C_X(p_1) = C_Y(p_2)} - \prob{C_X(p_1)\neq C_Y(p_2)}\\
    &=\EE\Big(p_1p_2 +(1-p_1)(1-p_2)\Big) - \EE\Big((1-p_1)p_2 +p_1(1-p_2)\Big)
    \end{align*}
    \begin{align*}
&=\mathbb{E}\Big(\left(\frac{1+\tilde{f}(X^d)}{2}\right)\left(\frac{1+\tilde{g}(Y^d)}{2}\right)+\left(\frac{1-\tilde{f}(X^d)}{2}\right)\left(\frac{1-\tilde{g}(Y^d)}{2}\right)
    \\&-\left(\frac{1-\tilde{f}(X^d)}{2}\right)\left(\frac{1+\tilde{g}(Y^d)}{2}\right)-\left(\frac{1+\tilde{f}(X^d)}{2}\right)\left(\frac{1-\tilde{g}(Y^d)}{2}\right)\Big)
    \\&= \mathbb{E}(\tilde{f}(X^d)\tilde{g}(Y^d))= \sum_{i,j\in [q-1]}\tilde{f}^*_i\tilde{g}^*_j\rho_{i,j}= \rho(\mathcal{X},\mathcal{Y},P_{XY}),
    \end{align*}
    where in the last equality, we have used the fact that $\tilde{f}_{0}=\mathbb{E}(\tilde{f}(X^d))=\mathbb{E}(U)=0$ and
    $\tilde{g}_{0}=\mathbb{E}(\tilde{g}(Y^d))=\mathbb{E}(V)=0$. 
%    \begin{align*}
%    \\&= \mathbb{E}(f(X^d)g(Y^d))= \mathbb{E}\Big(\big(\frac{X_1-\mathbb{E}(X)}{\sqrt{Var(X)}}\big)\big(\frac{Y_1-\mathbb{E}(Y)}{\sqrt{Var(Y)}}\big)\Big)
%    = \rho(\mathcal{X},\mathcal{Y},P_{X,Y}). 
%\end{align*}
As a result, from the definition of $\lambda$, we have that 
\begin{align*}
    P(U\neq V)= \frac{1-\lambda^2\rho(\mathcal{X},\mathcal{Y},P_{XY})}{2}= \frac{1-\rho(\mathcal{U},\mathcal{V},Q'_{U,V})}{2}.
\end{align*}
We have shown that $P_U(1)=Q_U(1)$, $P_V(1)=Q_V(1)$ and $P(U\neq V)=Q'(U\neq V)$. Hence, $P_{UV}=Q'_{UV}$.

Particularly, for BB-NISS, to find $\tilde{f}_1^*, \tilde{g}_1^*$, one needs to optimize $\tilde{f}_1\tilde{g}_1\rho$, subject to the following:
\begin{align*}
  &  -1\leq \tilde{f}_1\frac{1-\mathbb{E}(X)}{\sqrt{Var(X)}}\leq 1, \quad -1\leq \tilde{f}_1\frac{-1-\mathbb{E}(X)}{\sqrt{Var(X)}}\leq 1,\\ 
  & -1\leq \tilde{g}_1\frac{1-\mathbb{E}(Y)}{\sqrt{Var(Y)}}\leq 1, \quad -1\leq \tilde{g}_1\frac{-1-\mathbb{E}(Y)}{\sqrt{Var(Y)}}\leq 1.
\end{align*}
It is straightforward to see that $\frac{-\sqrt{Var(X)}}{1+|\mathbb{E}(X)|}\leq \tilde{f}_1\leq \frac{\sqrt{Var(X)}}{1+|\mathbb{E}(X)|}$. Consequently, to maximize $\tilde{f}^*_1\tilde{g}^*_1\rho$, we get $\tilde{f}(X^d)= \tilde{f}_1^*\phi_1(X_1)= \frac{\sqrt{Var(X_1)}}{1+|\mathbb{E}(X)|}\frac{X_1-\mathbb{E}(X)}{\sqrt{Var(X)}}= \frac{X_1-\mathbb{E}(X)}{1+|\mathbb{E}(X)|}$. Similarly, $\tilde{g}(Y^d)= \frac{Y_1-\mathbb{E}(Y)}{1+|\mathbb{E}(Y)|}$. 
This completes the proof.
$\qed$
\section{Proof of Corollary \ref{cor:28}}
\label{App:Cor:28}

    The proof follow by the Cauchy-Schwarz inequality which implies that 
    \[\sum_{\mathcal{S}\subseteq [d]}\fS \gS \rho^{|\mathcal{S}|}= \sum_{\mathcal{S}\subseteq [d]}(\fS \rho^{\frac{|\mathcal{S}|}{2}})(\gS \rho^{\frac{|\mathcal{S}|}{2}})\leq \sqrt{(\sum_{\mathcal{S}\subseteq [d]}f^2_{\mathcal{S}}\rho^{|\mathcal{S}|})(\sum_{\mathcal{S}\subseteq [d]}g^2_{\mathcal{S}}\rho^{|\mathcal{S}|})},\]

    with equality if and only if $(\fS ,\mathcal{S}\subseteq[d])$ is parallel to $(\gS ,\mathcal{S}\subseteq[d])$.
    So, 
\begin{align*}
 \rho_b(P_{XY},Q_{U},Q_{V})&=\sup_{d\in \mathbb{N}}\sup_{\substack{(\fS , \mathcal{S}\subseteq [d])\in\mathcal{F}(Q_U)\\(\gS , \mathcal{S}\subseteq [d])\in\mathcal{G}(Q_V)}}\sum_{\mathcal{S}\subseteq[d]} \fS \gS \rho^{|\mathcal{S}|}
 \\& \leq \sup_{d\in \mathbb{N}}\sup_{(\fS , \mathcal{S}\subseteq [d])\in\mathcal{F}(Q_U)}\sqrt{\sum_{\mathcal{S}\subseteq [d]}f^2_{\mathcal{S}}\rho^{|\mathcal{S}|}}\sup_{(\gS , \mathcal{S}\subseteq [d])\in\mathcal{G}(Q_V)}\sqrt{\sum_{\mathcal{S}\subseteq [d]}g^2_{\mathcal{S}}\rho^{|\mathcal{S}|}}.
\end{align*}
Note that since by assumption $Q_U=Q_V$, we have $\mathcal{F}(Q_U)=\mathcal{G}(Q_V)$.
So, 
\[\sup_{(\fS , \mathcal{S}\subseteq [d])\in\mathcal{F}(Q_U)}\sqrt{\sum_{\mathcal{S}\subseteq [d]}f^2_{\mathcal{S}}\rho^{|\mathcal{S}|}}=\sup_{(\gS , \mathcal{S}\subseteq [d])\in\mathcal{G}(Q_V)}\sqrt{\sum_{\mathcal{S}\subseteq [d]}g^2_{\mathcal{S}}\rho^{|\mathcal{S}|}}.\] 
As a result,
\begin{align*}
 \rho_b(P_{XY},Q_{UV})= \sup_{d\in \mathbb{N}}\sup_{(\fS , \mathcal{S}\subseteq [d])\in\mathcal{F}(Q_U)}{\sum_{\mathcal{S}\subseteq [d]}f^2_{\mathcal{S}}\rho^{|\mathcal{S}|}},
\end{align*}
where equality is achieved by noting that the optimal $(\fS ,\mathcal{S}\subseteq[d])$ is parallel to  the optimal $(\gS ,\mathcal{S}\subseteq[d])$ (in fact they are equal with each other).
$\qed$
\section{Proof of Theorem \ref{prop:dual}}
\label{app:th:dual}

We start with the following Lagrangian function for the optimization in 
\eqref{eq:primal}:
\begin{align*}
  &\mathcal{L}(\tilde{f}(\cdot),\tilde{g}(\cdot),\lambda_{{f}}^+(\cdot),\lambda_{{f}}^-(\cdot),\lambda_{{g}}^+(\cdot),\lambda_{{g}}^-(\cdot))
  = \sum_{s^d,t^d\in \mathbb{F}_q^d}\tilde{f}_{s^d}\tilde{g}_{t^d} \prod_{s,t\in \mathbb{F}_q}\rho_{s,t}^{n(s,t|s^d,t^d)}
  \\&-\frac{1}{q^d}\sum_{x^d\in \mathbb{F}_q^d}
  \lambda^+_{{f}}(x^d)(\tilde{f}(x^d)-1)
  -\frac{1}{q^d}\sum_{x^d\in \mathbb{F}_q^d}
  \lambda^-_{{f}}(x^d)(-\tilde{f}(x^d)-1)-
  \frac{1}{q^d}\sum_{y^d\in \mathbb{F}_q^d}
  \lambda^+_{{g}}(y^d)(\tilde{g}(y^d)-1)
  \\&-  \frac{1}{q^d}\sum_{y^d\in \mathbb{F}_q^d}
  \lambda^-_{{g}}(y^d)(-\tilde{g}(y^d)-1).
\end{align*}
The KKT optimality conditions are as follows:
\begin{itemize}[leftmargin=*]
    \item \textbf{Stationarity:}
    \begin{align}
\label{eq:stat_1}      
&\forall s^d\neq \mathbf{0}: \frac{\partial \mathcal{L}}{\partial \tilde{f}_{s^d}}=0 \Rightarrow \sum_{t^d\in \mathbb{F}_q^d}\tilde{g}_{t^d}\prod_{s,t}\rho_{s,t}^{n(s,t|s^d,t^d)}-\frac{1}{q^d}\sum_{x^d\in \mathbb{F}_q^d}(\lambda^+_{f}(x^d)-\lambda^-_{f}(x^d))\chi_{s^d}(x^d)=0\\
&    \label{eq:stat_2}      \forall s^d\neq \mathbf{0}: \frac{\partial \mathcal{L}}{\partial \tilde{g}_{s^d}}=0 \Rightarrow \sum_{t^d\in \mathbb{F}_q^d}\tilde{f}_{t^d}\prod_{s,t}\rho_{s,t}^{n(s,t|s^d,t^d)}-\frac{1}{q^d}\sum_{x^d\in \mathbb{F}_q^d}(\lambda^+_{g}(x^d)-\lambda^-_{g}(x^d))\chi_{s^d}(x^d)=0
    \end{align}

Now we apply the (uniform) Fourier expansion on each of the KKT coefficients $\lambda^+_{f}, \lambda^-_{f}, \lambda^+_{g}$ and $\lambda^-_{g}$. We use uniform Fourier as the marginals are uniform. The Fourier coefficients of the KKT coefficients are then equal to 
\begin{align*}
&\lambda^+_{f,s^d} =\frac{1}{q^d}\sum_{x^d\in \mathbb{F}_q^d}\lambda^+_{f}(x^d)\chi_{s^d},
\qquad 
\lambda^-_{f,s^d} =\frac{1}{q^d}\sum_{x^d\in \mathbb{F}_q^d}\lambda^-_{f}(x^d)\chi_{s^d},
\\& \lambda^+_{g,s^d} =\frac{1}{q^d}\sum_{x^d\in \mathbb{F}_q^d}\lambda^+_{g}(x^d)\chi_{s^d},
\qquad \lambda^-_{g,\mathcal{S}} =\frac{1}{q^d}\sum_{x^d\in \mathbb{F}_q^d}\lambda^-_{g}(x^d)\chi_{s^d}.
\end{align*}
Then, equations \eqref{eq:stat_1} and \eqref{eq:stat_2} can be rewritten as:
\begin{align}
&\sum_{t^d\in \mathbb{F}_q^d}\tilde{g}_{t^d}\prod_{s,t}\rho_{s,t}^{n(s,t|s^d,t^d)}={\lambda^+_{f,s^d}-\lambda^-_{f,s^d}}, \quad \forall s^d\neq \mathbf{0}.
\label{eq:stat_3}\\
&\sum_{t^d\in \mathbb{F}_q^d}\tilde{f}_{t^d}\prod_{s,t}\rho_{s,t}^{n(s,t|s^d,t^d)}=\lambda^+_{g,s^d}-\lambda^-_{g,s^d},\quad \forall s^d\neq \mathbf{0}.
\label{eq:stat_4}
\end{align}
Given $\lambda^+_{f,s^d}-\lambda^-_{f,s^d}$ and 
$\lambda^+_{g,s^d}-\lambda^-_{g,s^d}, s^d\in \mathbb{F}_q^d$, let us define $\bar{f}_{s^d}(\mathbf{\lambda}),\bar{g}_{s^d}(\mathbf{\lambda})$ as the solution to Equations \eqref{eq:stat_3} and \eqref{eq:stat_4}. 

\item \textbf{Primal Feasibility:} Using equations \eqref{eq:stat_3} and \eqref{eq:stat_4}, we have:
\begin{align*}
& \forall x^d\in \mathbb{F}_q^d: |\sum_{s^d\subseteq \mathbb{F}_q^d}\tilde{f}_s^d\chi_{s^d}|   \leq 1 \Rightarrow |2Q_U(1)-1 +\sum_{s^d\neq \mathbf{0}}\bar{f}_{s^d}(\mathbf{\lambda})|\leq 1
\\&  \forall x^d\in \mathbb{F}_q^d: |\sum_{s^d\in \mathbb{F}_q^d}\tilde{g}_{s^d} \chi_{s^d}|   \leq 1 \Rightarrow |2Q_V(1)-1 +\sum_{s^d\neq \mathbf{0}}\bar{g}_{s^d}(\mathbf{\lambda})|\leq 1
\end{align*}
\item \textbf{Dual Feasibility:} 
\begin{align*}
    &\forall x^d\in \mathbb{F}_q^d: \lambda^+_f(x^d)\geq 0\Rightarrow \sum_{s^d\in \mathbb{F}_q^d}\lambda^+_{f,s^d}\chi_{s^d}\geq 0,
    \\&\forall x^d\in \mathbb{F}_q^d: \lambda^-_f(x^d)\geq 0 \Rightarrow \sum_{s^d\in \mathbb{F}_q^d}\lambda^-_{f,s^d}\chi_{s^d}\geq  0,
    \\&\forall x^d\in \mathbb{F}_q^d: \lambda^+_g(x^d)\geq 0\Rightarrow \sum_{s^d\in \mathbb{F}_q^d}\lambda^+_{g,s^d}\chi_{s^d}\geq 0,
    \\&\forall x^d\in \mathbb{F}_q^d: \lambda^-_g(x^d)\geq 0\Rightarrow \sum_{s^d\in \mathbb{F}_q^d}\lambda^-_{g,s^d}\chi_{s^d}\geq 0.
    \end{align*}
\item \textbf{Complementary Slackness:}
\begin{align*}
 &   \sum_{x^d\in \mathbb{F}_q^d}
  \lambda^+_f(x^d)(\tilde{f}(x^d)-1)
  +
  \lambda^-_f(x^d)(-\tilde{f}(x^d)-1)
  \\&\qquad \qquad 
  +
  \lambda^+_g(y^d)(\tilde{g}(y^d)-1)
 +
  \lambda^-_g(y^d)(-\tilde{g}(y^d)-1)=0
  \\&\Rightarrow 
  \sum_{s^d}(\lambda^+_{f,s^d}-\lambda^-_{f,s^d})\tilde{f}_{s^d}
  +\sum_{s^d}(\lambda^+_{g,s^d}-\lambda^-_{g,s^d})\tilde{g}_{s^d}= 
  \lambda^+_{f,\phi}+
   \lambda^-_{f,\phi}+
    \lambda^+_{g,\phi}+
     \lambda^-_{g,\phi}.
\end{align*}
From Equations \eqref{eq:stat_3} and \eqref{eq:stat_4}, we have 
\begin{align*}
&\sum_{s^d} (\lambda^+_{f,s^d}-\lambda^-_{f,s^d})\tilde{f}_{s^d}
= (\lambda^+_{f,\mathbf{0}}-\lambda^-_{f,\mathbf{0}})\tilde{f}_{\mathbf{0}}
+ \sum_{s^d,t^d} \tilde{f}_{s^d}\tilde{g}_{t^d}\prod_{s,t}\rho_{s,t}^{n(s,t|s^d,t^d)}- \tilde{f}_\mathbf{0}\tilde{g}_\mathbf{0}
\\&\sum_{s^d} (\lambda^+_{g,s^d}-\lambda^-_{g,s^d})\tilde{g}_{s^d}
= (\lambda^+_{g,\mathbf{0}}-\lambda^-_{g,\mathbf{0}})\tilde{g}_{\mathbf{0}}
+ \sum_{s^d,t^d} \tilde{f}_{s^d}\tilde{g}_{t^d}\prod_{s,t}\rho_{s,t}^{n(s,t|s^d,t^d)}- \tilde{f}_\mathbf{0}\tilde{g}_\mathbf{0}
\end{align*}
So, the complementary slackness condition yields:
\begin{align*}
 &\sum_{s^d,t^d} \tilde{f}_{s^d}\tilde{g}_{t^d}\prod_{s,t}\rho_{s,t}^{n(s,t|s^d,t^d)}=
\\& 
\frac{1}{2}((1-\tilde{f}_{\mathbf{0}}) \lambda^+_{f,\mathbf{0}}+
   (1+\tilde{f}_{\mathbf{0}}) \lambda^-_{f,\mathbf{0}}+
    (1-\tilde{g}_{\mathbf{0}}) \lambda^+_{g,\mathbf{0}}+
     (1+\tilde{g}_{\mathbf{0}}) \lambda^-_{g,\mathbf{0}}+2\tilde{f}_{\mathbf{0}}\tilde{g}_{\mathbf{0}}).
\end{align*}
The proof is completed by noting that $1-\tilde{f}_{\mathbf{0}}= 2-2Q_U(1)=2Q_U(0)$, $1+\tilde{f}_{\mathbf{0}}= 2Q_U(1)$,$1-\tilde{g}_{\mathbf{0}}= 2-2Q_V(1)=2Q_V(0)$ and $1+\tilde{g}_{\mathbf{0}}= 2Q_V(1)$.
$\qed$
\end{itemize}

\section{Proof of Proposition \ref{prop:opt}}
\label{App:prop:opt}
    For a given $\tilde{f}$ and $\tilde{g}$, define $\mathcal{L}_{0,\alpha_0,\beta_0,\tilde{f},\tilde{g}}(P_{XY},Q_{U},Q_{V})$, $\mathcal{L}_{1,\alpha_1,\beta_1,\tilde{f},\tilde{g}}(P_{XY},Q_{U},Q_{V})$, $\mathcal{L}_{\lambda,\tilde{f},\tilde{g}}(P_{XY},Q_{U},Q_{V}), \lambda\in [0,1]$
    and  as follows:
    \begin{align*}
&\mathcal{L}_{0,\alpha_0,\beta_0,\tilde{f},\tilde{g}}(P_{XY},Q_{U},Q_{V})\triangleq \sum_{s^d,t^d\in \mathbb{F}_q^d} \left(\tilde{f}_{s^d} \tilde{g}_{{t}^d} \prod_{s,t\in\mathbb{F}_q}\rho_{s,t}^{n(s,t|s^d,t^d)}
 - \alpha_0 \tilde{f}^2_{s^d}
 -\beta_0 \tilde{g}^2_{t^d}\right)+\alpha_0+\beta_0,
 \\& \mathcal{L}_{1,\alpha_1,\beta_1,\tilde{f},\tilde{g}}(P_{XY},Q_{U},Q_{V})\triangleq \sum_{s^d,t^d\in \mathbb{F}_q^d} \left(\tilde{f}_{s^d} \tilde{g}_{{t}^d} \prod_{s,t\in\mathbb{F}_q}\rho_{s,t}^{n(s,t|s^d,t^d)}
 + \alpha_1 \tilde{f}^2_{s^d}
 +\beta_1 \tilde{g}^2_{t^d}\right)-\alpha_1-\beta_1.
 \\& \mathcal{L}_{\lambda,\tilde{f},\tilde{g}}(P_{XY},Q_{U},Q_{V})\triangleq  \lambda \mathcal{L}_{0,\alpha_0,\beta_0,\tilde{f},\tilde{g}}(P_{XY},Q_{U},Q_{V})+(1-\lambda)\mathcal{L}_{1,\alpha_1,\beta_1,\tilde{f},\tilde{g}}(P_{XY},Q_{U},Q_{V}).
\end{align*}
Note that for any $(\tilde{f},\tilde{g})$ on the boundary, the function $\mathcal{L}_{\lambda,\tilde{f},\tilde{g}}(P_{XY},Q_{U},Q_{V})= \lambda \mathcal{L}_{0,\alpha_0,\beta_0,\tilde{f},\tilde{g}}(P_{XY},Q_{U},Q_{V})+(1-\lambda)\mathcal{L}_{1,\alpha_1,\beta_1,\tilde{f},\tilde{g}}(P_{XY},Q_{U},Q_{V})$ is constant in $\lambda$ and for any $(\tilde{f},\tilde{g})$ inside the optimization search space, the function is strictly decreasing in $\lambda$. Furthermore, from condition $ii)$, we conclude that $(\tilde{f}^*,\tilde{g}^*)$ is the global optimum of $\mathcal{L}_{\lambda^*}(P_{XY},Q_{U},Q_{V})$. As a result, 
\begin{align*}
 &  \mathcal{L}_{1,\tilde{f},\tilde{g}}(P_{XY},Q_{U},Q_{V})\leq \mathcal{L}_{\lambda^*,\tilde{f},\tilde{g}}(P_{XY},Q_{U},Q_{V})
   \\&\qquad \qquad \qquad \leq
   \mathcal{L}_{\lambda^*,\tilde{f}^*,\tilde{g}^*}(P_{XY},Q_{U},Q_{V})=  \mathcal{L}_{1,\tilde{f}^*,\tilde{g}^*}(P_{XY},Q_{U},Q_{V}).
\end{align*}
This completes the proof. 
$\qed$

\section{Proof of Lemma \ref{lem:td}}
\label{App:lem:td}
Note that the input distribution can be parametrized as:
\begin{align*}
    P(X=i,Y=j)= \left(\frac{1+\rho}{4}\right)\mathbbm{1}(i=j)+\left(\frac{1-\rho}{4}\right)\mathbbm{1}(i\neq j).
\end{align*}
Consequently,
\begin{align*}
  &  \mathbb{E}(f(X^d)g(Y^d))= \sum_{x^d,y^d\in \{-1,1\}^d}f(x^d)g(y^d)P_{X^dY^d}(x^d,y^d)
  \\&=
    \sum_{x^d,y^d\in \{-1,1\}^d}f(x^d)g(y^d)\left(\frac{1+\rho}{4}\right)^{d-d_H(x^d,y^d)} \left(\frac{1-\rho}{4}\right)^{d_H(x^d,y^d)}
    \\&
    = \left(\frac{1+\rho}{4}\right)^d \sum_{x^d,y^d\in \{-1,1\}^d}f(x^d)g(y^d)\left(\frac{1-\rho}{1+\rho}\right)^{d_H(x^d,y^d)}
\end{align*}
Furthermore,
\begin{align*}
 &   \sum_{x^d,y^d\in \{-1,1\}^d}f(x^d)g(y^d)\left(\frac{1-\rho}{1+\rho}\right)^{d_H(x^d,y^d)}
 \\& 
= \sum_{\substack{x^d,y^d\in \{-1,1\}^d\\ f(x^d)=g(x^d)}}\beta(x^d,y^d)
- \sum_{\substack{x^d,y^d\in \{-1,1\}^d\\ f(x^d)\neq g(y^d)}}\beta(x^d,y^d)
\\& = 
2 \sum_{\substack{x^d,y^d\in \{-1,1\}^d\\ f(x^d)=g(x^d)}}\beta(x^d,y^d)
 - \sum_{\substack{x^d,y^d\in \{-1,1\}^d}}\beta(x^d,y^d)
= 2 \sum_{\substack{x^d,y^d\in \{-1,1\}^d\\ f(x^d)=g(y^d)}}\beta(x^d,y^d)
- 2^dC_{\rho}.
\end{align*}
Additionally, 
\begin{align}
   &\nonumber \sum_{\substack{x^d,y^d\in \{-1,1\}^d\\ f(x^d)=g(y^d)}}\beta(x^d,y^d)
    = 
    \sum_{\substack{x^d,y^d\in \{-1,1\}^d\\ f(x^d)=1,g(y^d)=1}}\beta(x^d,y^d)
    +
    \sum_{\substack{x^d,y^d\in \{-1,1\}^d\\ f(x^d)=-1,g(y^d)=-1}}\beta(x^d,y^d)
    \\&\label{eq:p30:1}
    =  \sum_{\substack{x^d,y^d\in \{-1,1\}^d\\ f(x^d)=1,g(y^d)=1}}\beta(x^d,y^d)
    -
     \sum_{\substack{x^d,y^d\in \{-1,1\}^d\\ f(x^d)=1,g(y^d)=-1}}\beta(x^d,y^d)
     +
      \sum_{\substack{x^d,y^d\in \{-1,1\}^d\\ g(y^d)=-1}}\beta(x^d,y^d)
\end{align}
Note that:
\begin{align}
\label{eq:p30:2}
&  \sum_{\substack{x^d,y^d\in \{-1,1\}^d\\ g(y^d)=-1}}\beta(x^d,y^d)
= \sum_{y^d:g(y^d)=-1}\sum_{x^d\in \{-1,1\}^d} \beta(x^d,y^d)= \sum_{y^d:g(y^d)=-1}C_{\rho}
= 2^dQ_V(1)C_{\rho}.
\end{align}
Consequently, from Equations \eqref{eq:p30:1} and \eqref{eq:p30:2}, we have: 
\begin{align}
\label{eq:QV}
       & \sum_{\substack{x^d,y^d\in \{-1,1\}^d\\ f(x^d)=g(y^d)}}\beta(x^d,y^d)
    =  \sum_{\substack{x^d,y^d\in \{-1,1\}^d\\ f(x^d)=1,g(y^d)=1}}\beta(x^d,y^d)
    -
     \sum_{\substack{x^d,y^d\in \{-1,1\}^d\\ f(x^d)=1,g(y^d)=-1}}\beta(x^d,y^d)
     +2^dQ_V(1)C_{\rho}
\end{align}
On the other hand,
\begin{align}
    &\nonumber
    \sum_{\substack{x^d,y^d\in \{-1,1\}^d\\ f(x^d)=1,g(y^d)=-1}}\beta(x^d,y^d)
    =\sum_{\substack{x^d,y^d\in \{-1,1\}^d\\ f(x^d)=1}}\beta(x^d,y^d)
    -
    \sum_{\substack{x^d,y^d\in \{-1,1\}^d\\ f(x^d)=1,g(y^d)=1}}\beta(x^d,y^d)
    \\&\label{eq:QU} 
    = 2^dQ_U(1)C_{\rho}-\sum_{\substack{x^d,y^d\in \{-1,1\}^d\\ f(x^d)=1,g(y^d)=1}}\beta(x^d,y^d). 
\end{align}
From Equations \eqref{eq:QV} and \eqref{eq:QU}, we conclude that:
\begin{align*}
     & \sum_{\substack{x^d,y^d\in \{-1,1\}^d\\ f(x^d)=g(y^d)}}\beta(x^d,y^d)
     =2\sum_{\substack{x^d,y^d\in \{-1,1\}^d\\ f(x^d)=1,g(y^d)=1}}\beta(x^d,y^d)
     - 2(Q_U(1)+Q_V(1))C_{\rho}. 
\end{align*}
% \begin{align*}
%  &   \sum_{x^d,y^d\in \{-1,1\}^d}f(x^d)g(y^d)\left(\frac{1-\rho}{1+\rho}\right)^{n(x^d\neq y^d)}= 
%  \\& 
%  \sum_{\substack{x^d,y^d\in \{-1,1\}^d\\ f(x^d)=1,g(y^d)=1}}\beta(x^d,y^d)
%  +
%  \sum_{\substack{x^d,y^d\in \{-1,1\}^d\\ f(x^d)=-1,g(y^d)=-1}}\beta(x^d,y^d) 
%  \\&-
%  \sum_{\substack{x^d,y^d\in \{-1,1\}^d\\ f(x^d)=1,g(y^d)=-1}}\beta(x^d,y^d)
%  -
%  \sum_{\substack{x^d,y^d\in \{-1,1\}^d\\ f(x^d)=-1,g(y^d)=1}}\beta(x^d,y^d)
% \end{align*}

% Note that:
% \begin{align*}
%    & \sum_{\substack{x^d,y^d\in \{-1,1\}^d\\ f(x^d)=1,g(y^d)=-1}}\beta(x^d,y^d)
%     = 
%     \\&
%     \sum_{\substack{x^d,y^d\in \{-1,1\}^d\\ f(x^d)=1,g(y^d)=-1}}\beta(x^d,y^d)
%     +
%     \sum_{\substack{x^d,y^d\in \{-1,1\}^d\\ f(x^d)=1,g(y^d)=1}}\beta(x^d,y^d)
% \end{align*}
$\qed$
\section{Proof of Lemma \ref{lem:cor_pres}}
\label{App:lem:cor_pres}
\noindent \textbf{Proof of Part i):}
\\ Note that $\mathbb{E}(f(X^d))= \mathbb{E}(\Xi_k(f(X^d)))$ since $P_{X^d}(x^d)= P_{X^d}(\xi_k(x^d))$ for all $x^d\in \{-1,1\}^d$. Similarly, $\mathbb{E}(g(Y^d))= \mathbb{E}(\Xi_k(g(Y^d)))$. Furthermore,
using Proposition \ref{prop:domin},  it suffices to show that $S(f,g)\preceq S(\Xi_k(f),\Xi_k(g))$. Let $S(f,g)=(n_1,n_2,\cdots,n_d)$ and $S(\Xi_k(f),\Xi_k(g))=(n'_1,n'_2,\cdots,n'_d)$. For a given $\ell\in [d]$, we have:
\begin{align*}
 &   \sum_{k=1}^\ell n_k = \sum_{\substack{x^d,y^d\in \{-1,1\}^d\\
 f(x^d)=g(y^d)=1}}
    \mathbbm{1}(d_H(x^d,y^d)\leq \ell)
\end{align*}
\begin{align*}
&= \frac{1}{4}
 \sum_{x^d,y^d\in \{-1,1\}^d}
     \mathbbm{1}(d_H(x^d,y^d)\leq \ell)\mathbbm{1}( f(x^d)=g(y^d)=1)
     \\&+\mathbbm{1}(d_H(\xi_k(x^d),y^d)\leq \ell)\mathbbm{1}( f(\xi_k(x^d))=g(y^d)=1)
    \\&+\mathbbm{1}(d_H(x^d,\xi_k(y^d))\leq \ell)\mathbbm{1}( f(x^d)=g(\xi_k(y^d))=1)    \\&+\mathbbm{1}(d_H(\xi_k(x^d),\xi_k(y^d))\leq \ell)\mathbbm{1}( f(\xi_k(x^d))=g(\xi_k(y^d))=1),
\end{align*}
where $\xi_k(\cdot)$ is the $k$th bit-flip operator. 
Let us define:
\begin{align}
\label{eq:dist}
  &  \gamma(\ell,f,g,x^d,y^d)\triangleq 
    \mathbbm{1}(d_H(x^d,y^d)\leq \ell)\mathbbm{1}( f(x^d)=g(y^d)=1)
     \\&\nonumber+\mathbbm{1}(d_H(\xi_k(x^d),y^d)\leq \ell)\mathbbm{1}( f(\xi_k(x^d))=g(y^d)=1)
    \\&\nonumber+\mathbbm{1}(d_H(x^d,\xi_k(y^d))\leq \ell)\mathbbm{1}( f(x^d)=g(\xi_k(y^d))=1)    \\&\nonumber+\mathbbm{1}(d_H(\xi_k(x^d),\xi_k(y^d))\leq \ell)\mathbbm{1}( f(\xi_k(x^d))=g(\xi_k(y^d))=1),
\end{align}
 It suffices to show that $  \gamma(\ell,f,g,x^d,y^d)
\leq    \gamma(\ell,f',g',x^d,y^d)$ for all $x^d,y^d\in \{-1,1\}$. Without loss of generality, assume that $x_k=y_k=-1$. Then,
\[d_H(x^d,y^d)=d_H(\xi_k(x^d),\xi_k(y^d))= d_H(x^d,\xi_k(y^d))-1= d_H(\xi_k(x^d),y^d)-1.\] 
We have the following cases:
\\\textbf{Case 1:} If $\gamma(\ell,f,g,x^d,y^d)=4$, then $d_H(x^d,y^d)\leq \ell-1$, and $f(x^d)=f(\xi_k(x^d))=g(y^d)=g(\xi_k(y^d))=-1$.
From Equation \eqref{eq:proj_rule}, $\gamma(\ell,f',g',x^d,y^d)=4$.
\\\textbf{Case 2:} If $\gamma(\ell,f,g,x^d,y^d)=2$, then
$d_H(x^d,y^d)\leq \ell-1$ and  $f(x^d)=f(\xi_k(x^d))=g(y^d)=-1, g(\xi_k(y^d))=1$ or $g(y^d)=g(\xi_k(y^d))=f(x^d)=-1, f(\xi_k(x^d))=1$ or $f(x^d)=f(\xi_k(x^d))=g(\xi_k(y^d))=-1, g(y^d)=1$ or $g(y^d)=g(\xi_k(y^d))=f(\xi_k(x^d))=-1, f(x^d)=1$. Then, from Equation \eqref{eq:dist},  we conclude that $\gamma(\ell,f',g',x^d,y^d)=2$.
\\\textbf{Case 3:} If $\gamma(\ell,f,g,x^d,y^d)=1$, then 
if $g(x^d)=f(y^d)=1$ or $g(\xi_k(x^d))=f(\xi_k(y^d))=1$ and $d_H(x^d,y^d)\leq \ell$, we have $\gamma(\ell,f',g',x^d,y^d)=1$. Otherwise, if $g(\xi_k(x^d))=f(y^d)=1$ or $g(\xi_k(x^d))=f(y^d)=1$ and $d_H(x^d,y^d)\leq \ell-1$, 
we have $\gamma(\ell,f',g',x^d,y^d)\leq 1$ due to Equation \eqref{eq:dist}. 
\\\textbf{Case 4:} If $\gamma(\ell,f,g,x^d,y^d)=0$, then either  i) $d_H(x^d,y^d)> \ell$, or ii) $f(x^d)=f(\xi_k(x^d))=-1$ or iii)  $g(y^d)=g(\xi_k(y^d))=-1$, or iv)
 $d_H(x^d,y^d)=\ell$ and $f(\xi_k(x^d))=g(y^d)=1$ or $f(x^d)=g(\xi_k(y^d))=1$. In sub-cases i), ii), and iii) we have $\gamma(\ell,f',g',x^d,y^d)=0$, for sub-case vi), we have $\gamma(\ell,f',g',x^d,y^d)=1$. 
\\This conclude the proof of Part i).

\noindent \textbf{Proof of Part ii):} 

\noindent 
 Note that $\mathbb{E}(f(X^d))= \mathbb{E}(\Pi_{\pi}(f(X^d)))$ since $P_{X^d}(x^d)= P_{X^d}(\pi(x^d))$ for all $x^d\in \{-1,1\}^d$. Similarly, $\mathbb{E}(g(Y^d))= \mathbb{E}(\Pi_{\pi}(g(Y^d)))$. 
Furthermore, $d_H(x^d,y^d)= d_H(\pi(x^d),\pi(y^d))$. As a result, $S(\Pi_{\pi}(f),\Pi_{\pi}(g))= S(f,g)$, and the proof follows from Proposition \ref{prop:domin}.
$\qed$

\section{Proof of Theorem \ref{th:IC_BB_NISS}}
\label{App:th:IC_BB_NISS}
\textbf{Proof of i)}
\\ To prove that $(f_{L,d},g_{L,d})_{d\in \mathbb{N}}$ achieve biased maximal correlation, we show that they achieve the maximal correlation among all functions with the same output marginals. That is, for a given $d\in \mathbb{N}$, we start with an arbitrary pair of functions $(f_d,g_d)$, and perform correlation-preserving operations on these function pairs to arrive at $(f_{L,d},g_{L,d})$. Then, by definition of correlation-preserving operators, we conclude that $(f_{L,d},g_{L,d})$ generates output correlation at least as high as $(f_d,g_d)$. 
\\The proof follows by Noetherian induction. To elaborate, for a given $d\in \mathbb{N}$, we parameterize the set of output marginals which can be generated by inputs of length $d$ by parameters $(n_u,n_v)$, where $P(U=1)=\frac{n_u}{2^d}$ and $P(V=1)= \frac{n_v}{2^d}, n_u,n_v\in \{0\}\cup [2^d]$. We cosnider the set of vectors $\mathcal{N}= \{(d,n_u,n_v)| d\in \mathbb{N},  n_u,n_v\in \{0\}\cup [2^d]\}$ equipped with the ordering relation 
\[(d,n_u,n_v)\prec_{ind}  \ (d',n'_u,n'_v)
\iff
(d< d')\lor (d=d'\land n_u< n'_u) \lor (d=d' \land n_u=n'_u \land n_v< n'_v).  
\]
We perform Noetherian induction on $\mathcal{N}$:
\\\textbf{Induction Basis.} We base the induction on proving the claim for $d=1$ and $n_u,n_v\in [2]$. That is, we prove the claim for all four cases directly to form the induction basis. 
If $d=n_u=n_v=1$, then there are four possible functions $(f(X),g(Y))$ achieving output marginals $Q_U(1)=Q_V(1)=\frac{1}{2}$.  The maximal correlation is achieved by $f(X)=\mathbbm{1}(X=-1)$ and $g(Y)=\mathbbm{1}(Y=-1)$ since this is a dominating pair of functions in the sense of Definition \ref{def:domin}. To see the later statement it suffices to note that $D_1 = \begin{bmatrix} 0 & 1 \\ 1 & 0 \end{bmatrix}$ as stated in Lemma \ref{lem:dist_mat}. For the cases where  $d=n_u=1, n_v=2$ and  $d=n_v=1, n_u=2$, the function $g(y)=1, y\in \{-1,1\}$ and $f(x)=1, x\in \{-1,1\}$ are deterministic, respectively, and the choice of $f(\cdot)$ and $g(\cdot)$ does not affect correlation. Lastly, for $d=1, n_u=n_v=2$, the only choice of simulating functions is $f(x)=g(y)=1, x,y\in \{-1,1\}$. This proves the induction basis. 
\\\textbf{Induction Step.} For a given $(d,n_u,n_v)$, we assume that the claim is proved for all $(d',n'_u,n'_v)\prec_{ind}(d,n_u,n_v)$. Let $f_d,g_d:\{-1,1\}^d\to \{-1,1\}$ be an arbitrary pair of functions with $|\{x^d| f_d(x^d)=1\}|=n_u$ and $|\{y^d| g_d(y^d)=1\}|=n_v$. We perform the following sequence of correlation-preserving operations:
\\\textbf{Step 0:} Let 
\begin{align*}
&n_{u,-1}\triangleq |\{x^d| f_d(x^d)=1, x_1=-1\}|,\quad  n_{u,1}\triangleq |\{x^d| f_d(x^d)=1, x_1=1\}|,
\\& n_{v,-1}\triangleq |\{y^d| g_d(y^d)=1, y_1=-1\}|,
\quad n_{v,1}\triangleq |\{y^d| g_d(y^d)=1, y_1=1\}|.
\end{align*}
Define $f_d^{(0)},g_d^{(0)}$ as follows: 
\begin{align*}
   & f_d^{(0)}(x^d) \triangleq 
    \begin{cases}
        f_{L,d-1}^{(-1)}(x_2^{d})\qquad &\text{ if } x_1=-1\\
        f_{L,d-1}^{(1)}(x_2^d)& \text{ if } x_1=1 
    \end{cases},
    \\&
    g_d^{(0)}(y^d) \triangleq 
    \begin{cases}
        g_{L,d-1}^{(-1)}(y_2^{d})\qquad &\text{ if } y_1=-1\\
        g_{L,d-1}^{(1)}(y_2^d)& \text{ if } y_1=1 
    \end{cases},
\end{align*}
where $x_2^d\triangleq (x_2,x_3,\cdots,x_d)$, $y_2^d\triangleq (y_2,y_3,\cdots,y_d)$, 
$(f_{L,d-1}^{(\alpha)}, g_{L,d-1}^{(\beta)})$ are lexicographic functions with input length $d-1$, associated with $Q_U,Q_V$, for $Q_U(1)= \frac{n_{u,\alpha}}{2^{d-1}}$ and $Q_V(1)= \frac{n_{v,\beta}}{2^{d-1}}$ and $\alpha,\beta\in \{-1,1\}$. We show using the induction hypothesis that the operator $\Gamma:(f_d,g_d)\mapsto (f^{(0)}_d,g^{(0)}_d)$ is correlation-preserving. To see this, define $f_{d-1}^{(\alpha)}(x_2^d)\triangleq  f_d(\alpha,x_2^d)$ and $g_{d-1}^{(\alpha)}(y_2^d)\triangleq  g_d(\alpha,y_2^d), \alpha\in \{-1,1\}$.  From Equation \eqref{eq:dist_1}, we have:
\begin{align*}
&    S(f_d,g_d)=\sum_{\alpha\in \{-1,1\}} T_d(S(f^{(\alpha)}_{d-1},g^{(\alpha)}_{d-1}))+ T_1(S(f^{(\alpha)}_{d-1},g^{(\bar{\alpha})}_{d-1})),
\\&
    S(f^{(0)}_d,g^{(0)}_d)=\sum_{\alpha\in \{-1,1\}} T_d(S(f^{(\alpha)}_{L,d-1},g^{(\alpha)}_{L,d-1}))+ T_1(S(f^{(\alpha)}_{L,d-1},g^{(\bar{\alpha})}_{L,d-1})),
\end{align*}
where $\bar{\alpha}=\alpha\oplus 1$ and $T_i:\mathbb{N}^{d-1}\to \mathbb{N}^d, i\in [d]$ is the insertion operator which inserts $0$ in the $i$th position of its input, i.e.,
\begin{align*}
    \forall a^{d-1}\in \mathbb{N}^{d-1}: b^d=T_i(a^{d-1}) \iff b_j= a_j\mathbbm{1}(j<i)+ a_{j-1}\mathbbm{1}(j\geq i), j\in [d]-\{i\} \text{ and } b_i=0.  
\end{align*}
Furthermore, from the induction hypothesis, we have $S(f^{(\alpha)}_{d-1},g^{(\beta)}_{d-1})\preceq S(f^{(\alpha)}_{L,d-1},g^{(\beta)}_{L,d-1}), \alpha,\beta\in \{-1,1\}$. We conclude that $\Gamma$ is correlation-preserving.

\noindent\textbf{Step 1:} We generate $f^{(1)}_d(X^d)= \Pi_{(1,d)}(f^{(0)}_d(X^d))$ and $g^{(1)}_d(Y^d)= \Pi_{(1,d)}(g^{(0)}_d(Y^d))$, where $(1,d)$ denotes the permutation on $[d]$ which switches $1$ and $d$ and fixed all other values. This is a correlation-preserving operation by Lemma \ref{lem:cor_pres}.

\noindent\textbf{Step 2:}  We generate $f^{(2)}_d(X^d)= \Xi_{1}(f^{(1)}_d(X^d))$ and $g^{(2)}_d(Y^d)= \Xi_{1}(g^{(1)}_d(Y^d))$. This is a correlation-preserving operation by Lemma \ref{lem:cor_pres}. 

Let 
\begin{align*}
&n^{(2)}_{u,\alpha}\triangleq |\{x^d| f^{(2)}_d(x^d)=1, x_1=\alpha\}|,\quad  
 n^{(2)}_{v,\alpha}\triangleq |\{y^d| g^{(2)}_d(y^d)=1, y_1=\alpha\}|, \alpha\in \{-1,1\}.
\end{align*}

Note that by construction 
\begin{align}
\label{eq:mid1}
    &n^{(2)}_{u,-1}= \lceil \frac{n_{u,-1}+n_{u,1}}{2}\rceil,\quad  n^{(2)}_{u,1}= \lfloor \frac{n_{u,-1}+n_{u,1}}{2}\rfloor
    \\&\label{eq:mid2}
    n^{(2)}_{v,-1}= \lceil \frac{n_{v,-1}+n_{v,1}}{2}\rceil, \quad n^{(2)}_{v,1}= \lfloor \frac{n_{v,-1}+n_{v,1}}{2}\rfloor\end{align}

\noindent \textbf{Step 3:}   We generate $f^{(3)}_d(X^d)= \Gamma(f^{(2)}_d(X^d))$ and $g^{(3)}_d(Y^d)= \Gamma(g^{(2)}_d(Y^d))$ as in Step 0 to generate piecewise lexicographic functions. This is a correlation-preserving operation as shown in Step 0.

\noindent \textbf{Step 3:} 
 We generate $f^{(4)}_d(X^d)= \Pi_{(1,2)}(f^{(3)}_d(X^d))$ and $g^{(4)}_d(Y^d)= \Pi_{(1,2)}(g^{(3)}_d(Y^d))$, where $(1,2)$ is the permutation on $[d]$ which switches $1$ and $2$ and fixes all other values. This is a correlation-preserving operation by Lemma \ref{lem:cor_pres}. 
 
 The proof of Part i) follows by considering four cases. We provide the complete proof for the first case, and an outline of the proof for the other three cases, which follow from the same line of argument:
 \\\textbf{Case 1: $n_u,n_v\leq 2^{d-1}$}
 \\ In this case, the output of the shuffling operation is as follows: 
\begin{align*}
 &   f^{(4)}_d(x^d)=
    \begin{cases}
       f^{(3)}(-1,-1,x_3^d)\qquad  &\text{ if } x_1=x_2=-1
    \\ f^{(3)}(1,-1,x_3^d)
        & \text{ if } x_1=-1,x_2=1\\
        -1& \text{ otherwise}
    \end{cases}
    \\&  g^{(4)}_d(y^d)=
    \begin{cases}
       g^{(3)}(-1,-1,y_3^d)\qquad  &\text{ if } y_1=y_2=-1
    \\ g^{(3)}(1,-1,y_3^d)
        & \text{ if } y_1=-1,y_2=1\\
        -1& \text{ otherwise}
    \end{cases},
\end{align*}
The pair of equalities hold since $n_u,n_v\leq 2^{d-1}$ and Equations \eqref{eq:mid1} and \eqref{eq:mid2} imply that $n^{(2)}_{u,\alpha}, n^{(2)}_{v,\alpha}\leq 2^{d-2}, \alpha\in \{-1,1\}$. Consequently, since $f^{(3)}$ is piecewise lexicographic, we conclude that $f^{(3)}(x^d)=-1$ if $x_1=-1,x_2=1$ or $x_1=1,x_2=1$. After the shuffling operation, this yields $f^{(4)}(x^d)= -1$ if   $x_1=1,x_2=-1$ or $x_1=1,x_2=1$. Similarly, $g^{(4)}(y^d)= -1$ if   $y_1=1,y_2=-1$ or $y_1=1,y_2=1$. Let us define $f^{(4)}_{d-1}(x_2^d)\triangleq f^{(4)}_d(-1,x_2^d)$ and $g^{(4)}_{d-1}(y_2^d)\triangleq g^{(4)}_d(-1,y_2^d)$. 
Note that $S(f^{(4)}_{d},g^{(4)}_{d})= S(f^{(4)}_{d-1},g^{(4)}_{d-1})$. On the other hand,
by the induction hypothesis, $S(f^{(4)}_{d-1},g^{(4)}_{d-1})\preceq S(f^{(4)}_{L,d-1},g^{(4)}_{L,d-1})$, where $f^{(4)}_{L,d-1},g^{(4)}_{L,d-1}$ are the lexicographic functions with the same output bias as that of $(f^{(4)}_{d-1},g^{(4)}_{d-1})$. Note that $f^{(4)}_{L,d}(x^d)= f^{(4)}_{L,d-1}(-1,x^d)$ and $g^{(4)}_{L,d}(y^d)= g^{(4)}_{L,d-1}(-1,y^d)$, so that the spectrum of $(f^{(4)}_{L,d}(x^d),g^{(4)}_{L,d}(y^d))$ is equal to that of $(f^{(4)}_{L,d-1},g^{(4)}_{L,d-1})$ concatenated with $n_d=0$. Consequently, $S(f_d^{(4)},g_d^{(4)})\preceq S(f_{L,d}^{(4)},g_{L,d}^{(4)})$. This completes the proof for Case 1. 
\\\textbf{Case 2:  $n_u\leq 2^{d-1}$, $n_v>2^{d-1}$}
 \\ In this case, the output of the shuffling operation is as follows: 
\begin{align*}
 &   f^{(4)}_d(x^d)=
    \begin{cases}
       f^{(3)}(-1,-1,x_3^d)\qquad  &\text{ if } x_1=x_2=-1
    \\ f^{(3)}(1,-1,x_3^d)
        & \text{ if } x_1=-1,x_2=1\\
        -1& \text{ otherwise}
    \end{cases}
    \\&  g^{(4)}_d(y^d)=
    \begin{cases}
       1\qquad  &\text{ if } y_1=-1
    \\ g^{(3)}(-1,1,y_3^d)
        & \text{ if } y_1=1,y_2=-1\\
        -1& \text{ otherwise}
    \end{cases},
\end{align*}
By the same argument as in the previous case, using the induction hypothesis $ f^{(3)}(-1,-1,x_3^d)$,  $f^{(3)}(1,-1,x_3^d)$, and $g^{(3)}(-1,1,y_3^d)$ can be substituted with corresponding lexicographic functions in a correlation-preserving manner. Since $g^{(4)}_d(-1,y_2^d)=1, y_2^d\in \{-1,1\}^{d-1}$,  the resulting functions after the substitution are lexicographic functions. 
\\\textbf{Case 3:  $n_u> 2^{d-1}$, $n_v\leq 2^{d-1}$}
\\ In this case, the output of the shuffling operation is as follows: 
\begin{align*}
 &   f^{(4)}_d(x^d)=
    \begin{cases}
       1 \qquad  &\text{ if } x_1=-1
    \\ f^{(3)}(-1,1,x_3^d)
        & \text{ if } x_1=1,x_2=-1\\
        -1& \text{ otherwise}
    \end{cases}
    \\&  g^{(4)}_d(y^d)=
    \begin{cases}
        g^{(3)}(-1,-1,y_3^d)\qquad  &\text{ if } y_1=y_2=-1
    \\ g^{(3)}(1,-1,y_3^d)
        & \text{ if } y_1=-1,y_2=1\\
        -1& \text{ otherwise}
    \end{cases},
\end{align*}
The proof follows by symmetry between Case 2 and Case 3, and following the same line of argument. 
\\\textbf{Case 4:  $n_u> 2^{d-1}$, $n_v> 2^{d-1}$}
\\\begin{align*}
 &   f^{(4)}_d(x^d)=
    \begin{cases}
       1 \qquad  &\text{ if } x_1=-1
    \\ f^{(3)}(-1,1,x_3^d)
        & \text{ if } x_1=1,x_2=-1\\
        -1& \text{ otherwise}
    \end{cases}
    \\&  g^{(4)}_d(y^d)=
    \begin{cases}
        1\qquad  &\text{ if } y_1=-1
    \\ g^{(3)}(-1,1,y_3^d)
        & \text{ if } y_1=1,y_2=-1\\
        -1& \text{ otherwise}
    \end{cases},
\end{align*}
In this case $f^{(3)}(-1,1,x_3^d)$ and $ g^{(3)}(-1,1,y_3^d)$ can be substituted by lexicographic functions in a correlation-preserving manner, using the induction hypothesis, yielding a pair of lexicographic function operating on sequences of length $d$. The proof follows by a similar line of argument as in Case 1. 
\\\textbf{Proof of Part ii)} 
\\ Let $U_d= f_{L,d}(X^d)$ and $V_d= g_{L,d}(Y^d), d\in \mathbb{N}$, following the result of Part i), it suffices to show that for a fixed $d\in \mathbb{N}$:
\begin{align}
\label{eq:temp}
\lim_{d'\to \infty} d_{TV}(P_{U_dV_d},P_{U_{d'},V_{d'}})\leq c2^{-d},
\end{align}
then it follows from the triangle inequality and taking $d'\to \infty$ that $d_{TV}(P_{U_dV_d},Q^{*}_{UV})\leq c2^{-d}$. To prove Equation \eqref{eq:temp}, let $d'>d$ and define $f_{d'}(x^{d'})\triangleq  f_{L,d}(x^{d}), x^{d'}\in \{-1,1\}^{d'}$, that is $f_{d'}(x^{d'})$ applies a lexicographic function on the first $d$ elements of $x^{d'}$. Then, it follows by construction that $|\{x^{d'}| f_{L,d'}(x^{d'})\neq f'_{d'}(x^{d'})\}|\leq 2^{d'-d}$. Consequently, $P(f_{L,d'}(X^d)\neq f'_{d'}(X^{d'}))\leq 2^{-d}$. Similarly, we define $g_{d'}(y^{d'})\triangleq  g_{L,d}(y^{d}), y^{d'}\in \{-1,1\}^{d'}$. Then, $P(g_{L,d'}(Y^d))\neq g'_{d'}(Y^{d'}))\leq 2^{-d}$. As a result:
\begin{align*}
    |P_{U_dV_d}(\alpha,\beta)- P_{U_{d'},V_{d'}}(\alpha,\beta)|\leq P(U_d\neq U_{d'})+P(V_d\neq V_{d'})\leq 2\times 2^{-d}, \quad \forall \alpha,\beta\in \{-1,1\}. 
\end{align*}
This completes the proof of Part ii). 
\\\textbf{Proof of Part iii)} From the proof of Proposition \ref{prop:Sol_BB_NISS}, it follows that the input complexity is upper-bounded by the summation of the number of samples needed to generate the biased maximal correlation distribution and the number of samples needed to generate the random coins. For symmetric input distributions, the number of samples needed to generate random coins with total variation distance less than $\epsilon$ is $O(\log{\frac{1}{\epsilon}})$. The reason is that to generate a coin with bias $p\in [0,1]$ one can take $d=\lceil\log{\frac{1}{\epsilon}}\rceil$ samples of the input $X^d$,  and generate the lexicographic function $f_{L,d}(X^d)$ associated with bias $Q_U(1)= \lceil 2^d p\rceil$ thus achieving a variation distance $2^{-d}<\epsilon$. Furthermore, from the proof of Part ii), the number of input samples needed to generate maximal correlation is $O(\log{\frac{1}{\epsilon}})$, and we have argued that the number of input samples needed to generate each random coin is also   $O(\log{\frac{1}{\epsilon}})$ yielding an upper-bound of $O(\log{\frac{1}{\epsilon}})$ on the input complexity. On the other hand, to derive a lower-bound on input complexity, given $d$ input samples with uniform marginals, let $\epsilon=2^{-d'}$ and the target marginal distributions be $Q_U(1)=Q_V(1)= \frac{1}{2^{d'}}$.  Note that using $d$ input samples, the distributions that can be generated at each terminal are quantized with step-size $2^{-d}$
So, the closest marginal distributions that can be generated are $Q_U(1)=0, Q_V(1)=0, Q_U(1)=2^{-d}$ and $Q_{V}(1)=2^{-d}$. Since by the triangle inequality, we have $d_{TV}(Q_{UV},P_{UV})\geq \max(d_{TV}(Q_U,P_U),d_{TV}(Q_V,P_V))$, it follows that  
$d_{TV}(Q_{UV},P_{UV})\geq 2^{-d}$. Hence, to achieve a total variation distance less than $\epsilon$, the agents need at least $2^{-d'}$ input samples. Since $d' =\log{\frac{1}{\epsilon}}$, we conclude that the input complexity is $\Omega(\log{\frac{1}{\epsilon}}$. This, along with the fact that input complexity is $O(\log{\frac{1}{\epsilon}})$ imlpies that input complexity is $\Theta(\log{\frac{1}{\epsilon}})$.$\qed$
\newpage

\bibliographystyle{unsrt}
  \bibliography{References}

\end{document}